%% file: main.tex
\documentclass[aps,prd,groupedaddress,nofootinbib,superscriptaddress,preprint,11pt,a4paper]{revtex4-1}

\usepackage[colorlinks=true,citecolor=blue,linkcolor=blue, breaklinks=true]{hyperref}
\usepackage{url}
\usepackage{array}
\usepackage{color}
\usepackage[pdftex]{graphicx}
\usepackage{textcomp}
\usepackage{amsmath}
\usepackage{mathtools}
\usepackage{amsfonts}
\usepackage{amssymb}
\usepackage{cancel}
\usepackage{multirow}
\usepackage[hmargin=1in, vmargin=1in]{geometry}
\usepackage{setspace}
\usepackage{epstopdf}
\usepackage{csquotes}
\usepackage{float}
\usepackage{notoccite}
\usepackage{xspace}
\usepackage[colorinlistoftodos]{todonotes}
\usepackage{orcidlink}

\input{mycommands}

\begin{document}

\setlength{\marginparwidth}{2cm}
\setstretch{1.0}

\preprint{IMSc/2025/04}

\title{Improving sensitivity of vectorlike top partner searches with jet substructure}

\author{Anupam Ghosh\,\orcidlink{0000-0003-4163-4491}\,}
\email{anupamg@rnd.iitg.ac.in}
\affiliation{Department of Physics, Indian Institute of Technology Guwahati, North Guwahati, 781039, Assam, India}

\author{Soumyadip Ghosh}
\email{sg22rs009@iiserkol.ac.in}
\affiliation{Department of Physical Sciences, Indian Institute of Science Education and Research Kolkata, Mohanpur, 741246, Nadia, India}

\author{Soureek Mitra\,\orcidlink{0000-0002-3060-2278}\,}
\email{soureekmitra@iisermohali.ac.in}
\affiliation{Department of Physical Sciences, Indian Institute of Science Education and Research, Knowledge City, Sector 81, S.~A.~S.~Nagar, Manauli PO 140306, Punjab, India}

\author{Tousik Samui\,\orcidlink{0000-0002-1485-6155}\,}
\email{tousiks@imsc.res.in}
% \email{tousiksamui@gmail.com}
\affiliation{The Institute of Mathematical Sciences, IV Cross Road, CIT Campus, Taramani, Chennai 600113, India}

\author{Ritesh K. Singh\,\orcidlink{0000-0001-7838-6191}\,}
\email{ritesh.singh@iiserkol.ac.in}
\affiliation{Department of Physical Sciences, Indian Institute of Science Education and Research Kolkata, Mohanpur, 741246, Nadia, India}

\begin{abstract}
Vectorlike quark partners appear in many BSM models and remain an important area of research, as they can offer insights into the electroweak symmetry breaking mechanism. 
In this work, we have focused on studying the production of a heavy vectorlike top partner in association with a SM top quark via chromomagnetic coupling and the four decay modes of the top partner, namely, bW, tZ, th, and tg. 
The signal has been studied in final states with one fat jet, at least one b-tagged jet, one lepton, and missing energy.
This study focuses on the extensive use of jet substructure techniques in jets clustered with fixed and dynamically varying radius to deal with events containing differently sized jets. 
Important kinematic information, along with jet substructure and event shape observables, has been used in a multivariate analysis to extract signal with high significance. 
A comparative study between fixed and dynamically varying radius clustering of jets is also presented, leading to an improvement in signal sensitivity in the highly boosted scenario.
\end{abstract}

\maketitle

%\listoftables
%\listoffigures
%\listoftodos
%\tableofcontents
%\clearpage
%\newpage

\section{\label{sec:intro}Introduction}\vspace{-2mm}
\input{intro}

\vspace{-5mm}
\section{\label{sec:signal_bkg}Signal and Background processes}
\input{signal_bkg}

\vspace{-3mm}
\section{\label{sec:jetalgo}Jet Algorithm and Substructure Methods}\vspace{-1mm}
\subsection{\label{ssec:drjetsalgo}Dynamic radius jet clustering algorithm}\vspace{-2mm}
\input{DRjetsAlgo}
% \vspace{-3mm}
\subsection{\label{ssec:jetsubstr}Jet substructure and parameter choices}%\vspace{-2mm}
\input{boostedjetparams}
%\clearpage
%\newpage
\vspace{-3mm}
\section{\label{sec:evtsel}Event Reconstruction and Selection}\vspace{-2mm}
\input{selection}

%\clearpage
%\newpage
\vspace{-3mm}
\section{\label{sec:result}Results}\vspace{-2mm}
\input{result}

\vspace{-3mm}
\section{\label{sec:summary}Summary and Outlook}\vspace{-2mm}
\input{summary}

\vspace{-3mm}
\section*{Acknowledgments}\vspace{-2mm}
\input{acknowledgement}

\vspace{-3mm}
\section*{Data Availability Statement}\vspace{-2mm}
Some of the data that support the findings of this article are openly available~\cite{ghosh_2026_18407474}. 
All simulated data corresponding to the findings of this article are not publicly available.
The data are available from the authors upon reasonable request.

%This study is based on a combination of simulated data generated using standard HEP packages and simulated data from CMS Open Data repository.  
%The details of the simulation process that support our findings are provided within the article. 
%The numerical data corresponding to Figs.~\ref{fig:Xsec-BR},  \ref{fig:significance},  and \ref{fig:upperLimits} have been deposited in Zenodo and are publicly available~\cite{ghosh_2026_18407474}.

%\clearpage
%\newpage
\input{ref.bbl}

%\clearpage
%\newpage
\appendix
\input{app}

\end{document}

%% file: mycommands.tex
\newcommand{\pt}{\ensuremath{\text{p}_{\text{T}}}\xspace}
\newcommand{\sqpt}{\ensuremath{\text{p}_{\text{T}}^{2}}\xspace}

\newcommand{\pthat}{\ensuremath{\hat{\text{p}}_{\text{T}}}\xspace}
\newcommand{\ptmiss}{\ensuremath{\text{p}_{\text{T}}^{\text{miss}}}\xspace}

\newcommand{\kt}{\ensuremath{\text{k}_{\text{T}}}\xspace}
\newcommand{\HT}{\ensuremath{\text{H}_{\text{T}}}\xspace}

\newcommand{\mT}{\ensuremath{\text{m}_{\text{T}}}\xspace}
\newcommand{\relIso}{\ensuremath{\text{I}_{\text{rel}}}\xspace}

\newcommand{\ttbar}{\ensuremath{\mathrm{t\bar{t}}}\xspace}
\newcommand{\TTW}{\ensuremath{\mathrm{t\bar{t}W}}\xspace}
\newcommand{\TTZ}{\ensuremath{\mathrm{t\bar{t}Z}}\xspace}

\newcommand{\dsum}{\displaystyle\sum}
\newcommand{\sumab}{\dsum_{a<b}}
\newcommand{\pta}{\ensuremath{{\rm p}_{{\rm T},a}}\xspace}
\newcommand{\ptb}{\ensuremath{{\rm p}_{{\rm T},b}}\xspace}
\newcommand{\dRab}{\ensuremath{\Delta \mathrm{R}_{ab}}\xspace}

%% file: intro.tex
The search for new physics beyond the standard model (BSM) remains one of the most important goals of current and future high-energy collider experiments. 
Despite many successes, the standard model (SM) of elementary particles leaves several fundamental questions unanswered, such as the nature of electroweak symmetry breaking and the stability of the electroweak vacuum~\cite{Giudice:2013yca, Arkani-Hamed:2012dcq, Alekhin:2012py, Degrassi:2012ry, Dawson:2010jx}, the origin of mass hierarchies~\cite{Froggatt:1978nt,Fritzsch:1999ee}, matter-antimatter asymmetry~\cite{Riotto:1999yt,Dine:2003ax}, the composition of dark matter~\cite{Clowe:2006eq,Sofue:2000jx}, etc. 
Many BSM models~\cite{Perelstein:2003wd, Matsedonskyi:2012ym, Chanowitz:1978mv} propose the existence of new resonances, such as vectorlike fermions~\cite{Aguilar-Saavedra:2013qpa, Falkowski:2013jya, Dawson:2012di, delAguila:1989rq}, that can provide critical insights into some of these open problems and offer experimentally verifiable predictions at the TeV scale~\cite{CMS:2024bni}.
Vectorlike fermions are directly related to the stability of the electroweak vacuum~\cite{Hiller:2022rla}.
A vectorlike top partner, being a colour triplet and having the same electric charge as the SM top quark, couples to the Higgs boson and alters the renormalization group evolution of the Higgs quartic coupling. 
Its Yukawa interaction contributes positively to the running of the Higgs quartic coupling, while its mixing with the SM top quark reduces the Yukawa contribution due to the top quark that typically drives the coupling negative. 
Together, these effects slow down or prevent the Higgs quartic coupling from turning negative at high energy scales, thus stabilizing the electroweak vacuum all the way up to the Planck scale~\cite{Hiller:2022rla}.
In addition, coloured vectorlike fermions can mediate interactions between dark matter and SM quarks as indicated in the phenomenological analyses of vectorlike portal dark matter models~\cite{Ghosh:2024boo,Colucci:2018vxz,Ghosh:2025agw,Ghosh:2022rta,Ghosh:2023xhs}.
Thus, identifying such BSM states would not only signify the discovery of new physics phenomena but also provide a profound understanding of the structure of fundamental interactions.  

Extraction of rare BSM signals from overwhelming SM backgrounds requires improved analysis techniques and theoretical modelling. 
State-of-the-art jet substructure methods~\cite{Larkoski:2017jix,Marzani:2019hun,Kogler:2018hem}, clustering algorithms~\cite{Salam:2010nqg, Sapeta:2015gee}, and machine learning techniques~\cite{Duarte:2024lsg, Radovic:2018dip} have become indispensable tools for identifying boosted heavy resonances and complex event topologies that are characteristic of various BSM signatures at the Large Hadron Collider (LHC). 
At the same time, access to high-quality data and simulation samples, such as CMS Open Data~\cite{CMS_Open_Data, CERN-OPEN-2020-013}, allows for a more reliable background model and realistic performance estimates.

The nature of modern BSM searches requires exploration at increasingly higher energy regimes.
At these high energies, fat jets frequently arise from the hadronic decays of boosted heavy particles such as the top quarks or the heavy bosons, after parton showering and hadronization.
One of the primary aspects of this paper is to assess the efficacy of the recently proposed ``dynamic-radius'' (DR) jet clustering algorithm~\cite{Mukhopadhyaya:2023rsb}, which is advantageous in the higher-energy regime where such fat jets are prominent.
The substructure of the fat jets originating from massive boosted particle decays differs significantly from that of narrow QCD jets~\cite{Sterman:1977wj} owing to their decay history.
The DR algorithm leverages this substructure -- more precisely, combinations of energy correlation functions~\cite{Larkoski:2013eya} to adaptively set the jet radius based on local kinematics, thus producing variable-sized jets in an event. 
Some recent studies have made use of the DR clustering algorithm with a zero pile up scenario~\cite{Kar:2022hxn,Mukhopadhyaya:2023akv,Ghosh:2025gue}.
However, understanding the performance of this algorithm in realistic experimental conditions remains extremely crucial. 

In this work, we investigate the production of a heavy vectorlike top partner ($T$) in association with a SM top quark, mediated through chromomagnetic interactions~\cite{Buchkremer:2013bha, Choudhury:2021nib, Balaji:2021lpr}.
The top partner decays through four main channels:  $\rm tZ$, $\rm th$, $\rm tg$, and $\rm bW$; each leading to distinct final state signatures.\footnote{In this article, we followed a convention to denote all SM particles in roman and BSM particles in italics. However, the associated fields of both SM and BSM particles are denoted in italics.} 
Among these, direct searches at the LHC have been performed primarily in the $\rm tZ$ and $\rm th$ channels, leading to stringent lower bounds on the mass of the top partner around a few TeV~\cite{ATLAS:2024xne, CMS:2023agg, CMS:2024qdd}. 
Furthermore, indirect constraints on the BSM nature of the $\rm tg$ coupling have been derived using \ttbar events~\cite{Hayreter:2015ryk, CMS:2022voq}. 
However, only a handful of direct searches in the $\rm bW$ decay mode have been performed at the LHC with full Run-II data~\cite{CMS:2022fck, CMS-PAS-B2G-23-009} due to its limited sensitivity.
The final state for this study consists of exactly one lepton, exactly one fat jet, at least one narrow jet that is b tagged, and missing transverse momentum (\ptmiss), capturing event topologies sensitive to the presence of vectorlike top partners.
The bW decay mode of the $T$ quark constitutes a significant contribution to this specific final state. 
The probe for the signal process is conducted using jets clustered using traditional fixed radius as well as the more recent DR approaches.   
We employ powerful jet substructure tools, namely Soft Drop (SD) grooming~\cite{Larkoski:2014wba}, $N$ subjettiness~\cite{Thaler:2010tr}, and energy correlation functions (ECFs)~\cite{Larkoski:2013eya} to better understand the nature and the behaviour of the fat jet in the event. 
In addition, we also studied several event shape variables, such as Sphericity~\cite{Bjorken:1969wi, Donoghue:1979vi}, Aplanarity~\cite{ATLAS:2012tch}, Fox-Wolfram moments ($H$)~\cite{Fox:1978vw, Bernaciak:2012nh}, etc., along with lepton and jet kinematic information in a multivariate analysis (MVA), to effectively distinguish signal events from the background.
The signal events are simulated with \textsc{MadGraph5\_aMC@NLO}~\cite{Alwall:2014hca,Frederix:2018nkq} together with \textsc{Pythia8}~\cite{Bierlich:2022pfr} for hadronization and parton showering and \textsc{Delphes}~\cite{deFavereau:2013fsa} for detector response modelling. 
For SM backgrounds, high-statistics Monte Carlo (MC) samples from CMS Open Data have been utilized for a reliable estimate.
This study also includes pile up simulation with a realistic value of the mean pile up interactions, considering the Run-II LHC conditions.

Before proceeding further, it should be noted that the main goal of this work is to explore the usefulness of the DR jet clustering in highly boosted final states and to compare its performance with standard fixed-radius clustering. 
Impact of various uncertainty sources, such as pile up contamination removal, jet energy scale and resolution, hadronization model, etc., on the DR algorithm requires dedicated in-depth studies that are beyond the scope of the present work. 
This study is intended as a proof of concept, focusing on qualitative improvements achievable with DR jet clustering using a purely statistical approach.
This is an initial step to demonstrate the applicability of the DR jet clustering, considering a well-motivated candidate BSM process involving a heavy top partner coupled strongly with the SM top quark via chromomagnetic coupling. 

The article is organized as follows. 
Section~\ref{ssec:sig_model} introduces the signal model based on chromomagnetic coupling between the vectorlike top partner and the SM top quark, while the description of signal and background simulation procedures is presented in  Sections~\ref{ssec:sig_simu} and \ref{ssec:bkgMC}, respectively. 
The DR algorithm and relevant jet substructure techniques are presented in Section~\ref{sec:jetalgo} together with a control study using \ttbar + jets events. 
Section~\ref{sec:evtsel} discusses the event reconstruction and selection strategy. 
In Section~\ref{sec:result}, we detail the multivariate analysis strategy, optimization methods, and present our main results. 
Finally, Section~\ref{sec:summary} provides a summary of the work and an outlook towards the future prospects. 
%\clearpage

%% file: signal_bkg.tex
\vspace{-2mm}
\subsection{\label{ssec:sig_model}Signal model}\vspace{-2mm}
The vectorlike $T$ quark is a triplet under $SU(3)$, and its electric charge is determined by its hypercharge and isospin structure. 
Additionally, it can couple with any generation of the SM quark. 
We assume the heavy top partner predominantly interacts with the third-generation SM quarks. 
The $T$ quark is a singlet under weak isospin and has the same electromagnetic charge as the top quark; hence, it mixes with the top quark. The general Lagrangian terms are given below~\cite{Buchkremer:2013bha, Choudhury:2021nib, Balaji:2021lpr}.
\begin{align}
\label{eq:TWb}
\mathcal{L}_{TWb} & =-\dfrac{g_w~\sin\theta_L}{\sqrt{2}}~ \overline{T}~\gamma^\mu P_L b W_\mu +h.c.,\\
 \label{eq:TZt}
\mathcal{L}_{TZt} & =\dfrac{g_w}{2 c_w}~ \overline{t}~\gamma^\mu \left(2 \sin\theta_L \cos\theta_L P_L +\sin\theta_R \cos\theta_R P_R \right)~ T  Z_\mu +h.c.,\\
\label{eq:Tht}
\mathcal{L}_{Tht} & =-s_w\dfrac{\sin\theta_R \cos\theta_R}{2 M_W}~ \overline{t}~\left(M_T  P_L +m_t P_R \right)~ T~h +h.c. 
\end{align}
Here, $t$ and $h$ denote the SM top quark and Higgs boson, respectively. $P_{L/R}=\dfrac{1\mp \gamma_5}{2}$ represent the left- and right-chiral projection operators. The mixing angle between the left-chiral (right-chiral) SM top quark and the left-chiral (right-chiral) $T$ quark is denoted by $\theta_L$ ($\theta_R$). The symbols $M_T$, $m_t$, and $M_W$ represent the masses of the $T$ quark, the SM top quark, and the $W$ boson, respectively. The $SU(2)_L$ gauge coupling is $g_w$; $s_w$ and $c_w$ are sine and cosine of the electroweak mixing angle.
The single production of the exotic $T$ quark in association with the SM top quark is possible at the LHC via gluon fusion. The effective Lagrangian of such interactions is as follows.
\begin{equation}
\label{eq.eff}
\mathcal{L}_{tg}=\dfrac{g_s}{\Lambda}~G^a_{\mu\nu}~\left[C_{tRL} \overline{T}\sigma^{\mu\nu} \tau^a P_L~t +C_{tLR} \overline{T}\sigma^{\mu\nu} \tau^a P_R~t \right] +h.c.    
\end{equation}
where $\sigma_{\mu\nu}=i/2~(\gamma_\mu\gamma_\nu-\gamma_\nu\gamma_\mu)$, and $g_s$ is the SM strong coupling constant. The notation $G^a_{\mu\nu}$ represents the gluon field strength tensor, and $\tau^a$ corresponds to the $SU(3)$ generators. The notation $\Lambda$ denotes the new physics scale, and $C_{tRL}$, $C_{tLR}$ are the coupling strengths. 
Corresponding to the above Lagrangian given in Eqs.~(\ref{eq:TWb})--(\ref{eq.eff}), the $T$ quark has four decay channels, namely bW, tZ, th, and tg. 
While the Eq.~(\ref{eq.eff}) primarily contributes to the production of the $T$ quark, it is also responsible for the tg decay mode.  
From Eqs.~(\ref{eq:TWb})--(\ref{eq:Tht}), one notices that the branching ratios for $T \to \rm tZ$, $T \to \rm bW$, and $T \to \rm th$ become negligibly small when the mixing angles are very small ($\theta_L, \theta_R \to 0$), and the $T$ quark decays predominantly into a top quark and a gluon. 
However, with increasing mixing between the $T$ quark and the top quark, the branching fractions into the electroweak channels $T \to \rm tZ$, $T \to \rm bW$, and $T \to \rm th$ increase significantly.

\vspace{-2mm}
\subsection{\label{ssec:sig_simu}Signal simulation}
\begin{figure}[hbpt]
\centering
\includegraphics[width=0.7\textwidth]{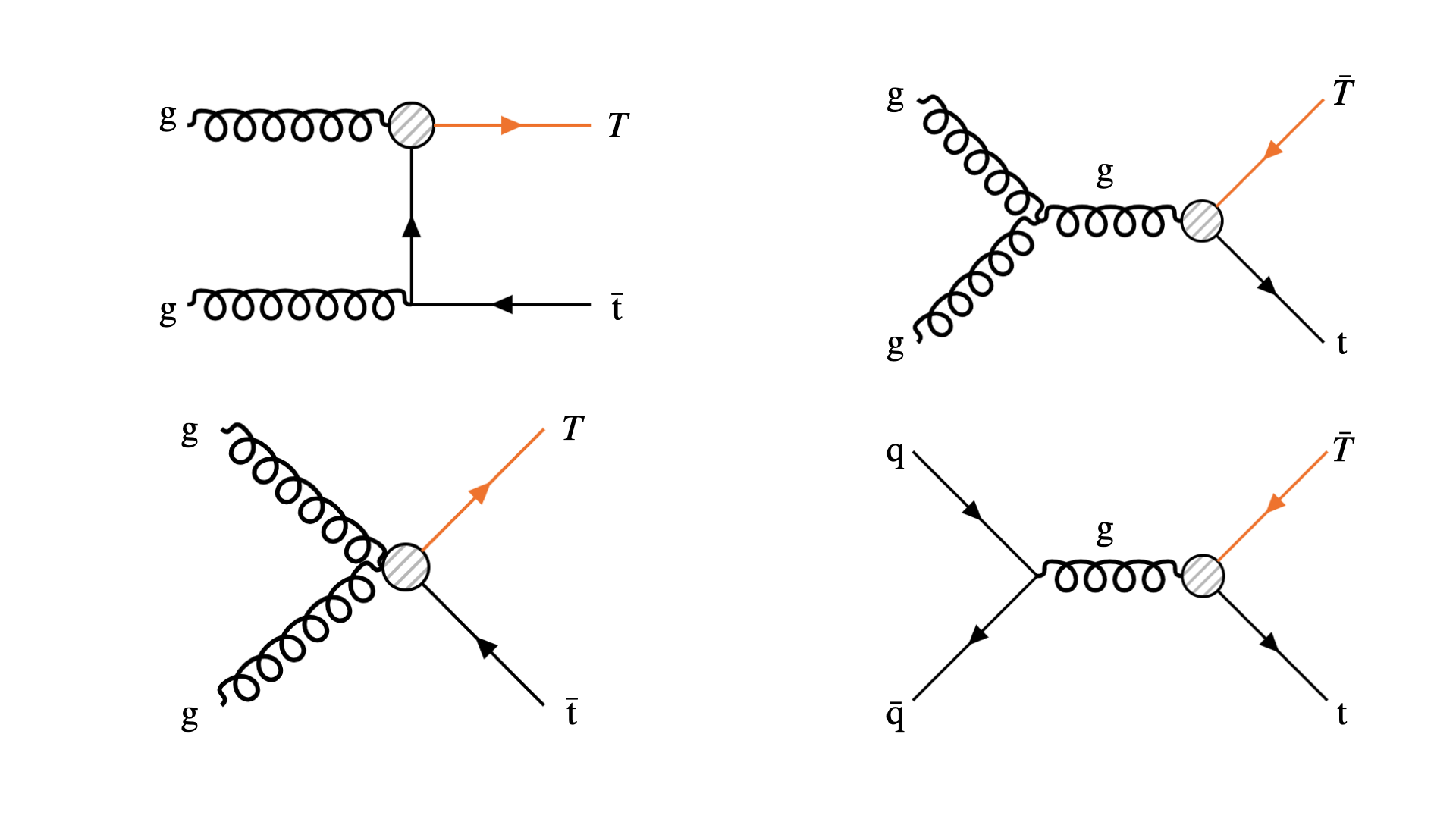}
\caption{\label{fig:feynman_diagram}Representative parton-level Feynman diagrams for the process $\mathrm{pp} \to T\,\bar{\mathrm{t}}\,/\,\bar T\,\mathrm{t}$.}
\end{figure}

\vspace{-1mm}
The process we are interested in is the associated production of an SM top quark and its vectorlike partner $T$ via the effective chromomagnetic interaction.
\begin{eqnarray}
\label{eq:signal_process_1}
\mathrm{pp} &\to & \mathrm{t}\,\overline{T} + X, \quad {\rm and}\\
\label{eq:signal_process_2}
\mathrm{pp} &\to& T\,\overline{\rm t} + X.
\end{eqnarray}
A set of representative parton-level Feynman diagrams corresponding to the process has been given in Figure~\ref{fig:feynman_diagram}.

\begin{figure}[hbpt]
\centering
\includegraphics[width=0.45\textwidth]{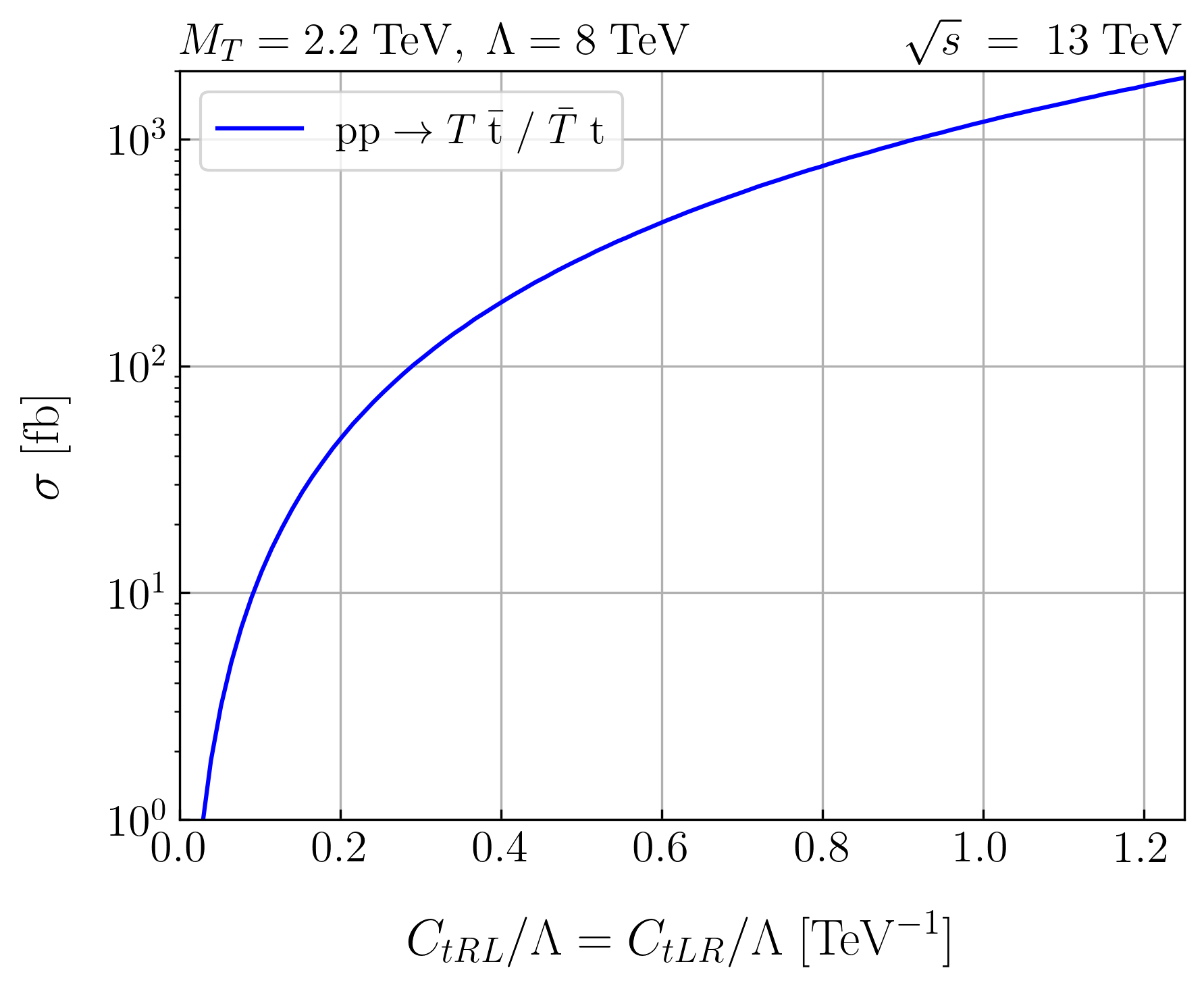}\hfill
\includegraphics[width=0.48\textwidth]{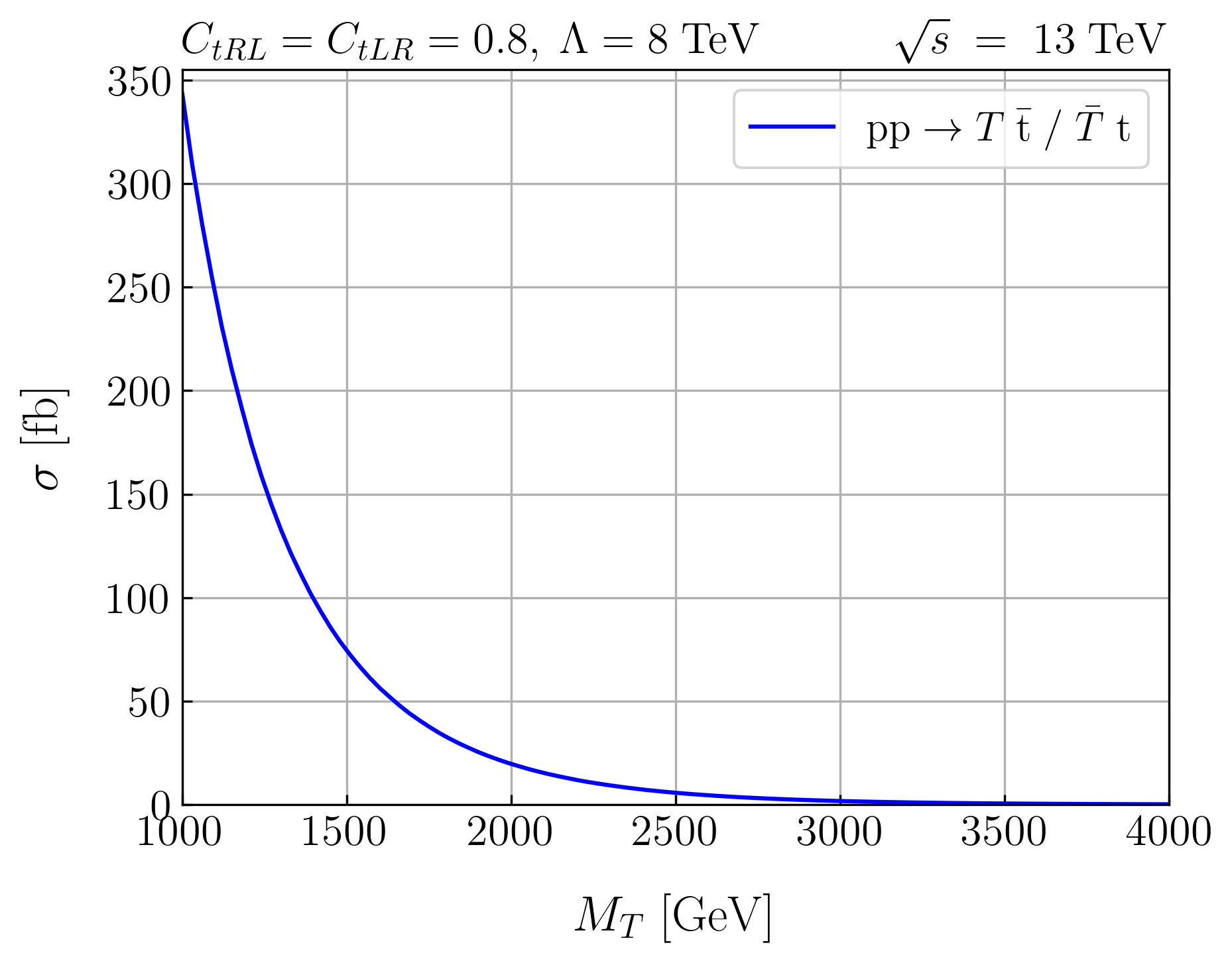}\\
\includegraphics[width=0.48\textwidth]{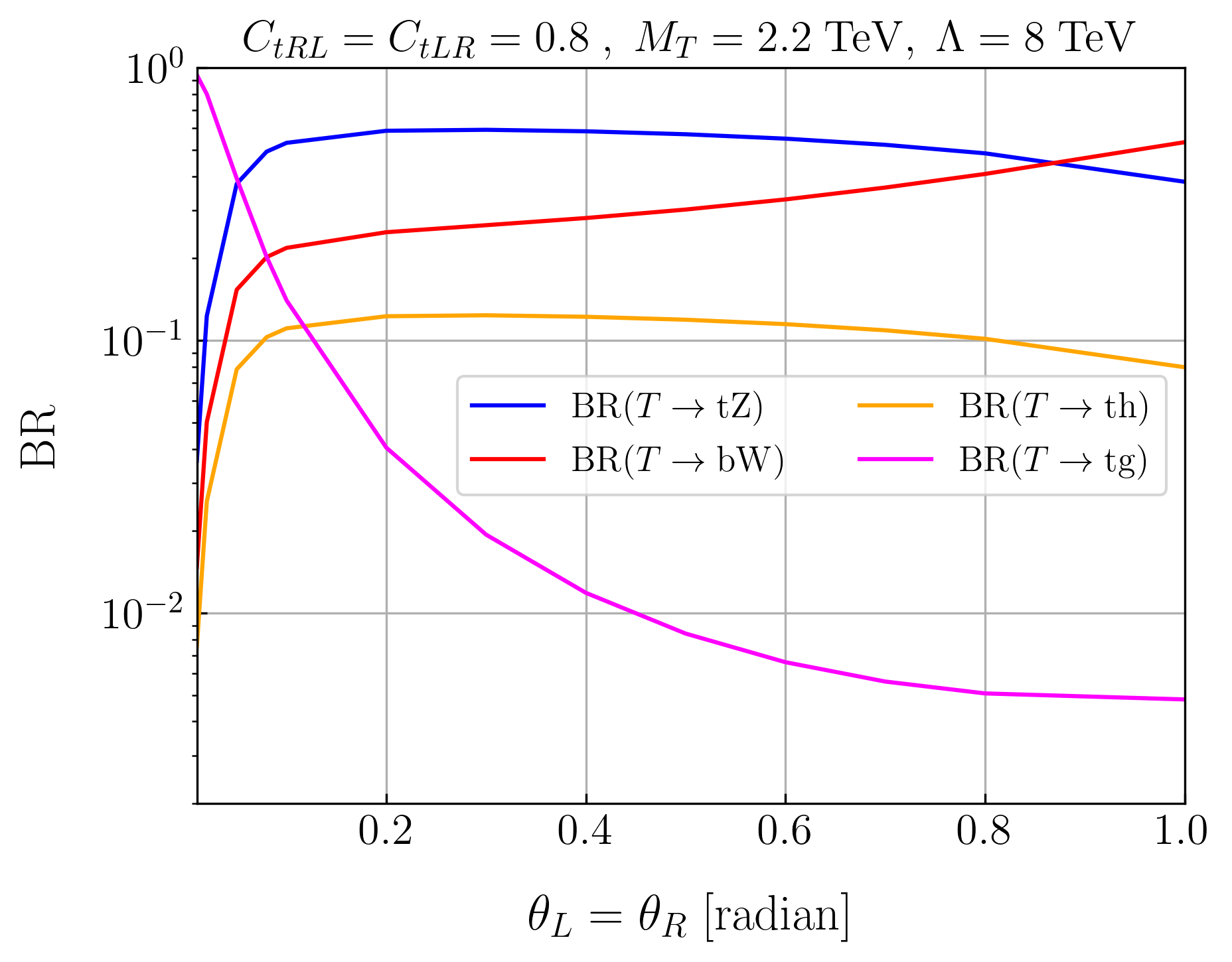}\\
\caption{\label{fig:Xsec-BR}Cross section vs.~coupling (upper left) and vs.~$T$ quark mass (upper right) are presented for $\mathrm{pp} \to T~\bar{\mathrm{t}}~/~\bar{T}~\rm t$ process at $\sqrt{s}~=~$13 TeV with new physics scale $\Lambda~=~8~$TeV, $M_T~=~2.2~$TeV (upper left), and $C_{tRL}~=~C_{tLR}~=~0.8$ (upper right). Branching ratios (BRs) vs. mixing angle for all $T$ decay modes (lower) are shown with $M_T~=~2.2~$TeV, $C_{tRL}~=~C_{tLR}~=~0.8$, and $\Lambda~=~8~$TeV.}
\end{figure}

To simulate the processes indicated by Eqs.~(\ref{eq:signal_process_1}) and (\ref{eq:signal_process_2}), we first implemented all the interaction terms defined in Eqs.~(\ref{eq:TWb})--(\ref{eq.eff}), along with the SM Lagrangian, using {\sc FeynRules}~\cite{Christensen:2008py, Alloul:2013bka} to generate a Universal FeynRules Output (UFO)~\cite{Degrande:2011ua} model file. 
This UFO model was then used in the {\sc MadGraph5\_aMC@NLO} framework to generate parton-level events at a centre-of-mass energy of $\sqrt{s}~=~13~$TeV with the parton distribution function (PDF) set to {\tt NNPDF30\_nlo\_nf\_5\_pdfas} PDFset~\cite{NNPDF:2014otw}.
The factorization and renormalization scales are set to the default dynamical scale in {\sc MadGraph5}, which corresponds to the transverse mass of the $2\to 2$ scattering process at leading order (LO). 
Figure~\ref{fig:Xsec-BR} shows the variation of the partonic cross section for $T$ quark production in association with a top quark at $\sqrt{s}~=~13~$TeV, along with the branching ratios of its various decay channels, for a fixed new physics scale of $\Lambda = 8$~TeV. 

%==================
%==================

To illustrate the collider analysis for this study, we consider a benchmark point that is consistent with existing experimental bounds. 
The strongest lower bound on $M_T$ is about 1.8~TeV~\cite{ATLAS:2024xne}. 
We further consider $\theta_{L}=\theta_{R}$ and $C_{tRL} = C_{tLR}$ to reduce the number of free parameters. 
The couplings $C_{tRL}$ and $C_{tLR}$ control the top-partner pair production cross section, which is restricted to $\mathcal{O}(1~{\rm fb})$~\cite{CMS:2024bni}, resulting in an upper limit on $\frac{C_{tRL}}{\Lambda} = \frac{C_{tLR}}{\Lambda} \lesssim 0.6~\text{TeV}^{-1}$. 
Furthermore, the mixing parameters $\theta_{L}$ and $\theta_{R}$ affect the Higgs boson branching ratio to gluon-gluon and di-photon modes. 
One finds $\theta_L=\theta_R \lesssim 0.25$ radian for $M_T=2.2$~TeV from Refs.~\cite{ATLAS:2025hhd, Dawson:2012di, Zhang:2025zkn}. 
Based on these factors, we choose the benchmark scenario as follows:

\begin{equation}
\label{eq:signal_params}
\frac{C_{tRL}}{\Lambda} = \frac{C_{tLR}}{\Lambda} = 0.1~\text{TeV}^{-1},\qquad \theta_{L}=\theta_{R} =0.06~\text{radian},\qquad \text{and}\quad \ M_{T} =2.2~{\rm TeV}.
\end{equation}

\begin{figure}[hbpt]
\centering
\includegraphics[width=0.98\textwidth]{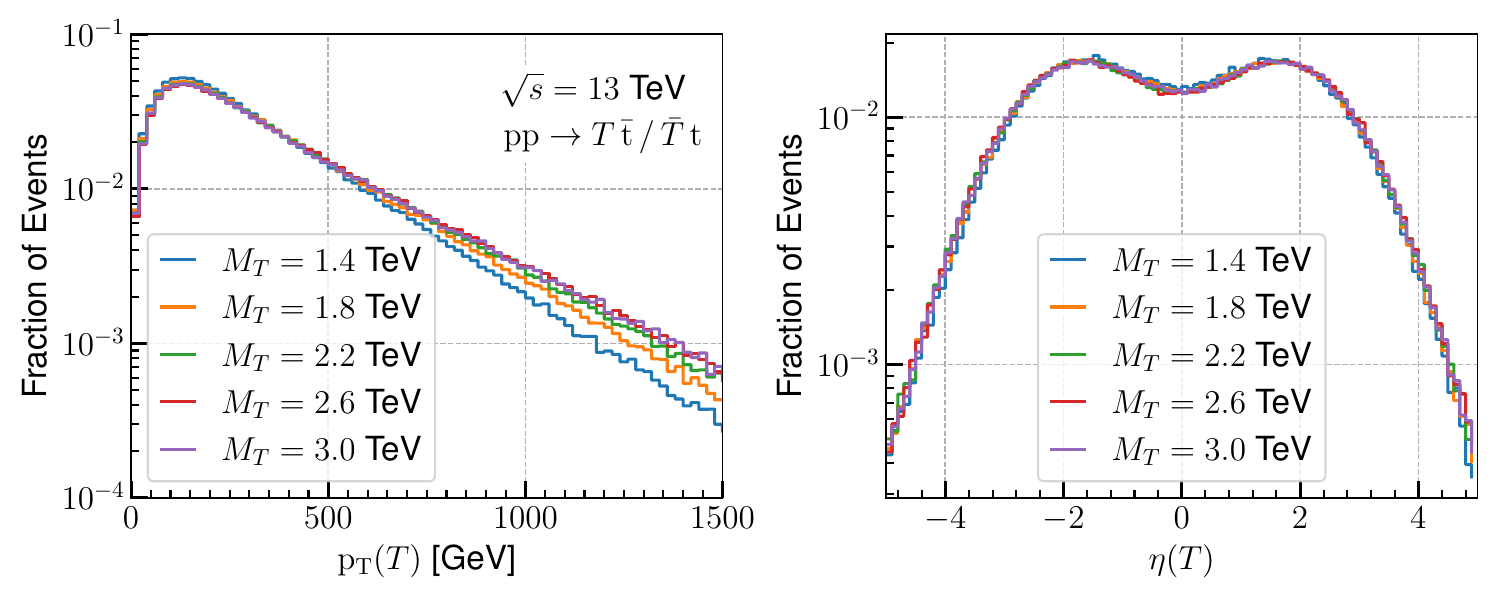}
\caption{\label{fig:top_partner_kin}Distributions of the top partner \pt (left) and $\eta$ (right) at parton level for different mass hypotheses for the process $\mathrm{pp} \to T\,\bar{\mathrm{t}}\,/\,\bar{T}\,\mathrm{t}$ at $\sqrt{s}=13$ TeV.}
\end{figure}

The parton-level kinematics of the produced top partner are shown in Fig.~\ref{fig:top_partner_kin} for different mass hypotheses.  
The $\text{p}_{\text{T}}(T)$ distribution peaks around $\text{p}_{\text{T}}(T) \sim 200~\text{GeV}$, with a rapidly falling tail at higher transverse momenta.
The higher tail of the $\pt$-spectrum of the $T$ quark has a distinct dependence on its mass. For the process pp $\to T~\mathrm{\bar t}~/~\bar{T}~\mathrm{t}$, the transverse momentum of the heavy top partner remains well below the cut-off scale $\Lambda = 8$~TeV throughout the analysed region, as shown in the left panel of Fig.~\ref{fig:top_partner_kin}. Thus, the effective field theory (EFT) description remains valid across the entire analysed phase space region in this work.
The cross section of the signal process, the $T$ quark decay width, and its branching fractions for all of the four decay modes are listed in Appendix~\ref{sec:massPoints} for different mass hypotheses.

The generated events are then passed through {\sc Pythia8}~\cite{Bierlich:2022pfr} for parton showering and hadronization. 
In the {\sc Pythia8} simulation, the {\sc CMS Tune MonashStar}~\cite{CMS:2015wcf}, based on Monash 2013 tune~\cite{Skands:2014pea}, has been used to better match CMS data, especially to model the underlying event, initial-state radiation, and beam remnants. We then incorporated the effects of pile up in the signal MC samples by overlaying minimum-bias event samples with the hard-scattered events with approximately $20$ average pile up interactions per event.
Finally, a fast detector simulation is performed on the pile up overlaid samples using {\sc Delphes\_v3.5.0}~\cite{deFavereau:2013fsa} that employs an overall detector geometry definition and a parametric model of average detector response\footnote{The choice of using {\sc Delphes} fast detector simulation for signal MC samples was made owing to limitations in the available computing and storage resources at our disposal for this study.}. 
The signal event simulation details are summarized in Table~\ref{tab:signalSim}.

\begin{table}[hbpt]
\centering
\caption{\label{tab:signalSim}Details of signal simulation tools and associated settings.}
%\resizebox{\textwidth}{!}{
\begin{tabular}{c|l}
\hline\hline
\multirow{2}{*}{BSM model} &  SM + Eqs.~(\ref{eq:TWb})--(\ref{eq.eff}) using \textsc{FeynRules}~\cite{Christensen:2008py,Alloul:2013bka} and \\
 & Universal FeynRules Output (UFO)~\cite{Degrande:2011ua}\\
\hline
\multirow{2}{*}{Matrix element} &  Eqs.~(\ref{eq:signal_process_1}) and (\ref{eq:signal_process_2}) using \textsc{MadGraph5\_aMC@NLO}~\cite{Alwall:2014hca}\\ 
 & at LO with $\sqrt{s} = 13$~TeV \\
\hline
PDF set & \texttt{NNPDF30\_nlo\_nf\_5\_pdfas}~\cite{NNPDF:2014otw} \\
\hline
Parton showering \& Hadronization & \textsc{Pythia8}~\cite{Bierlich:2022pfr} \\
\hline
\multirow{2}{*}{Underlying event} & \textsc{CMS Tune MonashStar}~\cite{CMS:2015wcf}, based on Monash \\ 
 & 2013 tune~\cite{Skands:2014pea}\\
\hline
Detector simulation model & \textsc{Delphes\_v3.5.0}~\cite{deFavereau:2013fsa} %with default CMS card
\\
\hline\hline
\end{tabular}
%}
\end{table}

Depending on the decay mode of the top partner into $\rm tZ$, $\rm th$, $\rm tg$, or $\rm bW$, the above processes could be realized into different event categories. To suppress a huge QCD multijet background at the LHC, we focus on the semi-leptonic channels requiring one lepton ($\ell\ =\ e,\mu$) in the final state. Therefore, the final states where one of the $\rm W$'s originating from either of the $\rm t$, $\rm {\bar t}$, $T$, or $\bar T$ decay leptonically are studied. The event sample for our search can be broadly classified into the following two categories:
\begin{eqnarray}
& {\rm Two\ fat\ jets\ category: } & 1 \ell + 2\ {\rm fat\ jets} + 1\ {\rm b-jet} + \ptmiss, \label{eq:onefj}\\
& {\rm One\ fat\ jet\ category: } & 1 \ell + 1\ {\rm fat\ jet} + \geqslant 1\ {\rm b-jet(s)} + \ptmiss. \label{eq:twofj}
\end{eqnarray}
Table~\ref{tab:event_topo} lists the different decay modes of the top partner and their possible contribution to each event category, applying the same selection criteria as defined in Section~\ref{sec:evtsel} on the lepton and jets.   
The signal fractions are larger in the two fat jets event category for all decay modes. 
However, in this event category, the \ttbar + X (X = W, Z, jets, etc.) backgrounds also have a significantly larger contribution, leading to a poor signal-to-background ratio and thus making it unsuitable as a search region for the signal process. 
Therefore, we focus on the one fat jet category where the contribution of the $\rm bW$ decay mode is relatively higher.  

\begin{table}[hbpt]
\centering
\vspace{-8pt}
\caption{\label{tab:event_topo}Simulated sample sizes (N$_{\text{Events}}$) for the signal process in different $T$ quark decay modes, along with their respective contributions to each event category as defined in Eqs.~(\ref{eq:onefj})and (\ref{eq:twofj}).}
% \resizebox{\textwidth}{!}{
\begin{tabular}{c|c|c|c|c}
\hline\hline
 & \multicolumn{4}{c}{Decay mode} \\ \cline{2-5}
 & $T\to \mathrm{t Z}$ & $T\to \mathrm{t h}$ & $T\to \mathrm{t g}$ & $T\to \mathrm{b W}$ \\
\hline\hline
 & & & &  \\[-8pt]
N$_{\text{Events}}$ (in millions) & $1.0$ & $1.0$ & $1.0$ & $0.6$ \\
 & & & &  \\[-8pt]
\hline
Two fat jets category & $73.00\%$ & $72.03\%$ & $76.11\%$ & $65.89\%$ \\
 & & & &  \\[-8pt]
One fat jet category & $27.00\%$ & $27.97\%$ & $23.89\%$ & $34.11\%$ \\
\hline\hline
\end{tabular}
% }
\end{table}

\vspace{-2mm}
\subsection{\label{ssec:bkgMC}Background processes}\vspace{-2mm}
Several SM processes that constitute the same final state as that of the signal are considered as backgrounds.
We list the relevant SM background processes for our study in Table~\ref{tab:bkgSamples}.
The background MC samples are obtained from the CMS Open Data portal~\cite{ttxj_full, tWjets_full, ttZjets_full, ttWjets_full, WZjets_full, WWjets_full, QCD_full}, while the cross sections are obtained from the available predictions at the highest perturbative order, with the exception of the QCD multijet process, where LO predictions are used.
The accuracies of the respective cross section predictions are also given, along with the number of MC events taken for the analysis, in the same table.
Among the MC samples, the dominant \ttbar + jets events were generated at NLO accuracy with up to two additional partons using {\sc MadGraph5\_aMC@NLO}~\cite{Alwall:2014hca} and FxFx merging~\cite{Frederix:2012ps}. 
For this background process, MC samples binned in \HT are stitched with the so-called inclusive sample according to the method described in Ref.~\cite{Ehataht:2021rkh} for better statistical modelling of the boosted topology. Table~\ref{tab:stitching_weights} summarizes the stitching weights applied to events lying in different \HT bins. 
The event samples for WZ + jets and \TTW processes were also generated at NLO accuracy with at most one extra parton using {\sc MadGraph5\_aMC@NLO}~\cite{Alwall:2014hca} and FxFx merging~\cite{Frederix:2012ps}. 
The tW/$\mathrm{\bar t}$W + jets were simulated at NLO~\cite{Re:2010bp} using {\sc POWHEG-BOX}~\cite{Nason:2004rx,Frixione:2007vw,Alioli:2010xd} with five flavour scheme~\cite{Maltoni:2012pa}, while the WW + jets events were produced at NLO accuracy~\cite{Nason:2013ydw} using {\sc POWHEG-BOX-v2}~\cite{Nason:2004rx,Frixione:2007vw,Alioli:2010xd}. The \TTZ process was generated using {\sc MadGraph5\_aMC@NLO} at LO with up to one additional jet, employing MLM merging scheme~\cite{MLM01, Hoeche:2005vzu, Mangano:2006rw}. 
The QCD multijet background was generated at LO using {\sc Pythia8}~\cite{Sjostrand:2007gs} generator. A combination of different QCD MC samples in non-overlapping \pthat bins above $15~$GeV is considered.
In all cases, {\tt NNPDF30\_nlo\_as\_0118}~\cite{NNPDF:2014otw}\footnote{The PDF predictions for signal and background MC samples use {\bf exactly same} number of quark flavours (5) and nominal value of the strong coupling constant at the Z-pole (0.118).} was chosen to be the default PDFset, and {\sc Pythia8}~\cite{Sjostrand:2007gs} was used for parton showering and hadronization; while {\sc CMS Tune Monash Star}~\cite{CMS:2015wcf} implemented in {\sc Pythia} is used to model the underlying event. 
The effect of pile up was simulated in the background MC samples by overlaying minimum-bias events with the hard-scattered events with an average of around $20$ pile up interactions per event.
All background MC samples undergo full simulation of the CMS detector based on {\sc Geant4}~\cite{GEANT4:2002zbu} that comprises of detailed geometry, alignment, and calibration information at the sub-detector level.

\begin{table}[hbpt]
\centering
\vspace{-8pt}
\caption{\label{tab:bkgSamples}List of relevant SM background processes along with their respective cross sections and MC sample sizes (N$_{\text{Events}}$). All MC samples are obtained from the CMS Open Data portal. The `cross section accuracy' refers solely to the order of the cross section calculation used for normalization correction, while `event generation accuracy' refers to the accuracy of the event generation at the matrix element level. References in the `process' column point to the official MC samples from the CMS Open Data portal, while those in the `cross section' column refer to therespective calculations. Here, NLO, NNLO, and NNLL denote next-to-leading order, next-to-next-to-leading order, and next-to-next-to-leading logarithmic accuracy, respectively, while EW stands for electroweak corrections.
}
\begin{tabular}{ccccc}
\hline\hline \\[-8pt]
\multirow{2}{*}{Process} & Event generation & Cross section & Cross section & N$_{\text{Events}}$\\
 & accuracy &  [pb] & accuracy & [million] \\ \\[-8pt]
\hline\hline \\[-8pt]
\ttbar + jets~\cite{ttxj_full} & NLO~\cite{Alwall:2014hca,Frederix:2012ps,Sjostrand:2007gs} & 833.900~\cite{Czakon:2011xx} & NNLO + NNLL & 117.70\\ 
\\[-8pt] 
tW/$\bar{\text{t}}$W + jets~\cite{tWjets_full} & NLO~\cite{Re:2010bp,Alioli:2010xd,Sjostrand:2007gs} & 79.300~\cite{Kidonakis:2021vob} & NNLO & 2.00 \\ \\[-8pt]
\TTZ~\cite{ttZjets_full} & LO~\cite{Alwall:2014hca,MLM01,Mangano:2006rw,Sjostrand:2007gs} & 0.863~\cite{Kulesza:2018tqz} & NLO + NNLL & 22.16 \\ \\[-8pt]
\multirow{2}{*}{\TTW~\cite{ttWjets_full}} & \multirow{2}{*}{NLO~\cite{Alwall:2014hca,Frederix:2012ps,Sjostrand:2007gs}} & \multirow{2}{*}{0.745~\cite{Buonocore:2023ljm}} & NNLO (QCD) & \multirow{2}{*}{0.83} \\
 & & & + NLO (EW) & \\ \\[-8pt]
\hline 
\\[-8pt]
WZ + jets~\cite{WZjets_full} & NLO~\cite{Alwall:2014hca,Frederix:2012ps,Sjostrand:2007gs} & 56.850~\cite{Grazzini:2016swo} & NNLO & 13.19\\ \\[-8pt]
WW + jets~\cite{WWjets_full} & NLO~\cite{Nason:2013ydw,Alioli:2010xd,Sjostrand:2007gs} & 1.371~\cite{Grazzini:2016ctr} & NNLO & 8.92 \\ \\[-8pt]
\hline
\\[-8pt]
QCD multijet~\cite{QCD_full} & LO~\cite{Sjostrand:2007gs} & $1.981 \times 10^9$ & LO & 97.19 \\ \\[-8pt]
\hline\hline
\end{tabular}
\end{table}

\begin{table}[hbpt]
\centering
\vspace{-16pt}
\caption{\label{tab:stitching_weights}Stitching weights for \ttbar + jets events in different \HT bins for 300~fb$^{-1}$ integrated luminosity, computed following the prescription in Ref.~\cite{Ehataht:2021rkh}.}
\begin{tabular}{c|ccccc}
\hline\hline \\[-8pt]
\multirow{2}{*}{} & \multicolumn{5}{c}{\HT bins [GeV]}  \\ \\[-8pt]
\cline{2-6} \\[-8pt]
 & [0,\ 600) & [600,\ 800) & [800,\ 1200) & [1200,\ 2500) & $\geqslant 2500$ \\ \\[-8pt]
\hline \\[-8pt]
\multirow{2}{*}{Stitching weights} & \multirow{2}{*}{$2.562$} & $2.550$ & $1.413$ & $9.180$ & $2.154$ \\
 & & $\times10^{-2}$ & $\times10^{-2}$ & $\times10^{-3}$ & $\times10^{-4}$ \\ 
\hline\hline
\end{tabular}
\end{table}

%% file: DRjetsAlgo.tex
The jet clustering algorithms are crucial steps in the analysis of events in a hadron collider experiment such as the LHC~\cite{Salam:2010nqg, Sapeta:2015gee}. 
The currently used algorithms are fixed-radius sequential recombination inclusive jet algorithms, namely the \kt (KT), anti-\kt (AK), and Cambridge-Aachen (CA)~\cite{Ellis:1993tq, Catani:1993hr, Cacciari:2008gp, Dokshitzer:1997in, Wobisch:1998wt} algorithms. 
These traditional algorithms take as input the list of four momenta of the particles from an event and successively recombine them into several jets. 
In their iterative steps, these algorithms calculate pair-wise ``\kt'' distance ($\text{d}_{\text{ij}}$) between $\text{i}^{\text{th}}$ and $\text{j}^{\text{th}}$ four momenta and ``beam'' distance ($\text{d}_{\text{iB}}$) for the $\text{i}^{\text{th}}$ four momentum, defined as
\begin{eqnarray}
&& \text{d}_{\text{ij}} = \min\left(\text{p}_{\text{T,\ i}}^{2n}\ ,\ \text{p}_{\text{T,\ j}}^{2n}\right)\ \frac{\Delta \text{R}_{\text{ij}}^{2}}{\text{R}^{2}},\ \text{and} \label{eqn:dij}\\
&& \text{d}_{\text{iB}} = \text{p}_{\text{T,\ i}}^{2n} \label{eqn:diB}
\end{eqnarray}
to decide whether to merge two four momenta or declare a particular four momentum as a final jet. 
Here, $\Delta \text{R}_{\text{ij}}$ is the distance between $\text{i}^{\text{th}}$ and $\text{j}^{\text{th}}$ four momenta in the pseudorapidity-azimuth ($\eta$-$\phi$) plane. 
The two four momenta are merged to give a new four momentum if $\text{d}_{\text{ij}}$ becomes the least among all the pair-wise ``\kt'' and beam distances. 
Otherwise, the $\text{i}^{\text{th}}$ four momentum corresponding to the least beam distance is declared as a final jet. 
Importantly, the algorithm has parameters $n$, which usually takes values $-1,\  0,\ \text{and}\ 1$ corresponding to AK, CA, and KT algorithms respectively, and $\text{R}$ as the distance measure in the $\eta$-$\phi$ plane.
These algorithms typically output jets of a fixed size, which is determined by the input parameter $\text{R}$ to the algorithm. 
In this article, the AK algorithm is chosen as the default since AK jets have a regular conical shape with a well-defined jet area~\cite{Cacciari:2008gp}.

The traditional jet clustering approaches mentioned above have the limitation of producing varying-sized jets in an event for a given value of $\text{R}$. 
Several interesting attempts have been made to modify the algorithms to accommodate jets of variable sizes, such as ``Variable R'' jets~\cite{Krohn:2009zg}, ``Fuzzy'' jets~\cite{Mackey:2015hwa}, ``SHAPER''~\cite{Ba:2023hix}, ``XCone''~\cite{Stewart:2015waa}, ``PAIReD'' jets~\cite{Mondal:2023law}, ``dynamic-radius'' jets~\cite{Mukhopadhyaya:2023rsb}, etc.
Here, we focus on the ``dynamic-radius'' jet clustering algorithm~\cite{Mukhopadhyaya:2023rsb}, which attempts to dynamically modify the jet radius during the evolution of each jet through the iterative steps of the algorithm. 
The approach is to start from an initial radius ${\rm R_0}$, the radius for each evolving ``protojet" gets modified depending on its internal structure during the process of clustering. 
The square of the additive modifier $\sigma_i$ to ${\rm R_0}$ for each evolving protojet at each interactive stage is given by
\begin{equation}
\label{eqn:var}
\sigma_{\rm i}^2 = \dfrac{\sumab \pta\,\ptb\ \dRab^2 }
{\sumab \pta\,\ptb } - \left(\dfrac{\sumab \pta\,\ptb\ \dRab}
{\sumab \pta\,\ptb }\right)^2, 
\end{equation}
where $a$ and $b$ run over the constituents of the proto-jet. 
So, the final radius for the $i^{\rm th}$ jet becomes $\mathrm{R_d} = \mathrm{R_0 + \sigma_i}$. 
The advantage one gets in this method is that the radius is no longer fixed; it gets adapted to the requirement of the jet that is being formed.
The radius modifier can also be expressed in terms of the ECF observable $\mathrm{C_1^{(\beta)}}$~\cite{Larkoski:2013eya} as
\begin{equation}
\mathrm{\sigma_i^2 = C_1^{(2)} - \left(C_1^{(1)}\right)^2 }
\end{equation}
For a splitting characterized by the opening angle ($\mathrm{\theta}$) between two hard cores, or between a single hard core and the dominant emission, the correlator behaves as $\mathrm{C_1^{(\beta)} \simeq \frac{E_2}{E_1}\theta^\beta}$, where $E_1$ and $E_2$ denote the energies of the hardest core and the associated emission, respectively. Consequently, the radius modifier scales as $\mathrm{\sigma_i \propto \theta}$ and is therefore directly sensitive to the decay or splitting opening angle. Furthermore, the two-point energy correlator behaves as $\mathrm{\sigma_i^2 \sim C_1^{(2)} \sim \frac{m^2}{p_T^2}}$ for the narrow QCD jets, where emissions are nearly collinear~\cite{Proceedings:2018jsb}. As a result, this radius modifier adapts according to the characteristic behaviour of corresponding jet types, making this clustering approach particularly effective. 
Thus, as expected, the radius modifier tends to become larger for top or heavy boson jets~\cite{ATLAS:2019kwg}; while on the other hand, it tends to be smaller for narrow jets like the ones appearing from light quarks or gluons. 
Since our study involves one fat jet along with narrow b-jets, the efficiency of the DR jet clustering is expected to be better. 

%% file: boostedjetparams.tex
Improving signal sensitivity over the backgrounds using jets and their substructures is the primary aim of this work. 
We present here a brief overview of the various jet substructure methods that are used in our analysis.
\begin{description}
\item[Soft drop grooming]{
To mitigate contamination from pile up, underlying events, etc., inside a jet, and improve the reliability of the jet substructure measurements, we have used the soft drop de-clustering method~\cite{Larkoski:2014wba}. 
This method primarily aims to remove soft and wide-angle radiations inside a jet while retaining the hard core of the jet. To briefly describe, the soft drop algorithm is applied on a jet that has been re-clustered using the CA algorithm (which orders by angle).
The essential step of this method is to remove soft and wide-angle subjets after de-clustering the jet into two subjets (labelled as i and j) by checking the following condition. 
\begin{equation}
\mathrm{\frac{\min(p_{T,i}, p_{T,j})}{p_{T,i} + p_{T,j}}} < z^{\rm SD}_{\text{cut}} \left( \mathrm{\frac{\Delta R_{ij}}{{\rm R}^{\rm SD}}} \right)^{\beta^{\rm SD}}    
\end{equation}
Here, $\mathrm{p_{T,i}}$ and $\mathrm{p_{T,j}}$ are the transverse momenta of the subjets, $\mathrm{\Delta R_{ij}}$ is the distance between them in the $\eta$-$\phi$ plane.  This process is repeated until the condition fails to be satisfied.
The radius ${\rm R}^{\rm SD}$ is the characteristic jet radius usually taken to be the radius parameter used during the clustering. In the case of the DR jet clustering algorithm, we have taken this parameter to be $\mathrm{R_d=R_0+\sigma}$, the final dynamically adjusted radius for each jet.
The two parameters of soft drop are (1) $z^{\rm SD}_{\text{cut}}$: the energy fraction threshold and (2) $\beta^{\rm SD}$: the exponent regulating the angular separation between the two subjets. 
In the special case of $\beta^{\rm SD} = 0$, the method reduces to the modified mass drop tagger~\cite{Butterworth:2008iy,Dasgupta:2013ihk}. 
The grooming then depends only on the momentum fraction $z$ and not on the angular separation.
We have used the value of $\beta^{\rm SD} = 0$ to keep our analysis aligned with the available jet substructure measurement~\cite{ATLAS:2019kwg}. 
We also check the alternative value of $\beta^{\rm SD} = 1$. 
For the energy fraction threshold parameter, a nominal value of $z_{\rm cut}^{\rm SD} = 0.1$ is used to match the studies performed in Refs.~\cite{Larkoski:2014wba, Larkoski:2014bia, Larkoski:2015lea, CMS:2017qlm}.
}
% \vspace{-8pt}
\item[$N$ subjettiness]{
The jet substructure method $N$ subjettiness\,\cite{Thaler:2010tr}, provides a measure of the consistency of a jet having $N$ subjets inside. 
In our event topology, we expect boosted top, W, or Z fat jet, which gives rise to a three- or two-prong structure.
Therefore, this method becomes particularly useful with the values of $N=3$ and $N=2$. 
The $N$-subjettiness observable captures these subjets by fitting $N$ subjet axes inside the fat jet. This is performed by minimizing the $N$ subjettiness measure~\cite{Thaler:2011gf} defined as
\begin{equation}
\label{eqn:nsubjettiness}
\mathrm{\tilde{\tau}_N^{(\beta^{\rm NS})} = \frac{1}{d_0}\sum_{i\in {\rm jet}} {p_T}_i} \min\left\{ \Delta {\rm R}_{{\rm i}, a_1}^{\beta^{\rm NS}}, \Delta {\rm R}_{{\rm i}, a_2}^{\beta^{\rm NS}}, \cdots, \Delta {\rm R}_{{\rm i}, a_N}^{\beta^{\rm NS}} \right\}, 
\end{equation}
where the sum runs over all constituents in the jet, and $\Delta {\rm R}_{{\rm i}, a_k}$ is the distance in the $\eta$-$\phi$ plane between constituent i and subjet axis $a_k$. The prefactor ${\rm 1/d_0}$ is a normalization factor. The minimization is performed over the choices of axes, and the final minimum value of $\tilde{\tau}_N$, denoted as $\tau_N$, represents the optimal measure of $N$ subjettiness. 
The exponent $\beta^{\rm NS}$ and the method used to determine the seed $N$ axes (e.g., exclusive \kt axes from an iterative algorithm) are parameters of choice. We have used the $N$ subjettiness method implemented in {\sc Fastjet Contrib}~\cite{Cacciari:2011ma}. 
In our study, we use $\beta^{\rm NS} = 1$, the seed axes obtained by {\sc WTA\_KT\_Axes} method~\cite{Bertolini:2013iqa,Larkoski:2014uqa,Salam_unpub}, and the ratios $\tau_{32} = \tau_3/\tau_2$ and $\tau_{21} = \tau_2/\tau_1$ to respectively identify three- and two-prong structures inside a fat jet.
}
\vspace{8pt}
\item[Energy correlation functions]{
These are a class of infrared and collinear safe observables proposed to study jet properties\,\cite{Larkoski:2013eya}. The first three lowest point functions take the form
\begin{eqnarray}
{\rm ECF}_1 (\beta^{\rm ECF}) &=& \mathrm{\sum_{i\,\in\,jet} p_{T,i}}\ ,\label{eqn:ecf1}\\
{\rm ECF}_2 (\beta^{\rm ECF}) &=& \mathrm{\sum_{i < j \,\in\, jet} p_{T,i}\,p_{T,j}\,\Delta R_{ij}^{\beta^{\rm ECF}}}\ , \label{eqn:ecf2}\\
{\rm ECF}_3 (\beta^{\rm ECF}) &=& \mathrm{\sum_{i < j < k \,\in\, jet} p_{T,i}\,p_{T,j}\,p_{T,k} \left(\Delta R_{ij}\,\Delta R_{ik}\,\Delta R_{jk}\right)^{\beta^{\rm ECF}}}\ ,\label{eqn:ecf3}
\end{eqnarray}
where the sum is over the jet constituents and the exponent $\beta^{\rm ECF}$ is a parameter that gives the weight to the angular separation between the jet constituents. These functions help distinguish between jets originating from different partonic configurations. In the case of top quark study, a better discrimination from QCD jets is obtained for $\beta^{\rm ECF} = 1.0$\,\cite{Larkoski:2013eya}. 
We therefore choose to use $\beta^{\rm ECF}=1.0$ as a nominal value and $\beta^{\rm ECF}=0.5$ for an alternate value.
Instead of using the ECFs as defined in Eqs.~(\ref{eqn:ecf1})--(\ref{eqn:ecf3}) directly, their ratios are better suited as they remove the \pt dependence. The following ratios~\cite{ATLAS:2019kwg} are studied.
\begin{eqnarray}
&\quad\ e_2 = \displaystyle\frac{{\rm ECF}_2}{{(\rm ECF}_1)^2}\ , \qquad\qquad &e_3 = \frac{{\rm ECF}_3}{({\rm ECF}_1)^3}\ ,\\
&\quad C_2 = \displaystyle \frac{e_3}{(e_2)^2}\ ,\qquad \text{and}\ \qquad\ &D_2 = \frac{e_3}{(e_2)^3}\ .
\end{eqnarray}
}
\end{description}

\begin{figure}[hbpt]\vspace{-10pt}
\centering
\includegraphics[width=0.48\textwidth]{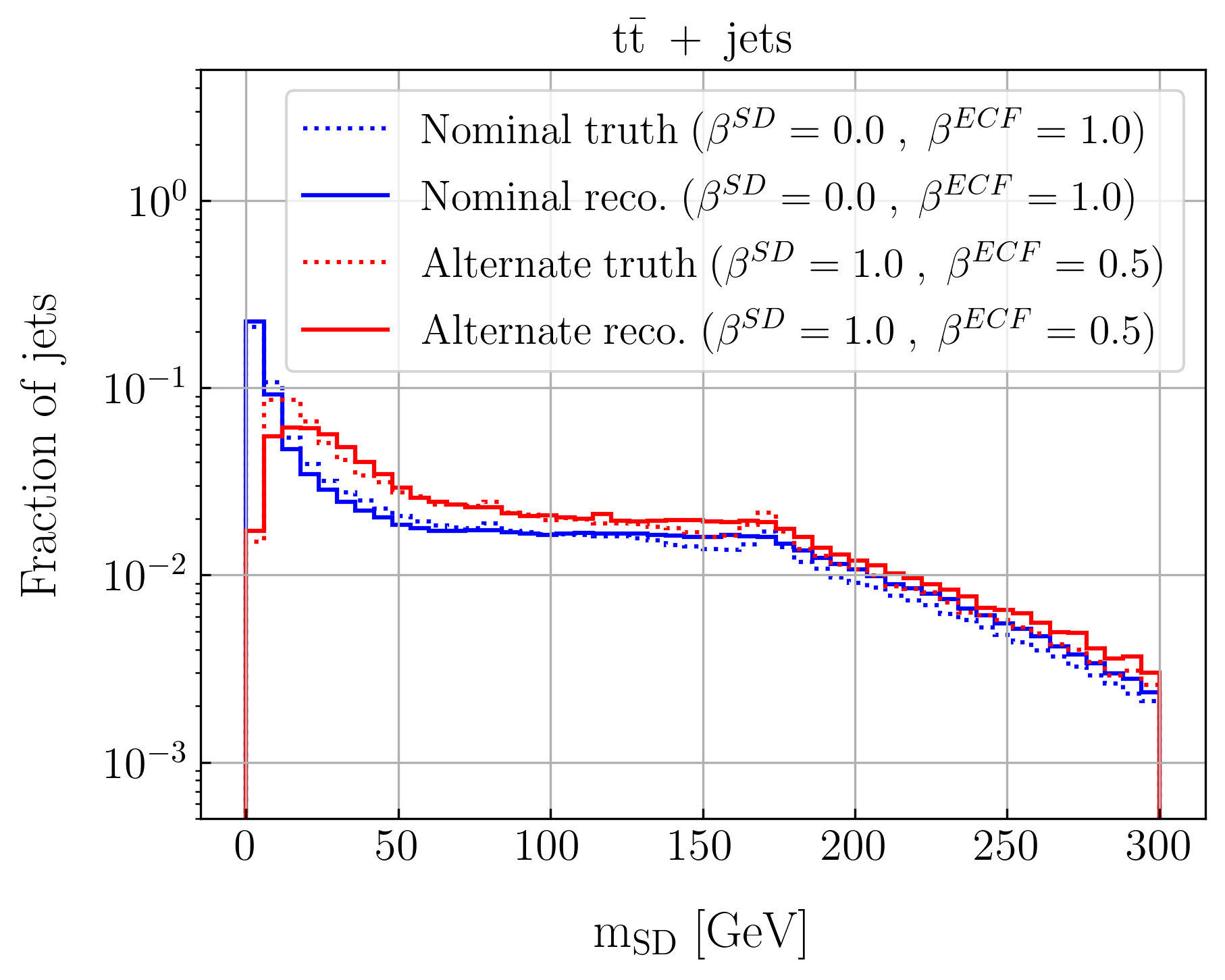}\hfill
\includegraphics[width=0.48\textwidth]{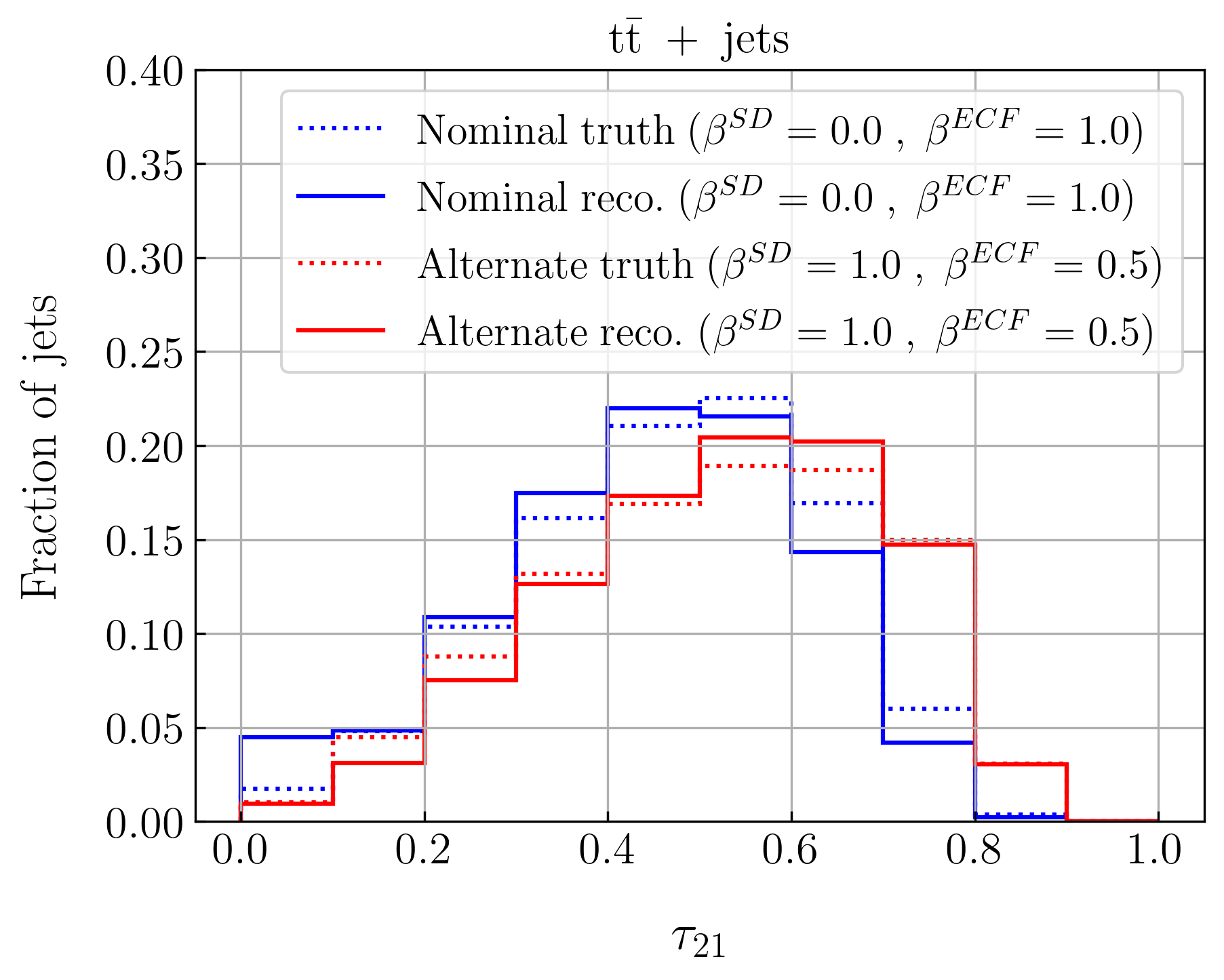}\\
\includegraphics[width=0.48\textwidth]{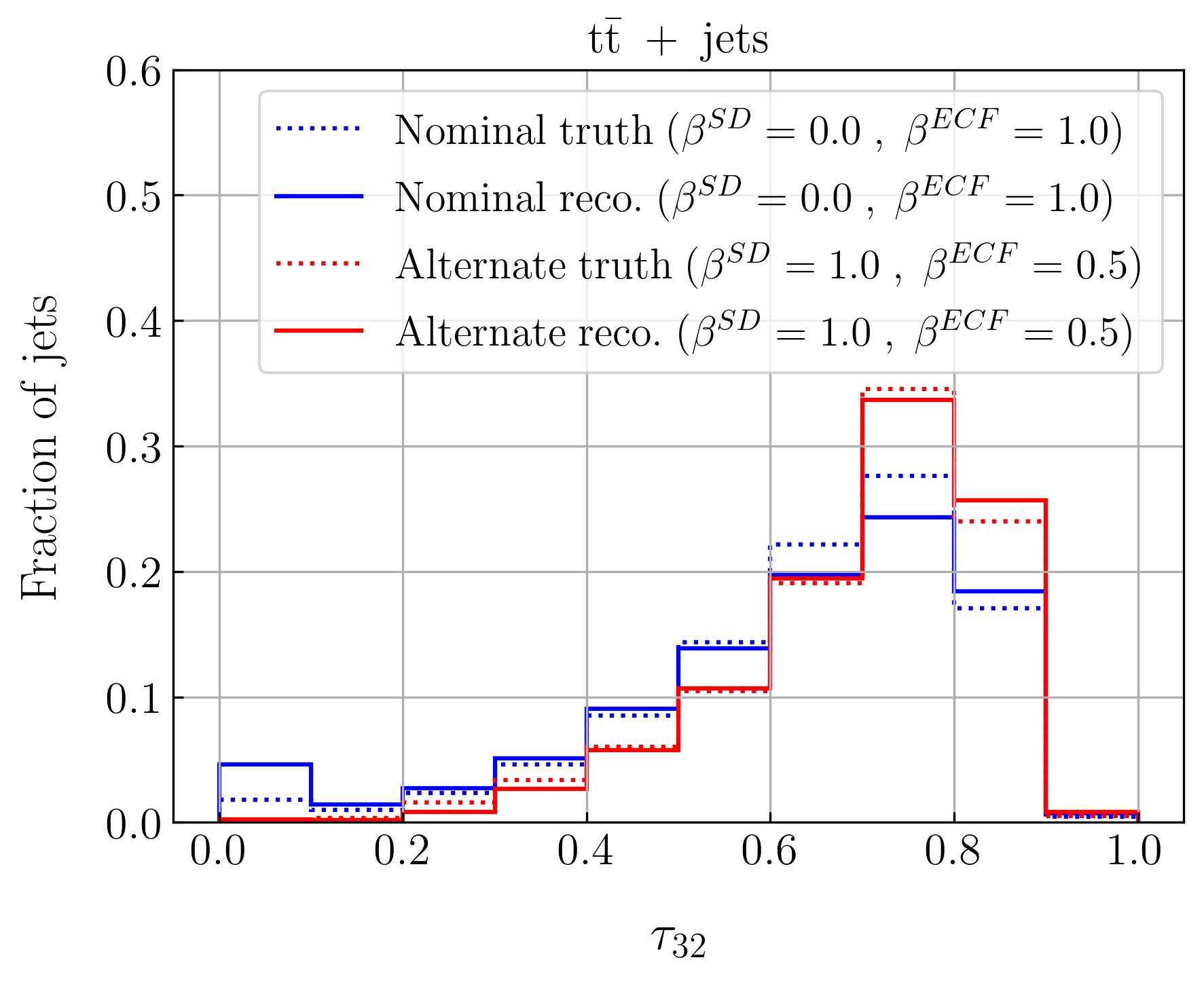}\hfill
\includegraphics[width=0.48\textwidth]{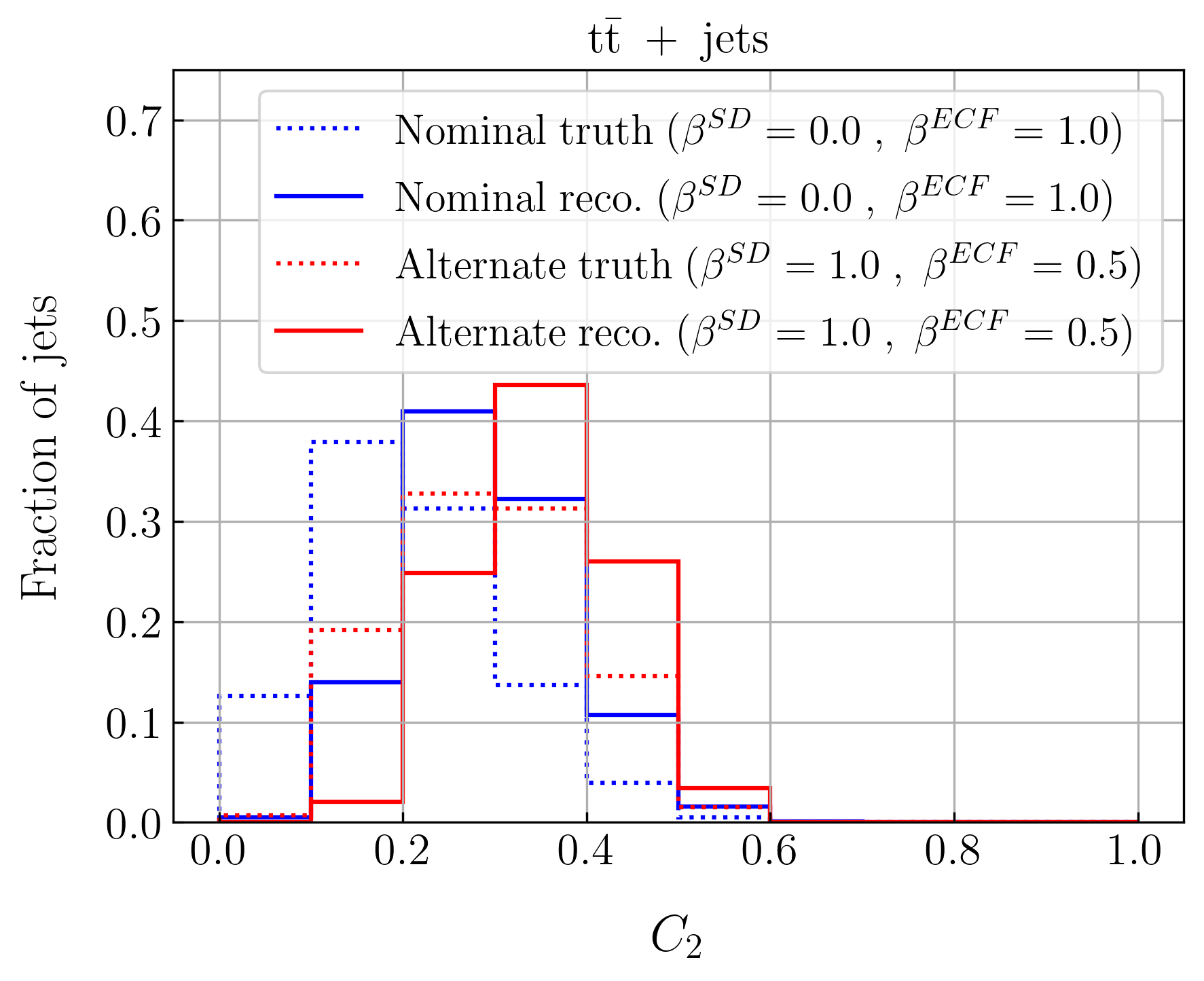}\\
\includegraphics[width=0.48\textwidth]{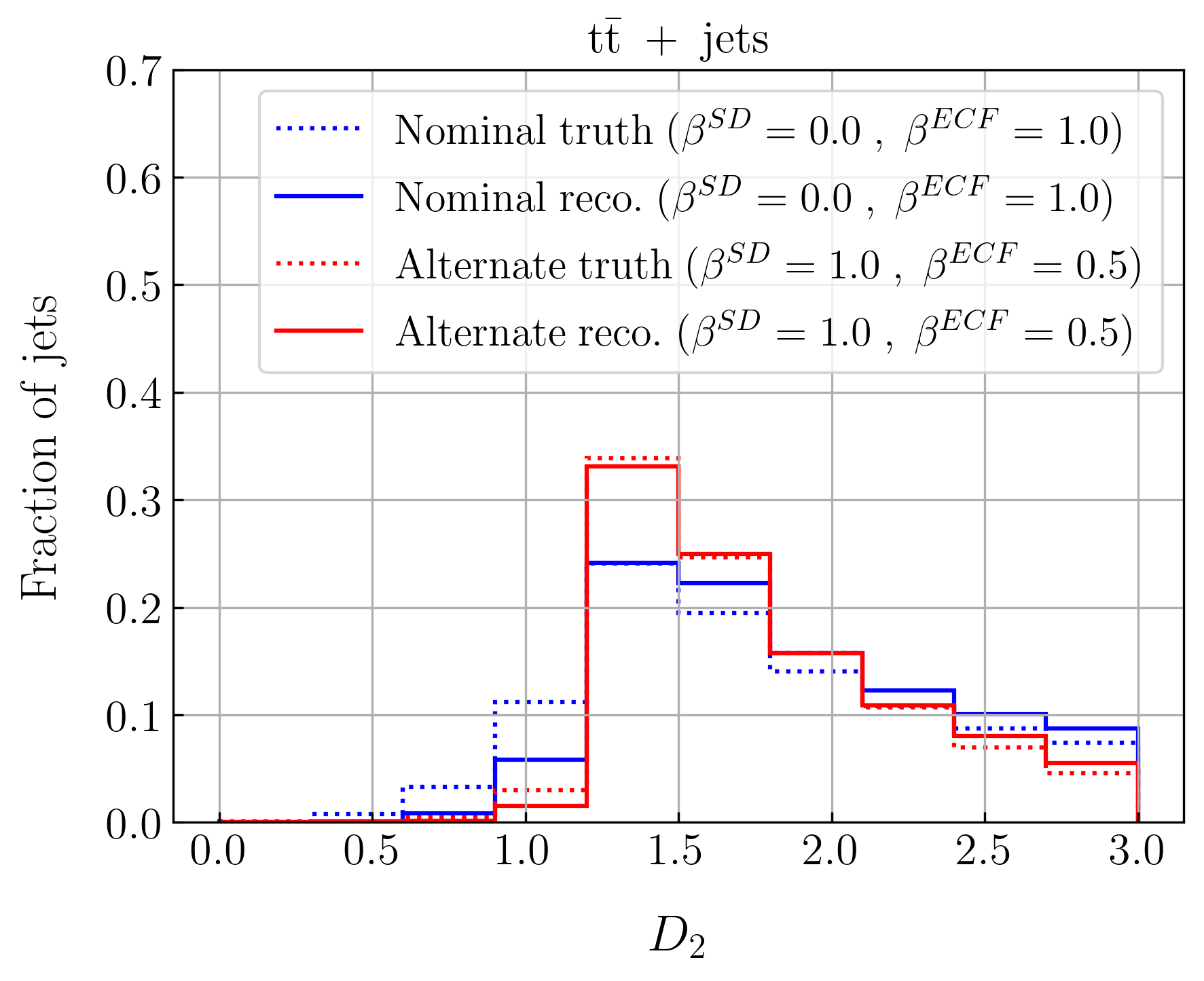}\hfill
\includegraphics[width=0.48\textwidth]{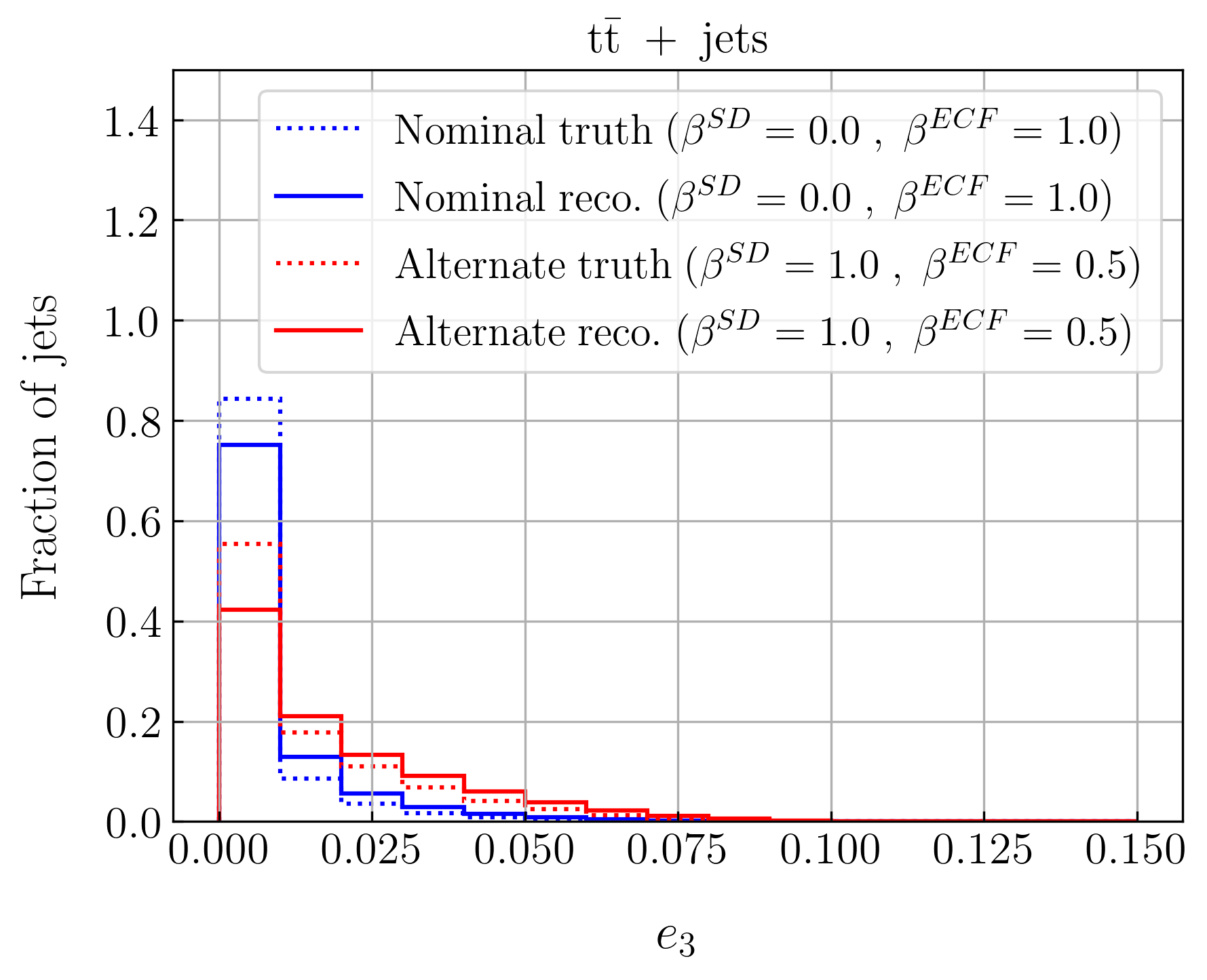}\\
\vspace{-8pt}
\caption{\label{fig:substr1}Comparisons of Soft Drop mass (upper-left), $\tau_{21}$ (upper-right), $\tau_{32}$ (middle-left), and the ECF ratios $C_{2}$ (middle-right), $D_{2}$ (lower-left), and $e_{3}$ (lower-right) distributions of the leading anti-\kt fat jet corresponding to nominal and alternate values of substructure parameters as indicated in Table~\ref{tab:boosted_params} for dynamic-radius jet clustering algorithm at truth and reconstructed levels in \ttbar + jets events.}\vspace{-14pt}
\end{figure}

\begin{figure}[hbpt]
\centering
\includegraphics[width=0.48\textwidth]{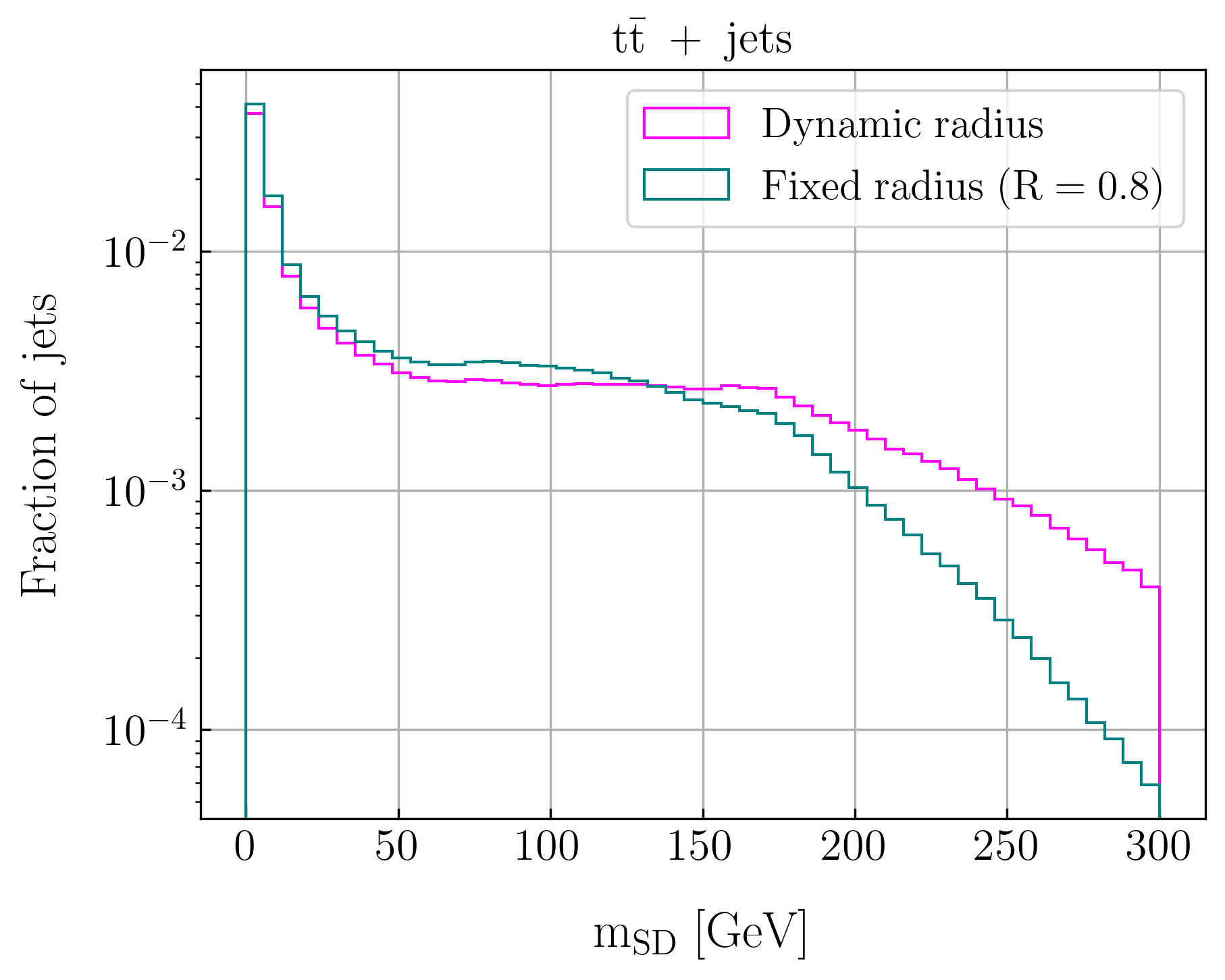}\hfill
\includegraphics[width=0.47\textwidth]{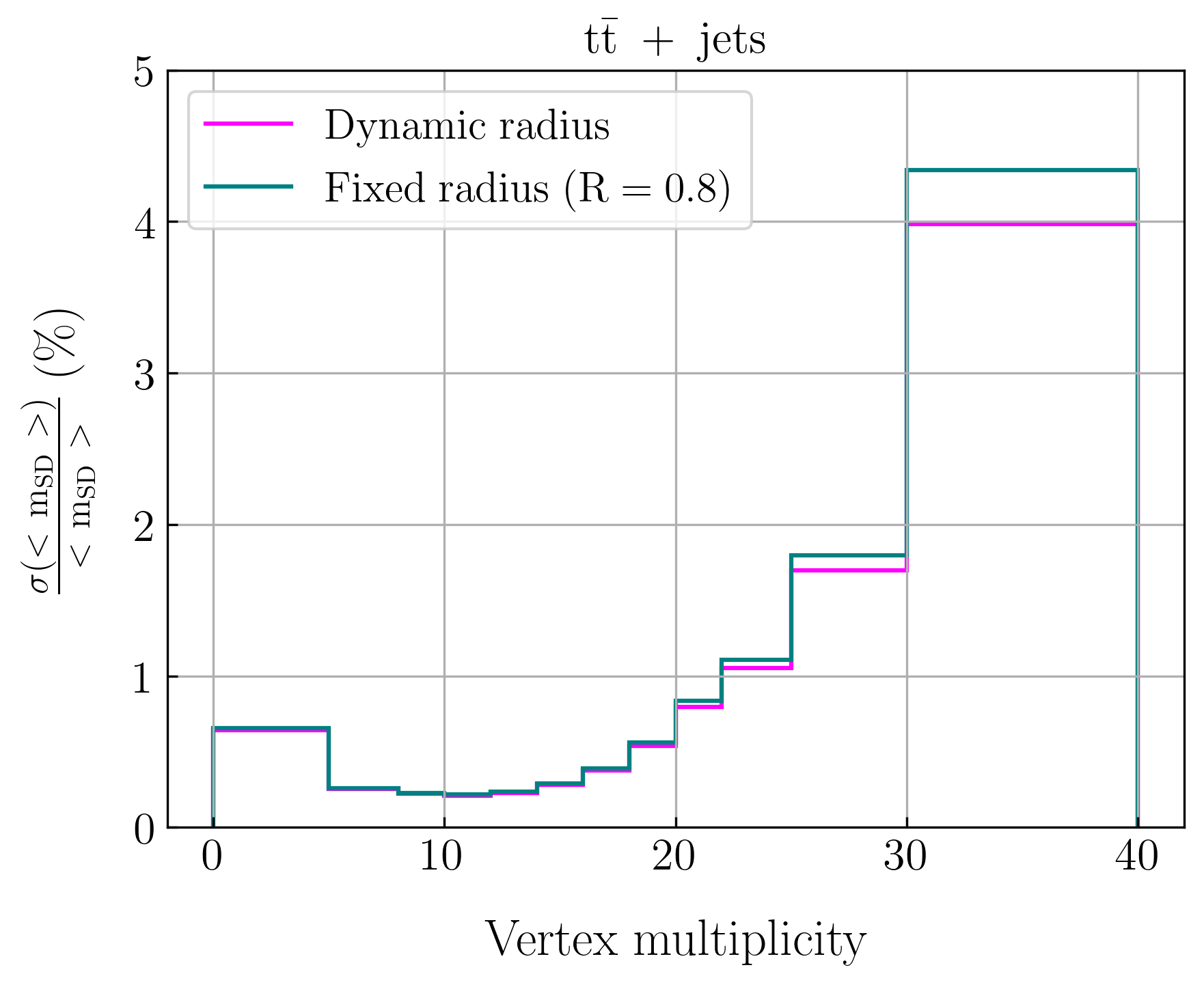}\\
\includegraphics[width=0.48\textwidth]{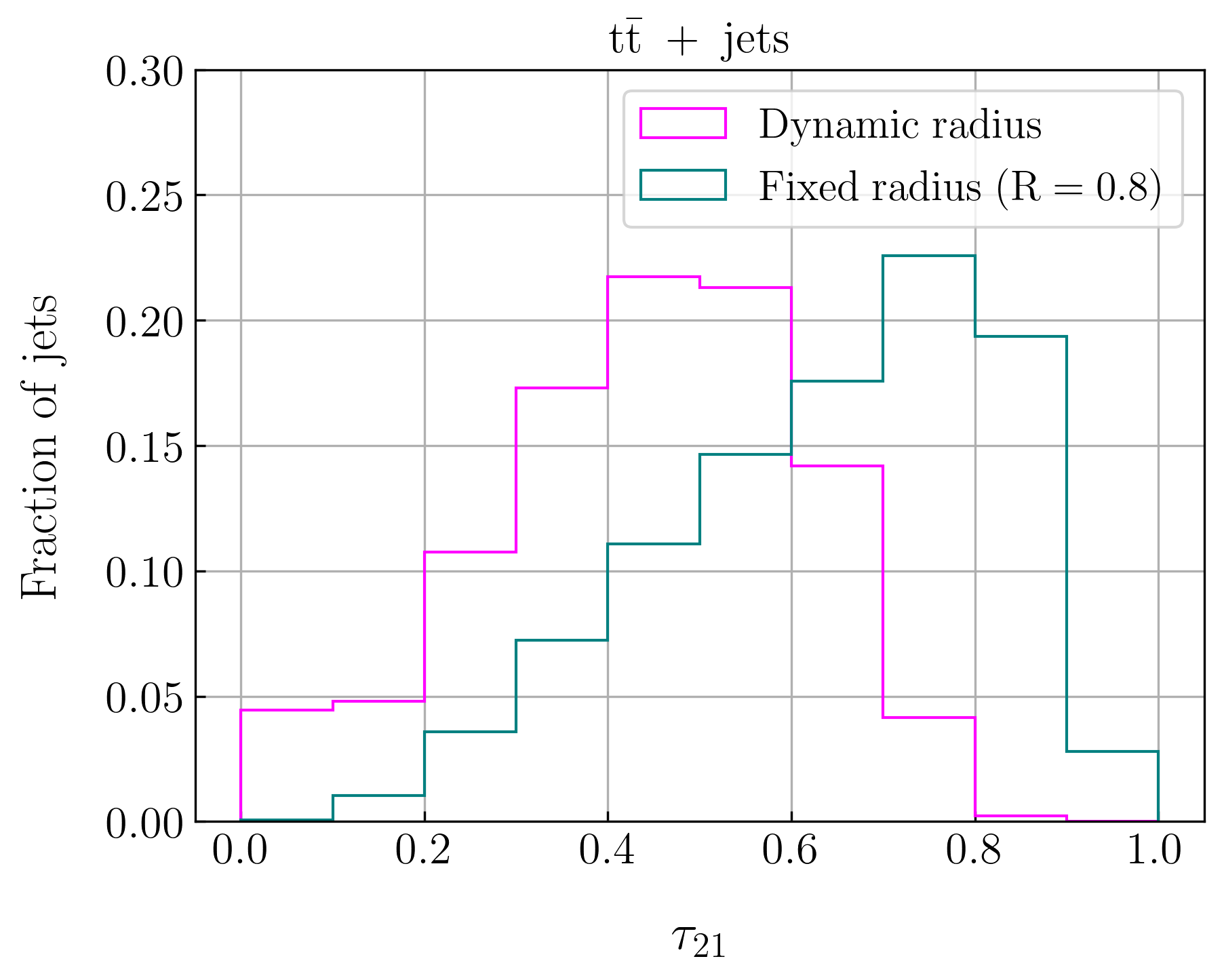}\hfill
\includegraphics[width=0.48\textwidth]{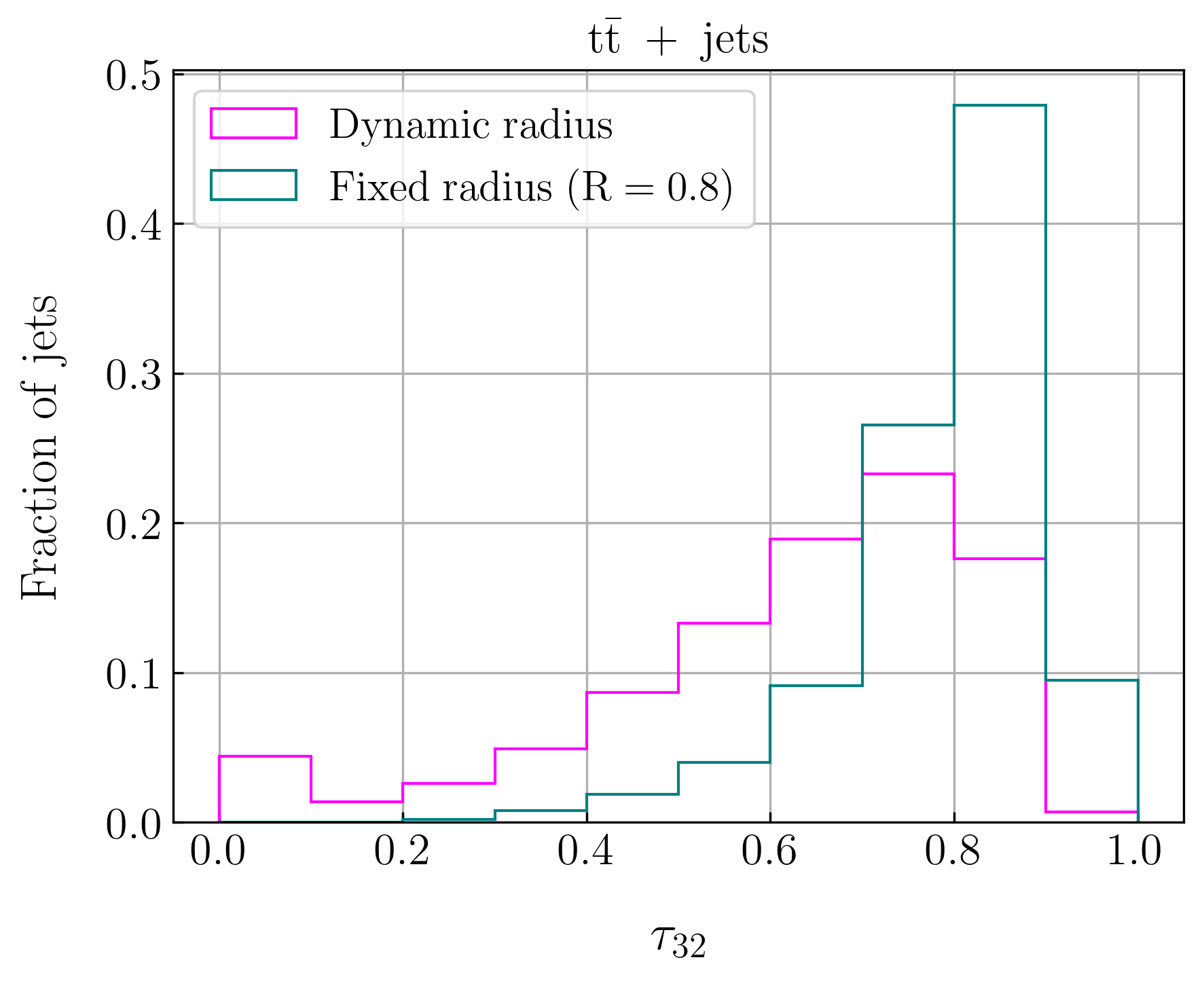}\\
\caption{\label{fig:substr2}Comparisons of soft drop mass (upper-left) and its resolution in bins of vertex multiplicity (upper-right), $\tau_{21}$ (lower-left), and $\tau_{32}$ (lower-right) distributions of the reconstructed leading anti-\kt fat jet between fixed-radius (${\rm R}=0.8$) and dynamic-radius jet clustering algorithms in \ttbar + jets events are presented. The nominal choice of substructure parameters as indicated in Table~\ref{tab:boosted_params} has been used for the dynamic-radius algorithm to obtain these plots.}
\end{figure}

The DR jet algorithm, along with the jet substructure methods, has been implemented within {\sc Delphes} and {\sc CMS SoftWare version 7\_6\_7} as appropriate plugins. 
The choices of parameters for the jet substructure methods described above for the DR jets, including both nominal and alternate values used for variations, are summarized in Table~\ref{tab:boosted_params}. 
The impact of these choices is studied by comparing the shapes of different observables sensitive to the respective substructure methods, as shown in Figure~\ref{fig:substr1}, both at truth and reconstructed levels. 
As seen in the figure, the distributions for nominal and alternate parameter choices look quite similar, suggesting that the overall analysis remains fairly robust against these variations.
In addition, substructure observables of the DR jets are also compared with the fixed-radius counterpart at the reconstructed level, and are shown in Figure~\ref{fig:substr2}. 
Only the nominal values of the substructure parameters are considered for this comparison.

\begin{table}[hbpt]
\centering
\caption{\label{tab:boosted_params} Parameter choices for jet substructure methods applied to the DR fat jets. ``Nominal Values" correspond to those used in the final analysis, while ``alternate Values" are considered for the control study shown in Figure~\ref{fig:substr1}.}\vspace{8pt}
%\resizebox{\textwidth}{!}{
\begin{tabular}{cccc}
\hline\hline \\[-8pt]
\multirow{2}{*}{Method} & \multirow{2}{*}{Parameter} &  Nominal & Alternate \\
& & Values & Values \\ \\[-8pt]
\hline \hline \\[-8pt] 
\multirow{2}{*}{Soft drop~\cite{Larkoski:2014wba}} & $\beta^{\text{SD}}$  & {\bf 0.0}~\cite{ATLAS:2019kwg} & 1.0 \\
 & $z_{\text{cut}}^{\text{SD}}$ & {\bf 0.1}~\cite{ATLAS:2019kwg} & 0.1 \\ \\[-8pt]
\hline \\[-8pt]
\multirow{2}{*}{$N$ subjettiness~\cite{Thaler:2010tr}~~} & $\beta^{\text{NS}}$ & {\bf 1.0}~\cite{ATLAS:2019kwg} & 1.0 \\
& ~~Seed Axes~\cite{Stewart:2015waa}~~ & ~~\textsc{WTA\_KT\_Axes}~\cite{Larkoski:2014bia}~~ & ~~\textsc{WTA\_KT\_Axes}~~ \\ \\[-8pt]
\hline \\[-8pt]
ECF~\cite{Larkoski:2013eya} &$\beta^{\text{ECF}}$ & {\bf 1.0}~\cite{ATLAS:2019kwg} & 0.5 \\ \\[-8pt]
\hline\hline
\end{tabular}
%}
\end{table}

The upper-left panel of Fig.~\ref{fig:substr2} shows the distribution of soft drop mass ($m_{\rm SD}$) for both fixed- and dynamic-radius algorithms. The upper-right panel in Fig.~\ref{fig:substr2} shows the variation of the resolution of the average m$_{\rm SD}$ of the leading fat jet with the number of vertices. From the figure, we see that the DR clustering consistently performs better than the fixed-radius method, especially at the large vertex multiplicities. The radius modifier, defined in Eq.~(\ref{eqn:var}), is the $\mathrm{p_T}$-weighted standard deviation of the two-particle angular distance in the $\eta$-$\phi$ plane among the jet constituents. Thus, the impact of the particles coming from pile-up vertices is partially cancelled because those are uncorrelated to the particles originating from the primary vertex constituting the jet-core.
Figure~\ref{fig:substr2} also demonstrates that the distributions of $\tau_{21}$ (lower-left) and $\tau_{32}$ (lower-right) for the DR clustering are more shifted towards $0$ compared to the fixed-radius clustering, indicating better reconstruction of the hadronically decaying boosted W bosons and top quarks in the former case with a fat jet~\cite{Thaler:2010tr}.

%% file: selection.tex
Various key features of the event reconstruction and selection are presented below.

\begin{figure}[hbpt]\vspace{-10pt}
\centering
\includegraphics[width=0.48\textwidth]{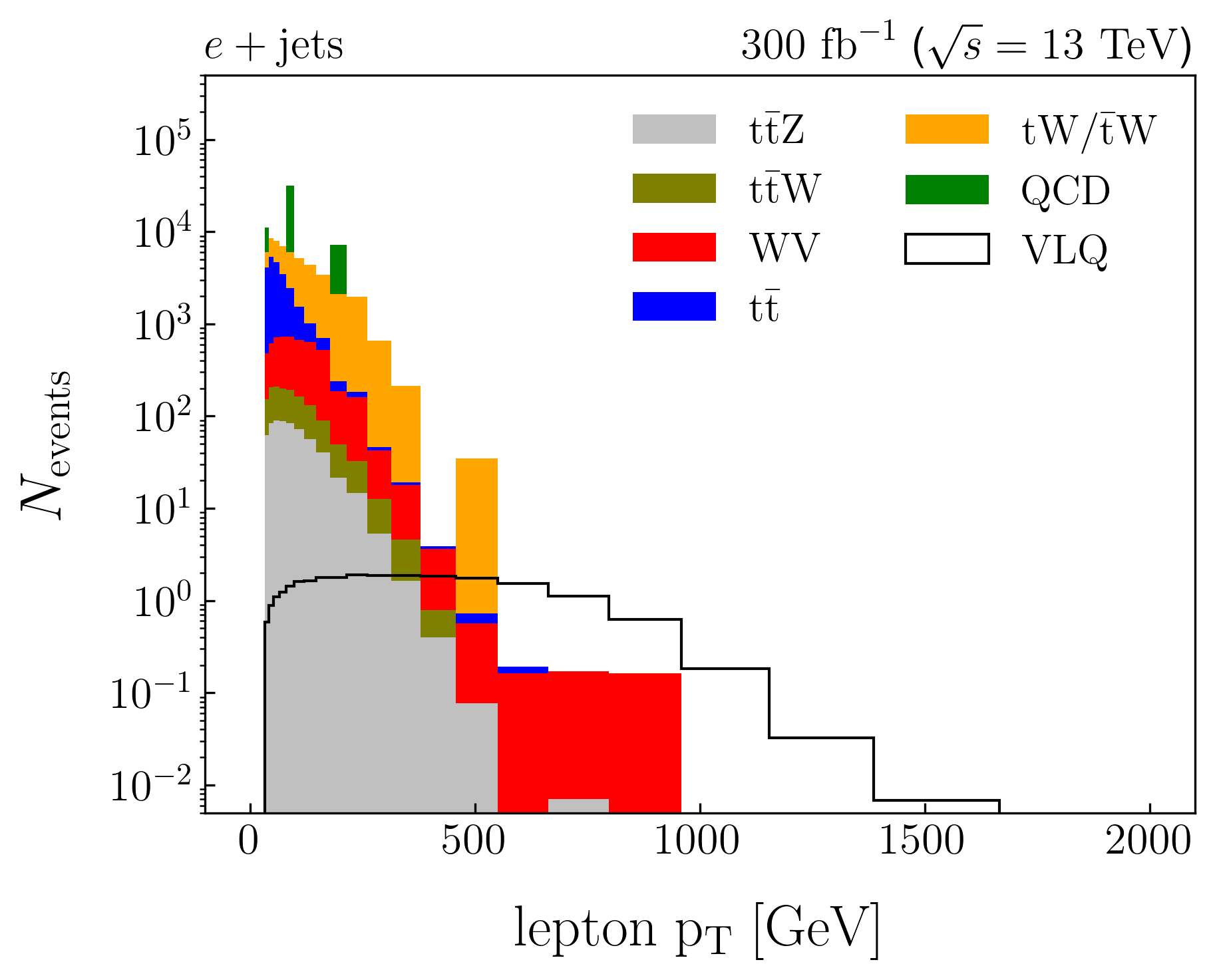}\hfill
\includegraphics[width=0.48\textwidth]{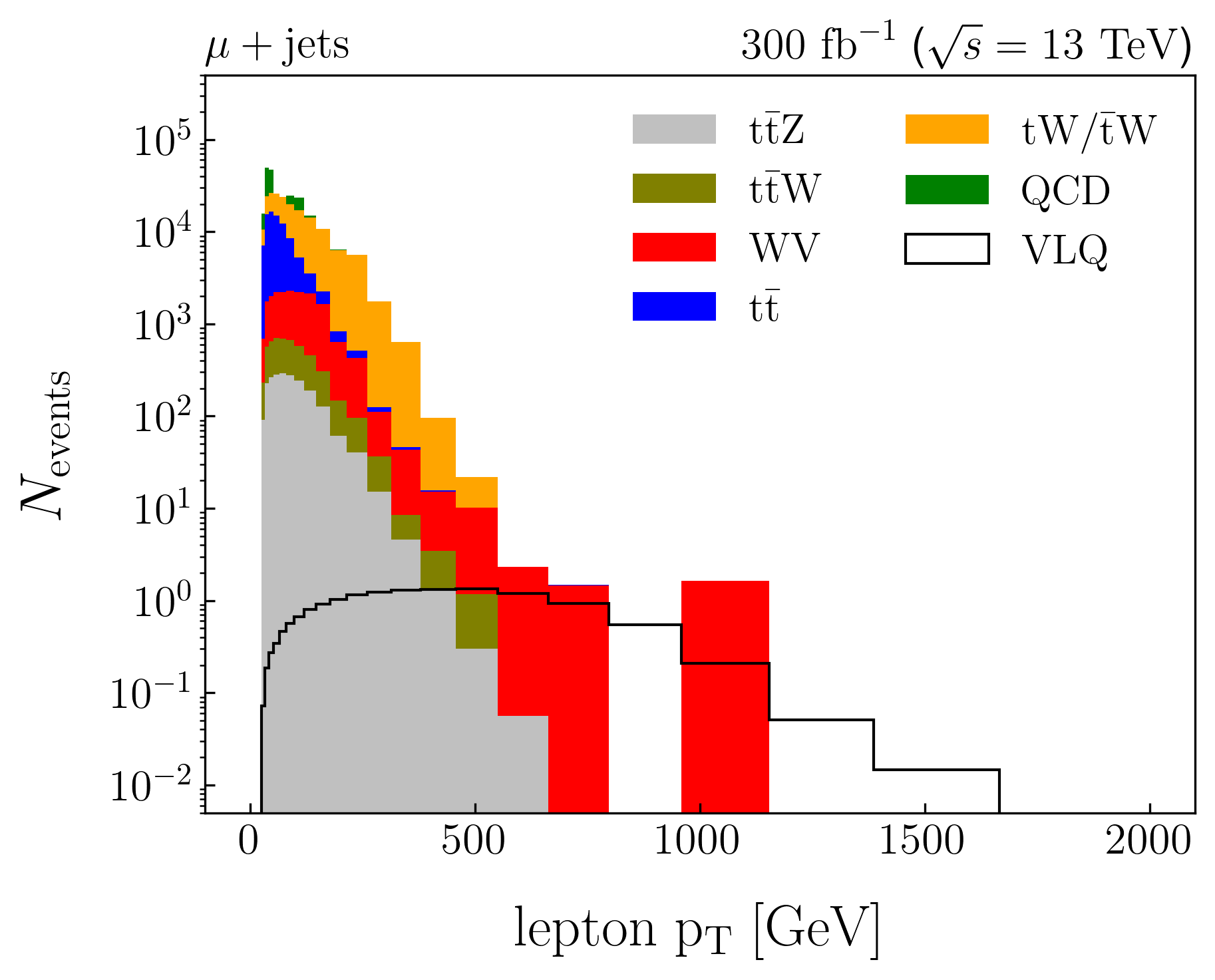}\\
\includegraphics[width=0.48\textwidth]{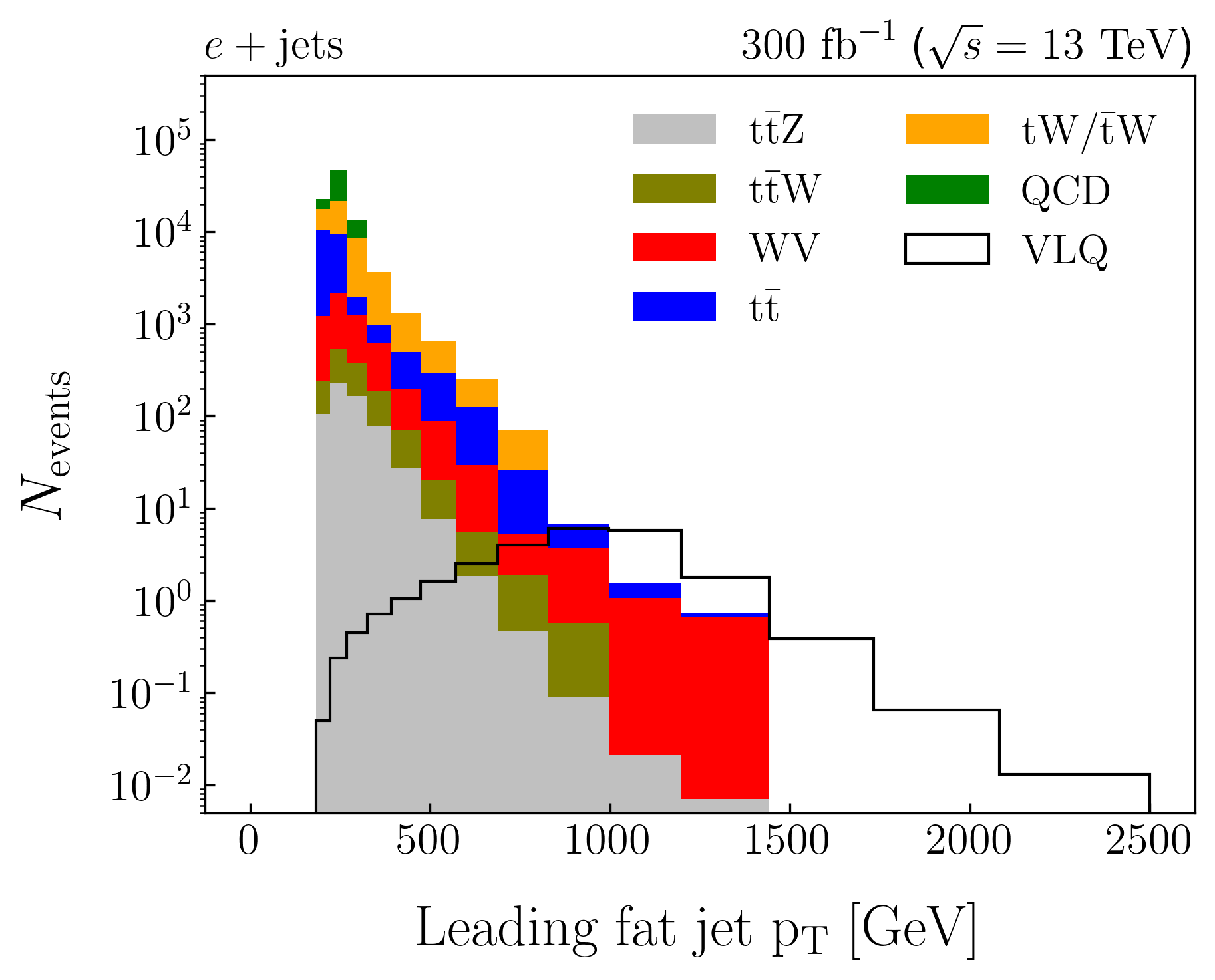}\hfill
\includegraphics[width=0.48\textwidth]{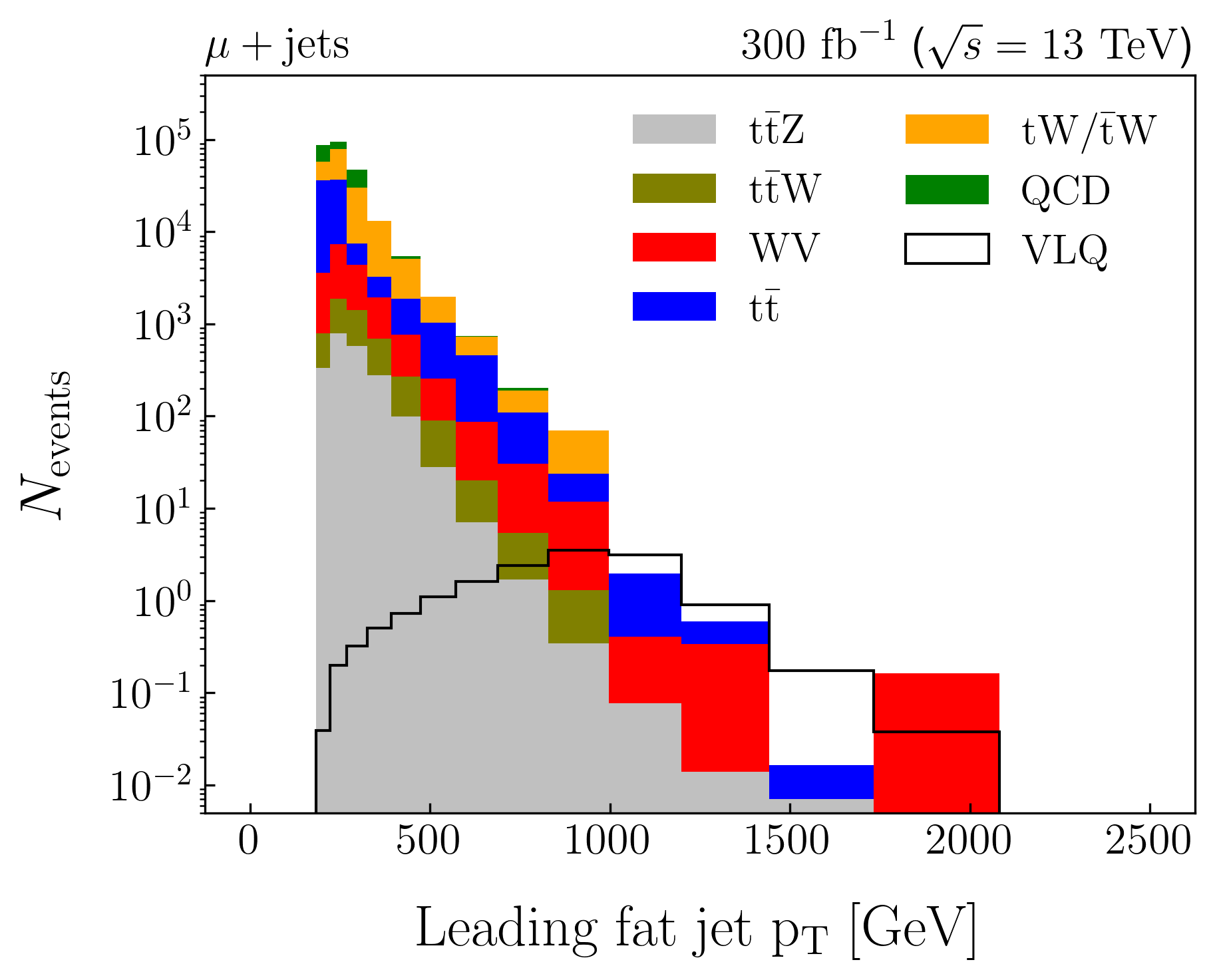}\\
\includegraphics[width=0.48\textwidth]{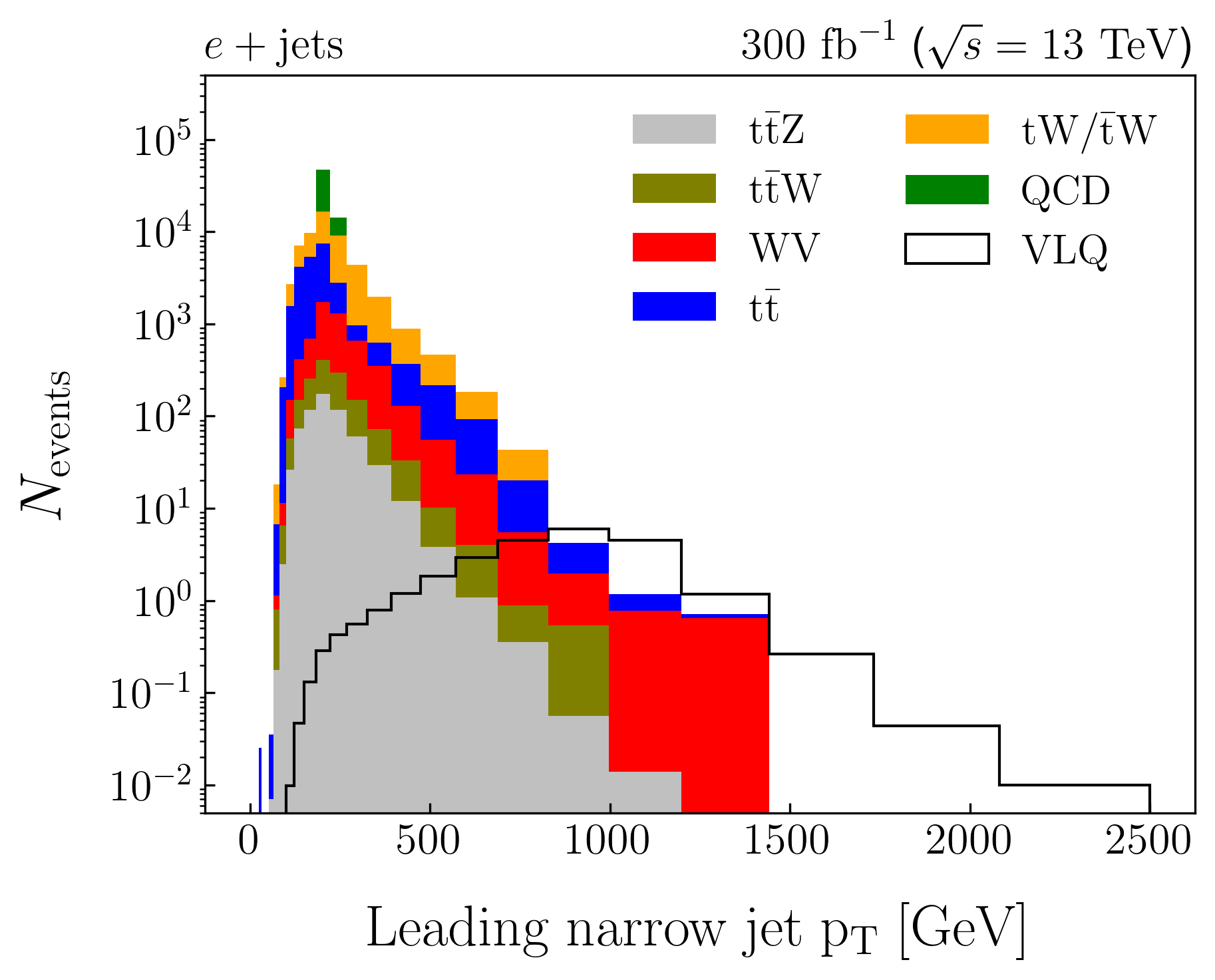}\hfill
\includegraphics[width=0.48\textwidth]{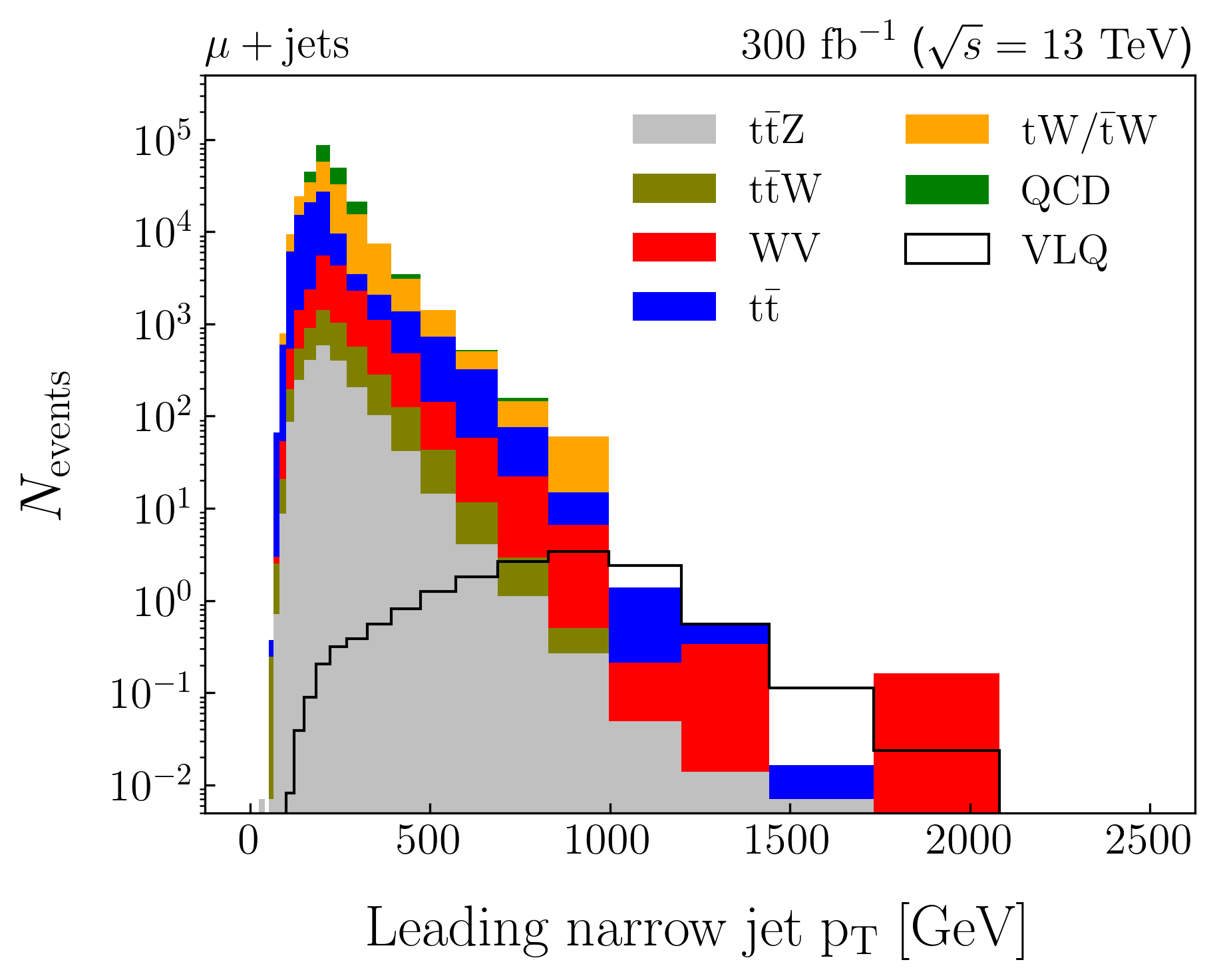}\\
\caption{\label{fig:1fj_plots_1}Distributions of lepton \pt (upper), fat jet \pt (middle), leading narrow jet \pt (lower), in the $e$ + jets (left) and $\mu$ + jets (right) final states.  In the legends, VLQ denotes the signal process with a vectorlike top partner quark, while the others correspond to the SM backgrounds as given in Table~\ref{tab:event_yields}.}
\end{figure}

\begin{figure}[hbpt]\vspace{-10pt}
\centering
\includegraphics[width=0.48\textwidth]{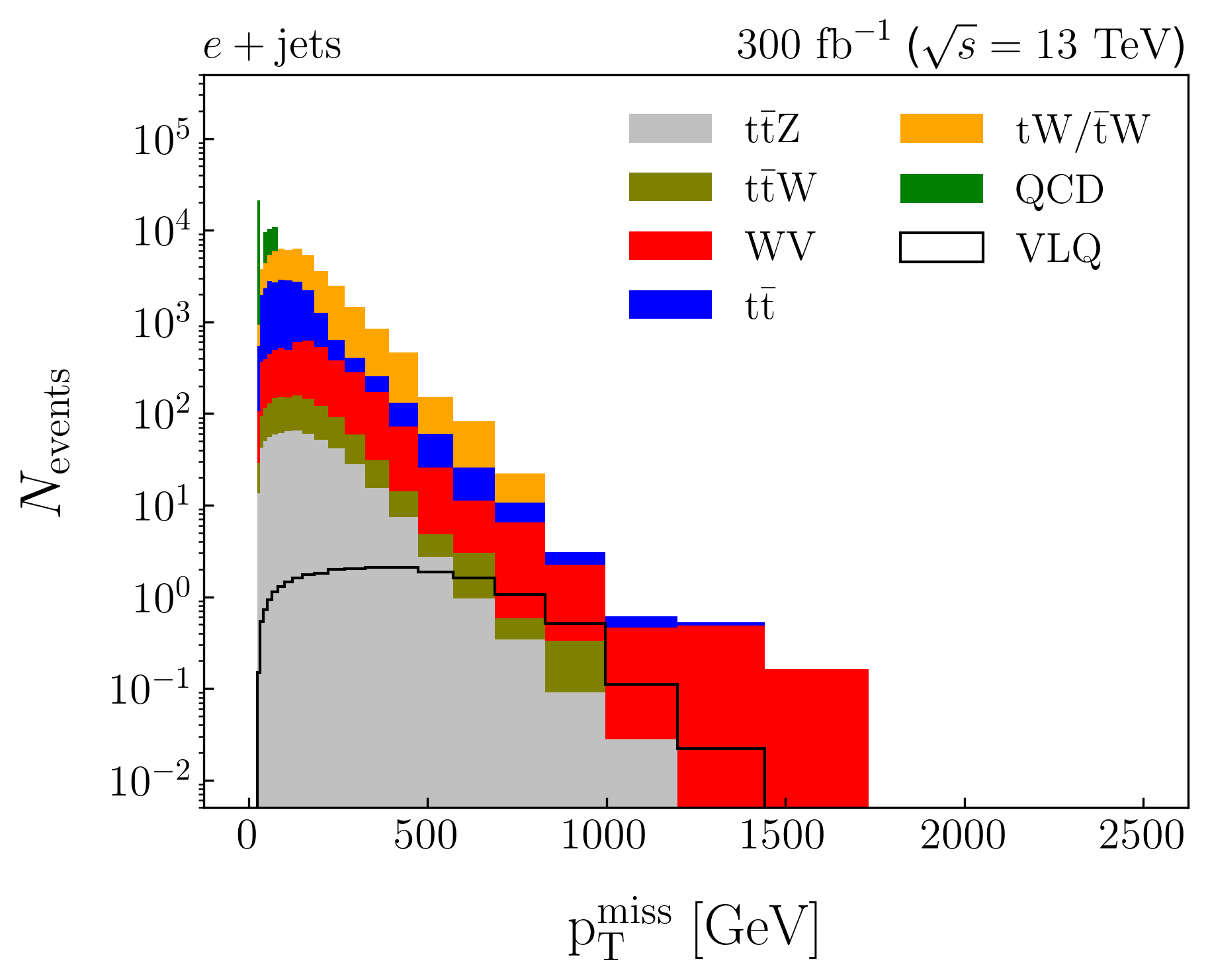}\hfill
\includegraphics[width=0.48\textwidth]{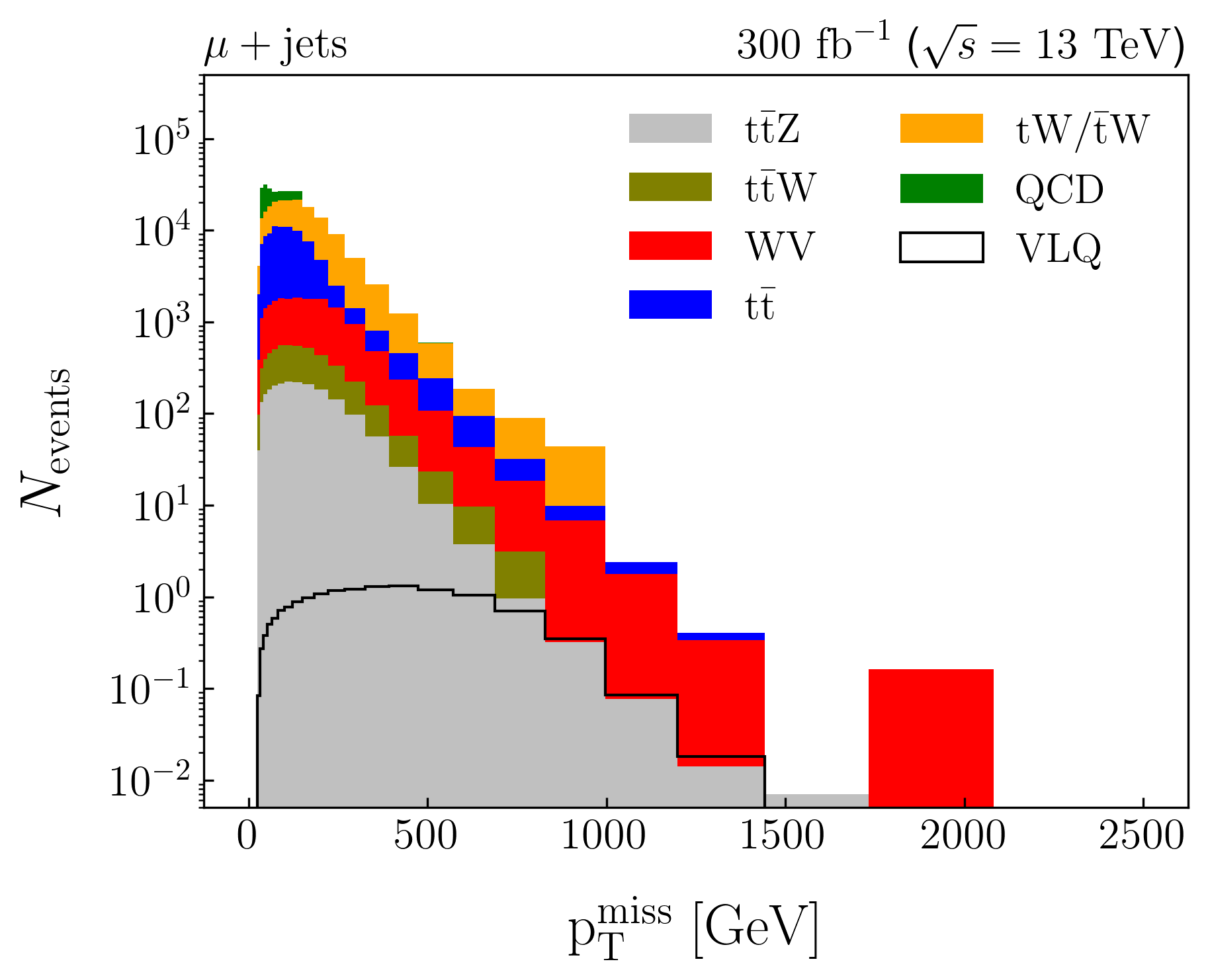}\\
\includegraphics[width=0.48\textwidth]{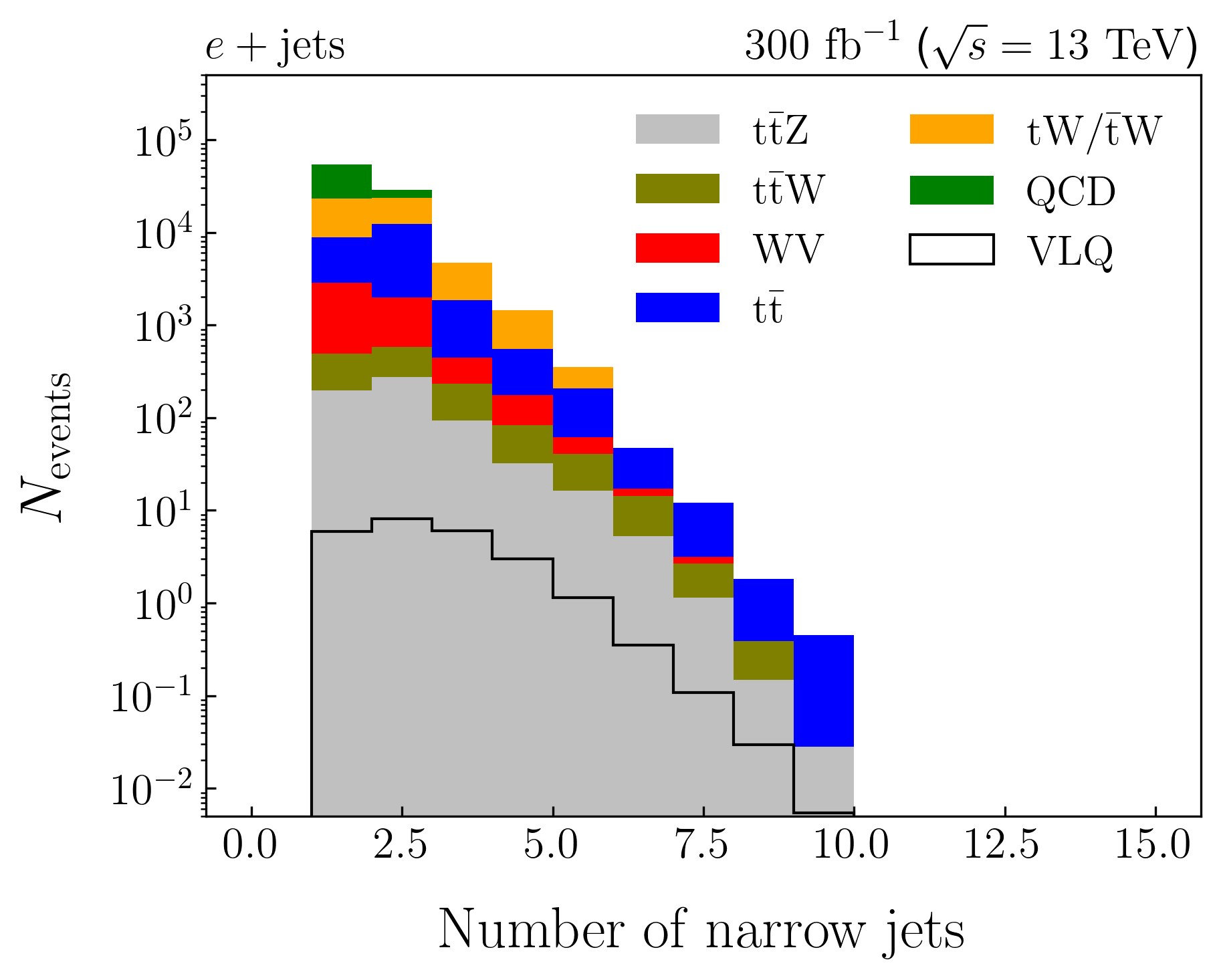}\hfill
\includegraphics[width=0.48\textwidth]{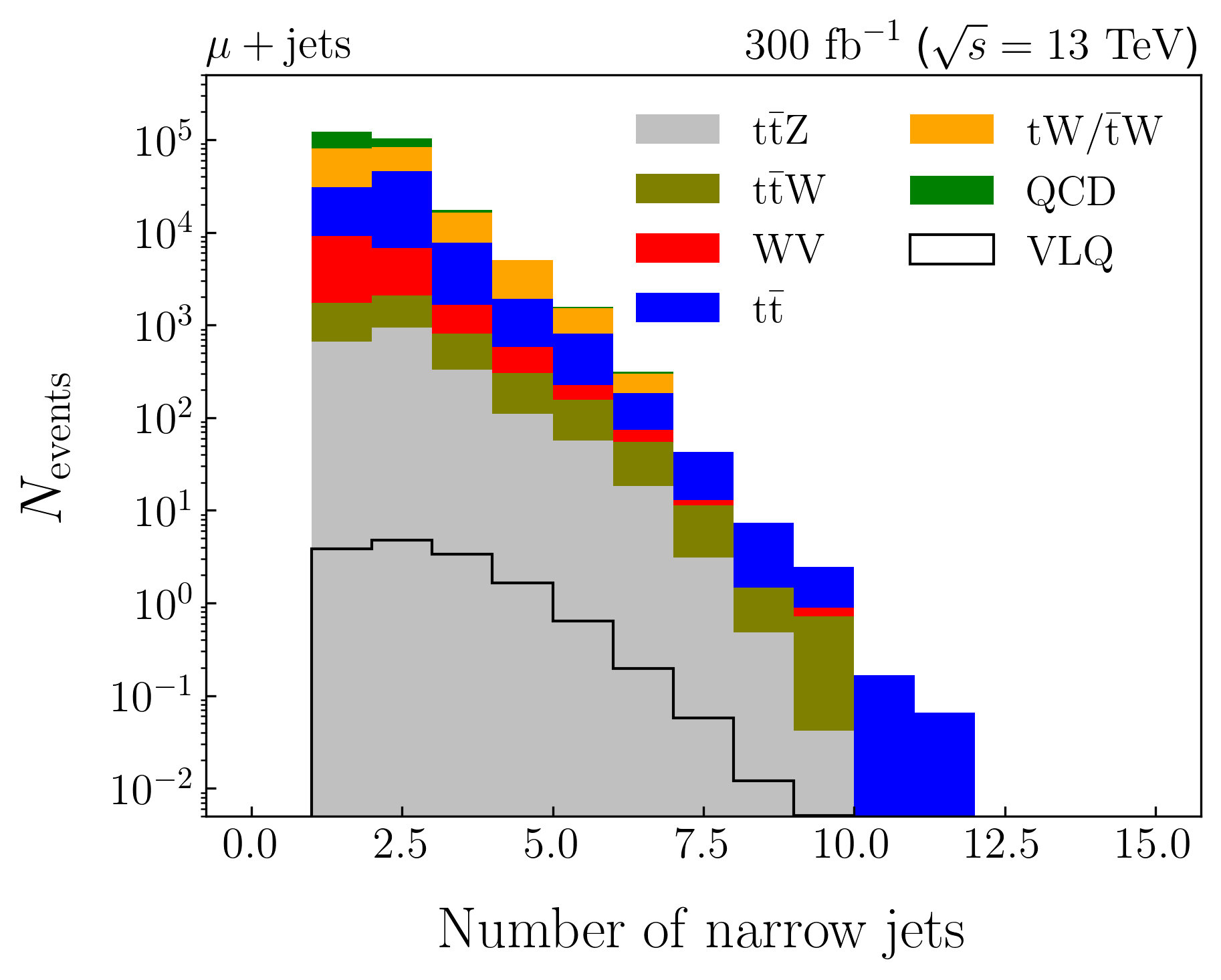}\\
\includegraphics[width=0.48\textwidth]{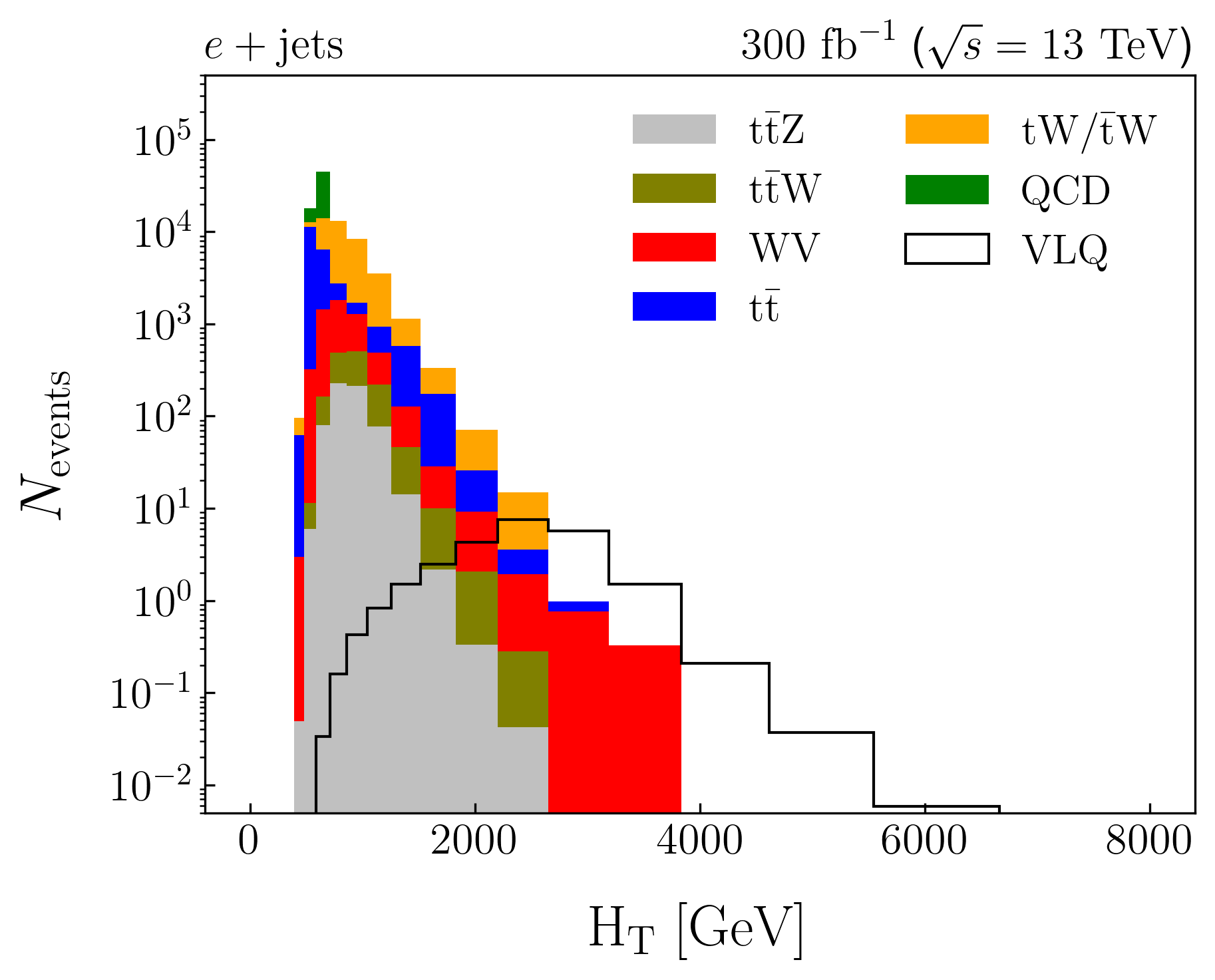}\hfill
\includegraphics[width=0.48\textwidth]{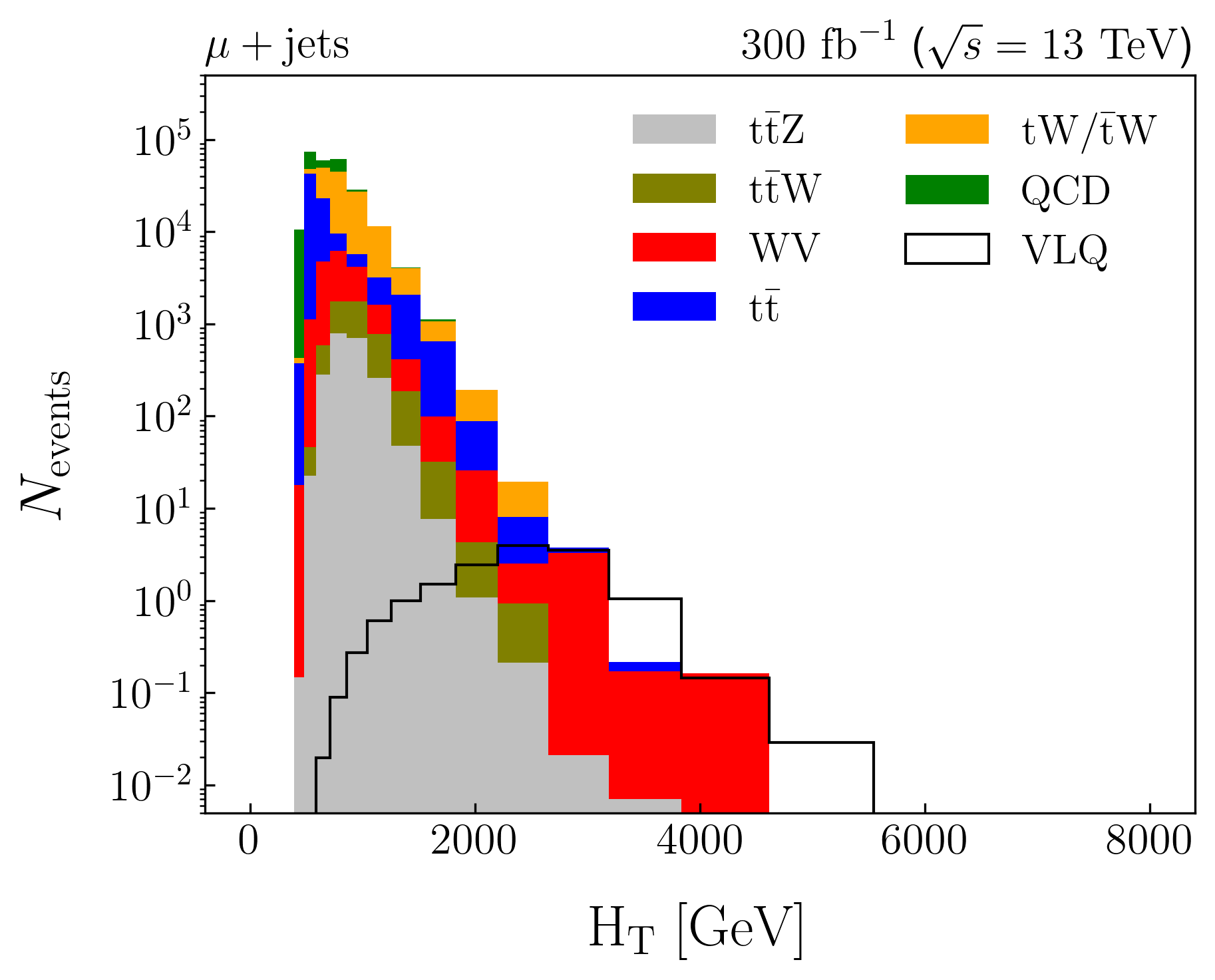}\\
%\hfill\\
\caption{\label{fig:1fj_plots_2}Distributions of \ptmiss (upper), number of narrow jets (middle), and H$_{\text{T}}$ (lower) in the $e$ + jets (left) and $\mu$ + jets (right) final states.  In the legends, VLQ denotes the signal process with a vectorlike top partner quark, while the others correspond to the SM backgrounds as given in Table~\ref{tab:event_yields}.}
\end{figure}

\begin{figure}\vspace{-10pt}
\includegraphics[width=0.48\textwidth]{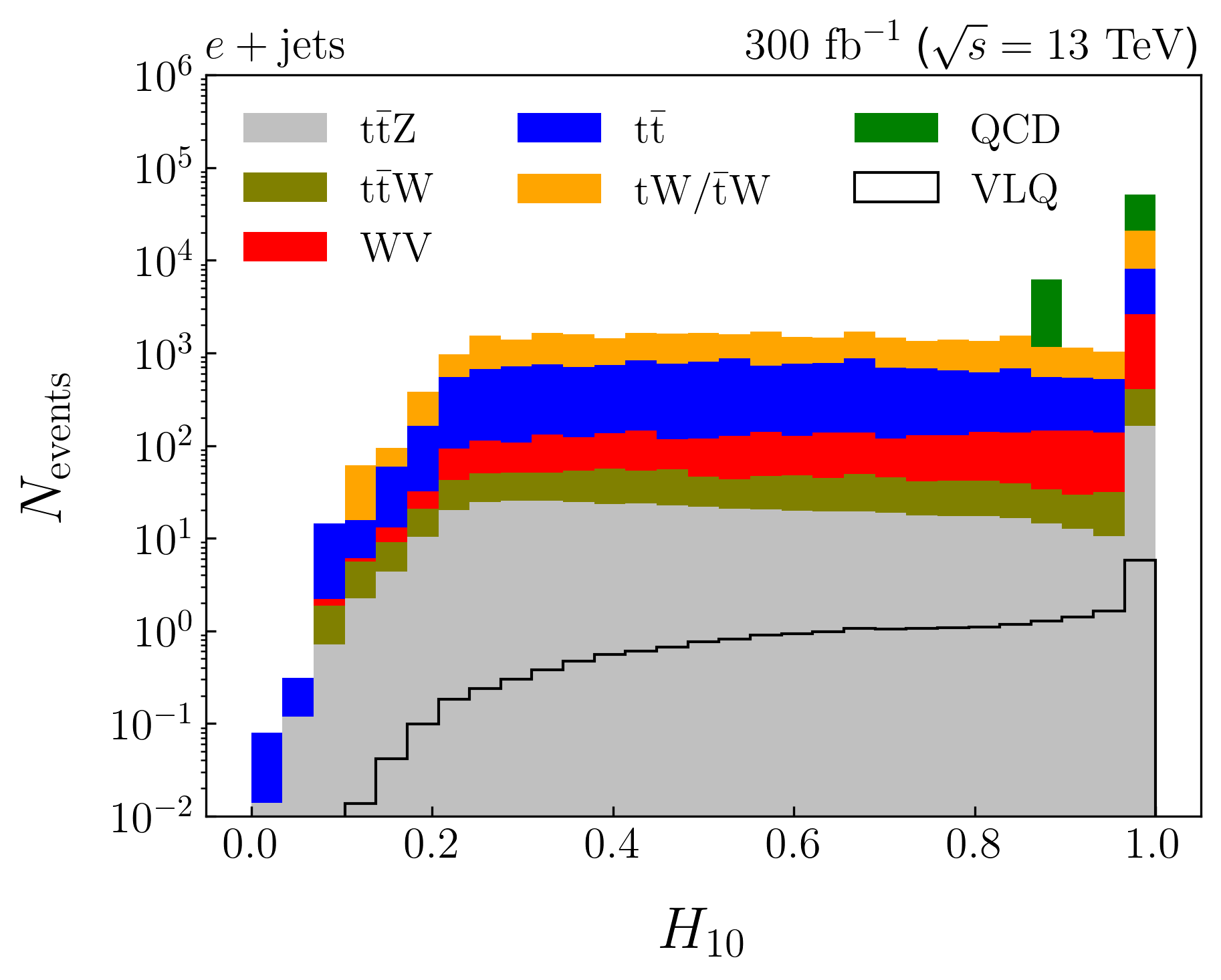}\hfill
\includegraphics[width=0.48\textwidth]{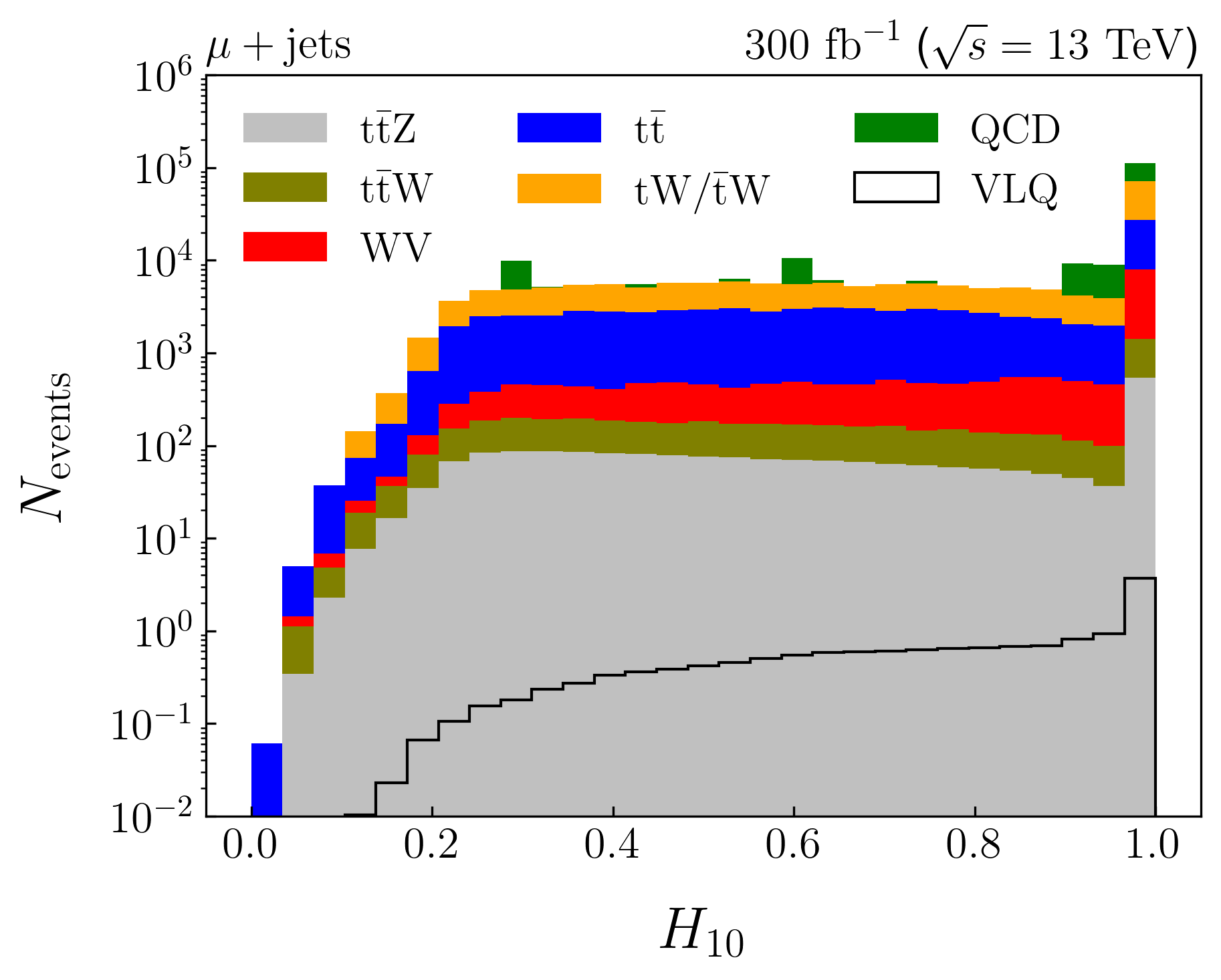}\\
\includegraphics[width=0.48\textwidth]{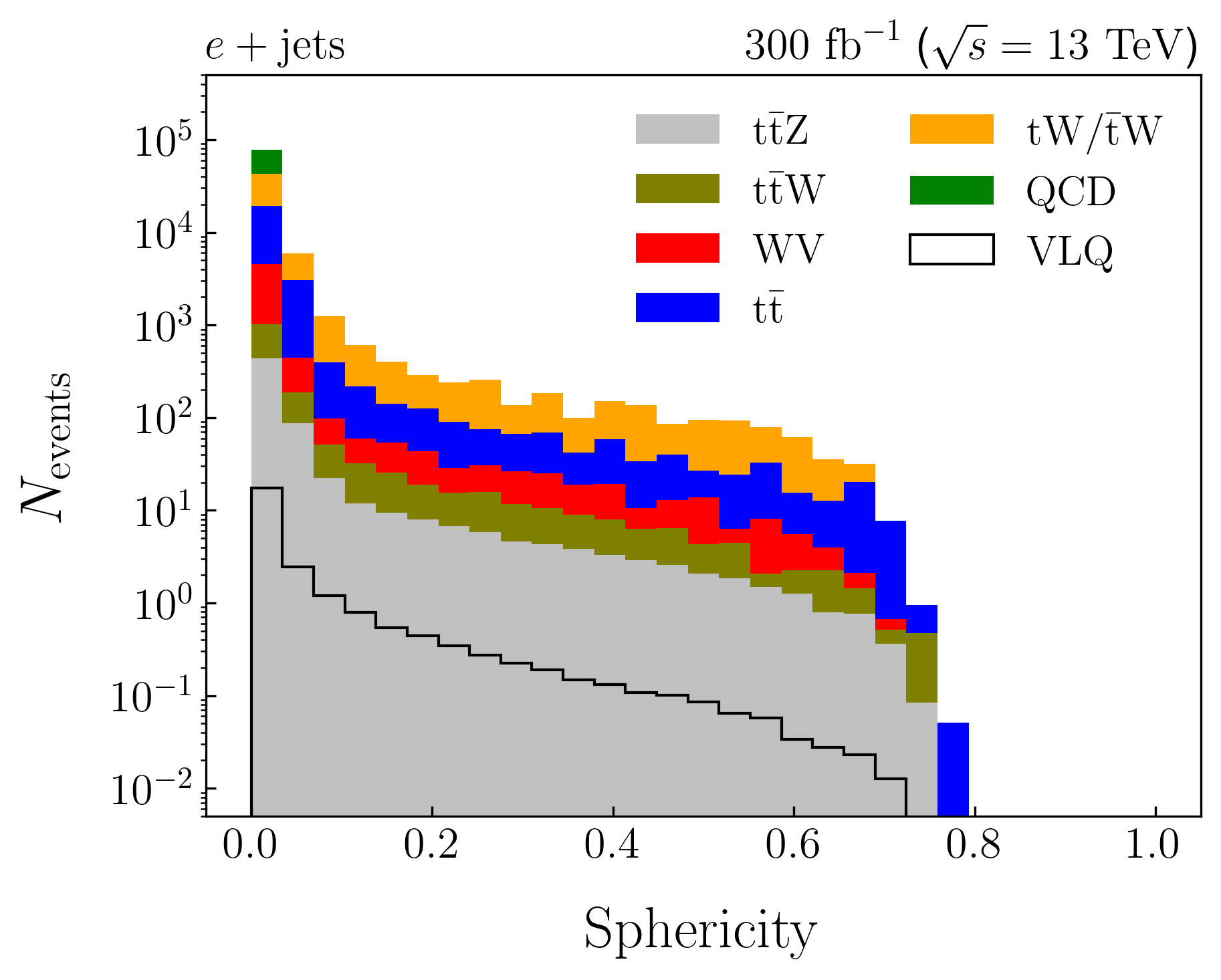}\hfill
\includegraphics[width=0.48\textwidth]{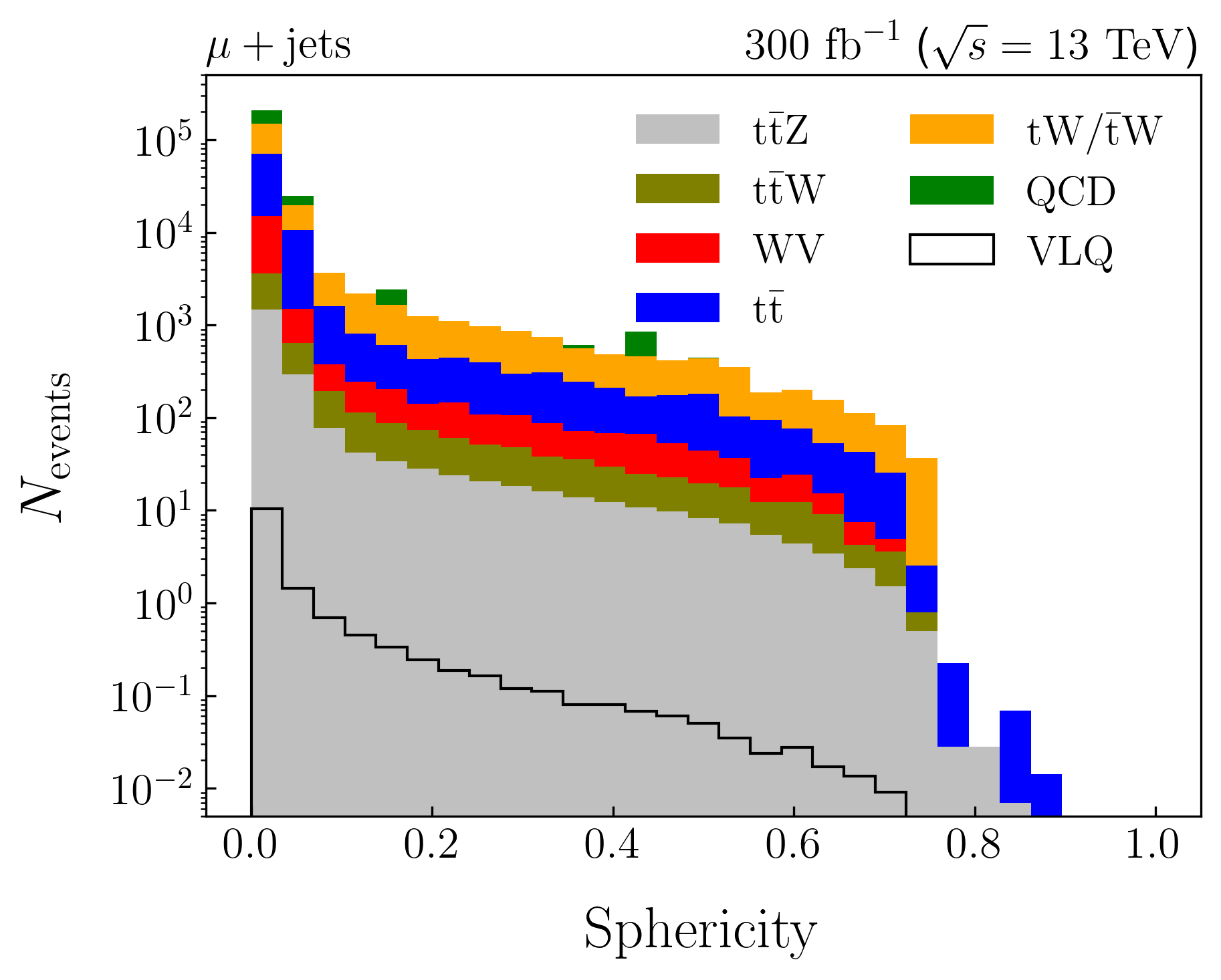}\\
\includegraphics[width=0.48\textwidth]{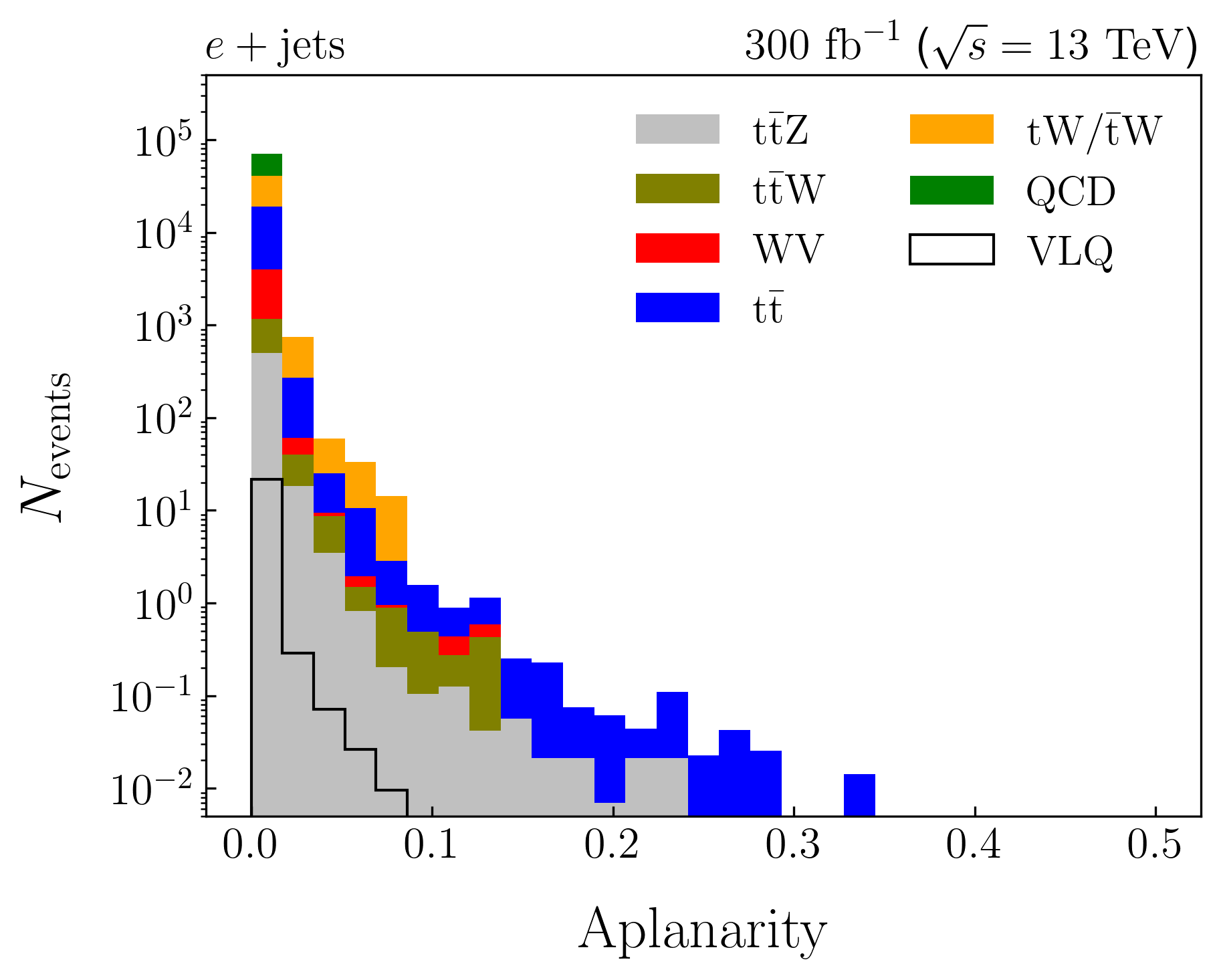}\hfill
\includegraphics[width=0.48\textwidth]{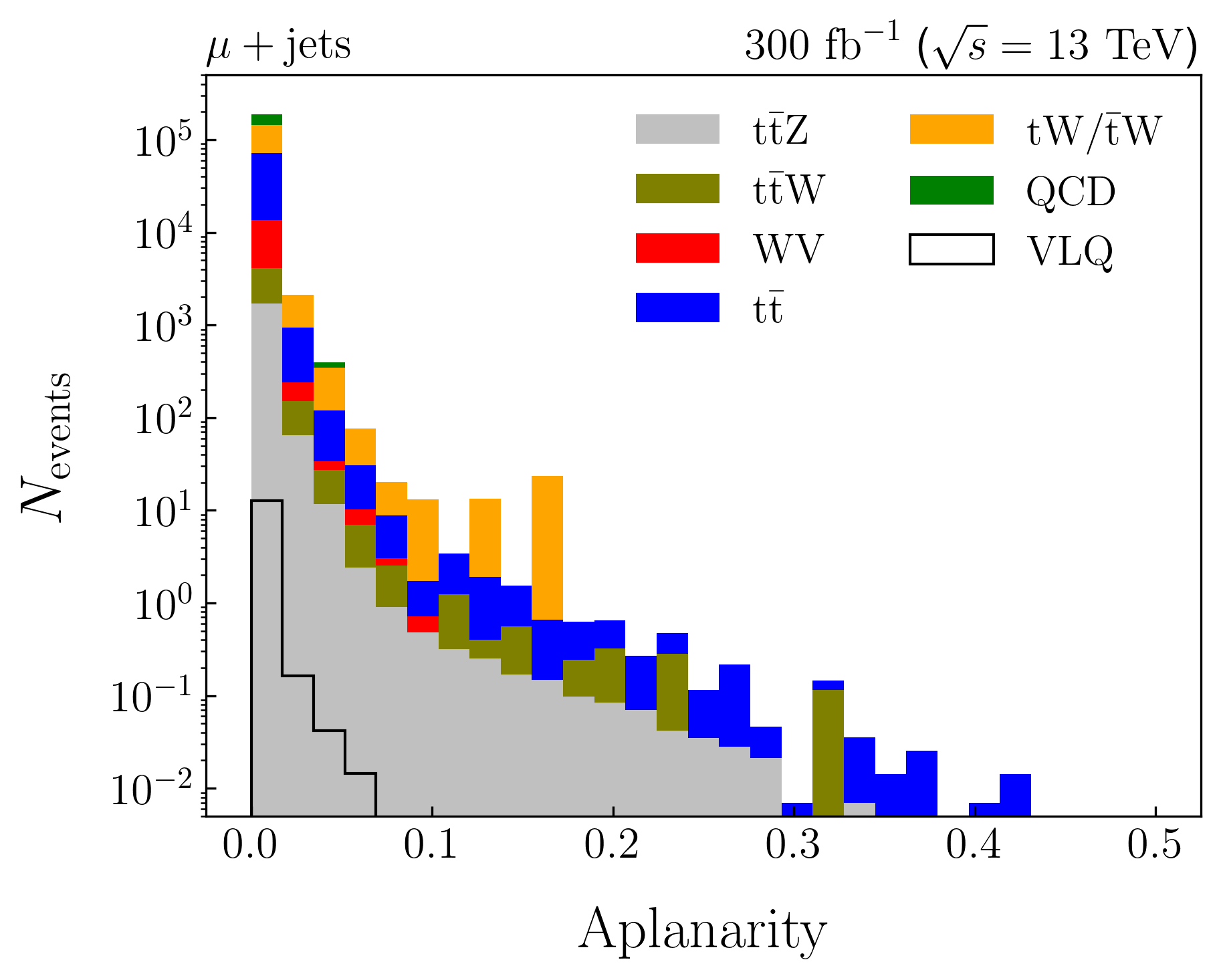}
\caption{\label{fig:1fj_evtshape}Distributions of event shape observables, namely, $10^{\text{th}}$ order Fox--Wolfram moment (upper), Sphericity (middle), and Aplanarity (lower), in the $e$ + jets (left) and $\mu$ + jets (right) final states.  In the legends, VLQ denotes the signal process with a vectorlike top partner quark, while the others correspond to the SM backgrounds as given in Table~\ref{tab:event_yields}.}
\end{figure}

\begin{description}
\item[Selection criteria]{ 
Events having a minimum of one reconstructed pp interaction vertex are retained for this study.
The vertices must be reconstructed from at least four tracks that have a longitudinal distance $|\text{d}_\text{z}| < 24$ cm and a radial distance $|\text{d}_\text{xy}| < 2$ cm from the nominal interaction point. 
If multiple vertices are found in an event, the one with the highest value of summed \sqpt of clustered particles is taken as the primary pp interaction vertex. In contrast, the others are identified as the pile up vertices. 
Individual particles are reconstructed via the particle-flow algorithm~\cite{CMS:2017yfk}.
Muons (Electrons) are required to have $\pt \geqslant 30\ (35)$ GeV, $|\eta| \leqslant 2.4\ (2.1)$ and to pass the ``tight'' identification criteria~\cite{CMS:2018rym, CMS:2020uim}.
The selected muon and electron candidates must pass the criterion on relative isolation ($\relIso$) to be within $6\%$ of their respective transverse momentum. 
Additionally, the selected electron candidate is required to satisfy $|\text{d}_\text{xy}| \leqslant 0.05\ (0.1)$~cm and $|\text{d}_\text{z}| \leqslant 0.1\ (0.2)$~cm relative to the primary vertex in the barrel (endcap) region of the electromagnetic calorimeter (ECAL).
Events containing additional muons (electrons) with $\pt \geqslant 10\ (15)$ GeV, $|\eta| \leqslant 2.4\ (2.5)$, and passing the ``loose'' (``veto'') identification criteria are rejected. 
In such cases, the criteria on the additional lepton isolation are relaxed to $\relIso \leqslant 20\%$ for muons and $\relIso \leqslant 18\%\ (16\%)$ for electrons in the barrel (endcap) ECAL region.
Standard jets are reconstructed using the AK clustering algorithm~\cite{Cacciari:2008gp} with radii of 0.4 and 0.8 for the narrow (AK4) and fat (AK8) jets, respectively, as implemented in the {\sc FastJet} package~\cite{Cacciari:2011ma} with the particle-flow candidates as inputs. 
The effect of additional tracks and energy deposits in calorimeters from pile up on the AK4 and AK8 jet momenta is mitigated by discarding tracks identified to be originating from pile up vertices, as well as by applying an offset correction to account for residual neutral pile up contributions~\cite{Cacciari:2007fd,CMS:2020ebo}. 
``Tight lepton veto'' identification criteria~\cite{CMS-PAS-JME-16-003} are applied to suppress jets arising from spurious sources, such as electronic noise in the calorimeters and jets containing ``tight''-ly identified muons or electrons.    
The combined secondary vertex tagging algorithm version 2 ({\sc CSVv2})~\cite{CMS:2017wtu} is used to identify AK4 jets originating from bottom quarks or antiquarks.
Candidate events are required to have {\bf exactly one} AK8 jet with \pt $\geqslant 200~$GeV and within $|\eta| \leqslant 4.0$, as mentioned in Section~\ref{ssec:sig_simu}.
In addition to the selected AK8 jet, events must have {\bf at least one} non-overlapping AK4 jet with $\pt \geqslant 30$ GeV, $|\eta| \leqslant 2.5$ and passing the ``medium'' working point defined on the {\sc CSVv2} score.
Finally, events should pass the $\ptmiss \geqslant 30~$GeV criterion.

We note that a full detector simulation with {\sc Geant4} with CMS detector specification has been used in the background MC samples.
However, for the signal MC samples, a fast detector simulation using {\sc Delphes} has been used.
According to Ref.~\cite{deFavereau:2013fsa}, the overall agreement in the average performance of {\sc Delphes} with the full simulation of the CMS detector based on {\sc Geant4} is reasonably good for different high-level physics objects, especially in the phase space region of interest.
Furthermore, the measured efficiencies of lepton (both selected and vetoed leptons)~\cite{CMS:2018rym, CMS:2020uim} and jet~\cite{CMS-PAS-JME-16-003, CMS:2017wtu} reconstruction are incorporated into the {\sc Delphes} for signal events in bins of their respective $\pt$ and $\eta$. 
In case of backgrounds, to mitigate observed differences with collision data; respective efficiency corrections are applied~\cite{CMS:2018rym, CMS:2020uim, CMS-PAS-JME-16-003, CMS:2017wtu}.
Table~\ref{tab:event_yields} lists signal and background yields after all event selection steps for the two lepton flavours, while Figures~\ref{fig:1fj_plots_1} -- \ref{fig:1fj_evtshape} show the distributions of observables related to lepton, jet, and event kinematics as indicated in Table~\ref{tab:mvaInputs_1fj}.

\begin{table}[hbpt]\vspace{-8pt}
\centering
\caption{\label{tab:event_yields}Event yields of signal and background processes for the two lepton flavours after applying all selection criteria for an integrated luminosity of 300 fb$^{-1}$.}
%\resizebox{\textwidth}{!}{
\begin{tabular}{c|c|c|c|c|c|c|c|c}
\hline\hline
& \multicolumn{6}{c|}{} & & \\[-10pt]
\multirow{4}{*}{Final state} & \multicolumn{6}{c|}{Bkg. processes} & \multirow{4}{*}{Total Bkg.} & \multirow{4}{*}{Signal}\\ 
 & \multicolumn{6}{c|}{} & & \\[-8pt] \cline{2-7}
 & & & & & & & & \\[-8pt]
 & \ttbar & tW~/$~\bar{\text{t}}$W & \multirow{2}{*}{\TTZ} & \multirow{2}{*}{\TTW} & WV + jets  & QCD  &  &  \\
 & + jets & + jets & & & (V$~=~$W or Z) & multijet &  & \\
\hline\hline
 & & & & & & & & \\[-10pt]
$e~$+ jets & 18313 & 29544 & 620 & 825 & 4077 & 35634 & 89013 & 25 \\
& & & & & & & & \\[-10pt]
$\mu~$+ jets &  69083 & 99413 & 2112 & 3035 & 13243 & 63309 & 250196 & 15 \\
\hline\hline 
\end{tabular}
%}
\end{table}
}
\item[Reconstruction of dynamic-radius jets]{
In addition to the AK4 and AK8 jets, we reconstruct the DR anti-\kt jets (see Section~\ref{ssec:drjetsalgo}) for the {\bf selected events}\footnote{The jet selection criteria are applied to AK4 and AK8 jets, and {\bf not to} dynamic radius jets. This choice is made primarily due to the absence of optimized pile up contamination removal, jet identification working points, jet energy resolution and correction, etc., for the dynamic radius jets.} using the particle-flow candidate information as input. 
The DR algorithm is used to form two jet collections per event, corresponding to ${\rm R_0}$ values of $0.3$ and $0.7$ for narrow and fat jets, respectively. 
The narrow jets that do not overlap with any fat jet in the event are considered for further analysis. 
The {\sc CSVv2} score of the nearest AK4 jet is assigned as the b-tagging discriminant value for each narrow jet.
}
\item[Signal event reconstruction with dynamic-radius jets]{ 
It is important to understand the impact of the DR clustering method on the signal event reconstruction, especially in the boosted regime.
A simple study is performed for this purpose by applying the following selections on the jet substructure observables shown in Figure~\ref{fig:substr2}, to identify jets due to hadronically decaying boosted W or Z bosons and top quarks in the simulated signal events.
\begin{itemize}
\item{{\bf Boosted W- or Z-like jet selection criteria:} m$_{\text{SD}}\in [60.0, 100.0]~$GeV,\ $\tau_{21} < 0.5$,\ and fat jet $\pt \geqslant 900~$GeV.}
\item{{\bf Boosted toplike jet selection criteria:} m$_{\text{SD}}\in [150.0, 200.0]~$GeV,\ $\tau_{32} < 0.5$,\ and fat jet $\pt \geqslant 900~$GeV.}
\end{itemize}
The results are presented in Table~\ref{tab:eff} in terms of the selection efficiencies of the boosted W or Z bosons and top quarks for the two clustering methods at high fat jet \pt. 
Higher selection efficiencies are observed for both the boosted W- or Z-like and toplike jets in the case of the DR clustering method, indicating better reconstruction of the daughters of the $T$ quark in the boosted regime; thus improving the overall signal event reconstruction in this region.
The adaptive approach adopted by the DR clustering (see Section~\ref{ssec:drjetsalgo}) helps in capturing all daughters of the boosted W or Z boson or the top quark originating from the $T$ quark decay (see Table~\ref{tab:event_topo}) within the fat jet by optimizing the jet size.
\begin{table}[hbpt]\vspace{-8pt}
\centering
\caption{\label{tab:eff}Selection efficiencies of boosted W- or Z-like and toplike jets for fat jet \pt $\geqslant 900~$GeV corresponding to fixed- and dynamic-radius clustering methods in simulated signal events.}
\begin{tabular}{c|c|c|c}
\hline\hline
\multirow{2}{*}{Process} & Clustering & \multicolumn{2}{c}{Selection efficiency} \\ \cline{3-4}
  & method & Boosted W or Z boson & Boosted top quark \\
\hline\hline
 & & & \\[-10pt]
\multirow{2}{*}{Signal} & Fixed radius & $38.3\%$ & $46.5\%$ \\
  & Dynamic radius & {\bf 41.1\%} & {\bf 47.5\%} \\
\hline\hline
\end{tabular}
\end{table}
}
\end{description}

%% file: result.tex
\subsection{\label{ssec:mva}Multivariate analysis}\vspace{-2mm}
Several observables are combined into boosted decision tree (BDT) discriminants to effectively separate signal from the collective SM backgrounds listed in Table~\ref{tab:bkgSamples}. 
The input observables to the BDT discriminants are listed in Table~\ref{tab:mvaInputs_1fj}.
The most important aspects of the BDT analysis are discussed in detail below.

\begin{table}[hbpt]\vspace{-10pt}
\centering
\caption{\label{tab:mvaInputs_1fj} List and description of input observables to the BDT discriminants.}
\resizebox{\textwidth}{!}{
\begin{tabular}{c|c|l}
\hline\hline
Type & Observable & Description \\ 
\hline\hline
\multirow{2}{*}{Lepton kinematics}& p$_{\text{T},\ell}$ & Lepton \pt \\
& $|\eta_{\ell}|$ & Lepton $|\eta|$ \\
\hline
\multirow{5}{*}{Jet kinematics} & p$_{\text{T},\text{j}}$ & \pt of the leading narrow jet that is not part of any fat jet \\
& $|\eta_{\text{j}}|$ & $|\eta|$ of the leading narrow jet that is not part of any fat jet \\
& m$_{\text{j}}$ & Mass of the leading narrow jet that is not part of any fat jet \\
& p$_{\text{T},\text{fj}}$ & \pt of the leading fat jet \\
& $|\eta_{\text{fj}}|$ & Leading fat jet $|\eta|$ \\
\hline
\multirow{3}{*}{Jet substructure} & m$_{\text{SD}}$ & Soft Drop mass of the leading fat jet  \\
& $\tau_{21}$ & 2-to-1 subjettiness ratio of the leading fat jet \\
& $\tau_{32}$ & 3-to-2 subjettiness ratio of the leading fat jet \\
\hline
 & & \\[-10pt]
\multirow{8}{*}{Event kinematics} & \ptmiss & Missing transverse momentum \\
& \mT & Transverse mass of the lepton--\ptmiss system \\
& \HT & Scalar \pt sum of non-overlapping narrow and fat jets \\
& N$_{\text{jets}}$ & Number of narrow jets that are not part of the fat jet \\
& N$_{\text{b-jets}}$ & No. of narrow b-jets that are not part of any fat jet \\
& $H_{10}$ & tenth order Fox--Wolfram moment~\cite{Fox:1978vw, Bernaciak:2012nh} \\
& S & Sphericity obtained from the eigenvalues of the momentum tensor~\cite{ATLAS:2012tch} \\
& A & Aplanarity obtained from the eigenvalues of the momentum tensor~\cite{ATLAS:2012tch} \\
\hline\hline
\end{tabular}
}
\end{table}

\begin{figure}[hbpt]\vspace{-10pt}
\centering
\includegraphics[width=0.47\textwidth]{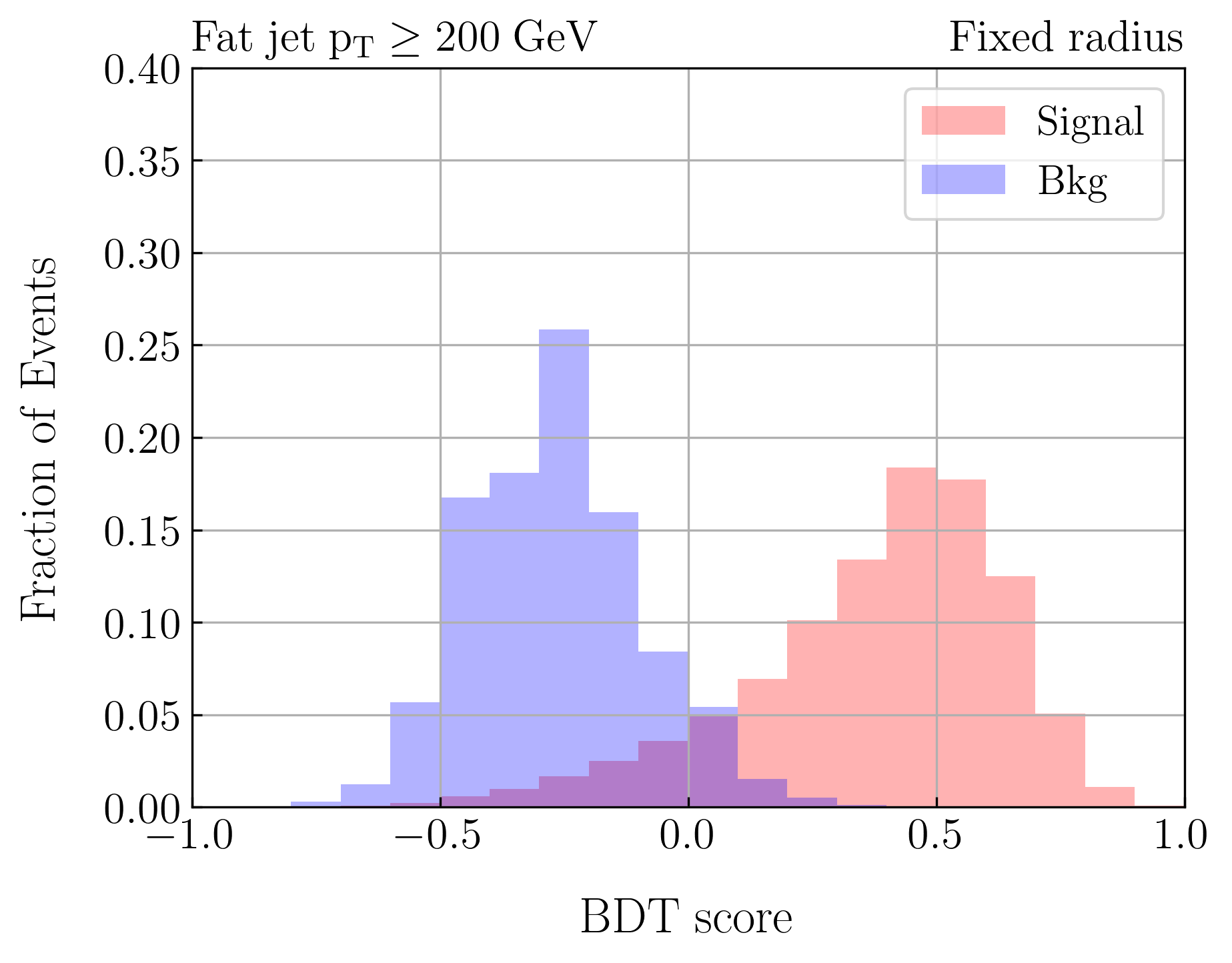}\hfill
\includegraphics[width=0.48\textwidth]{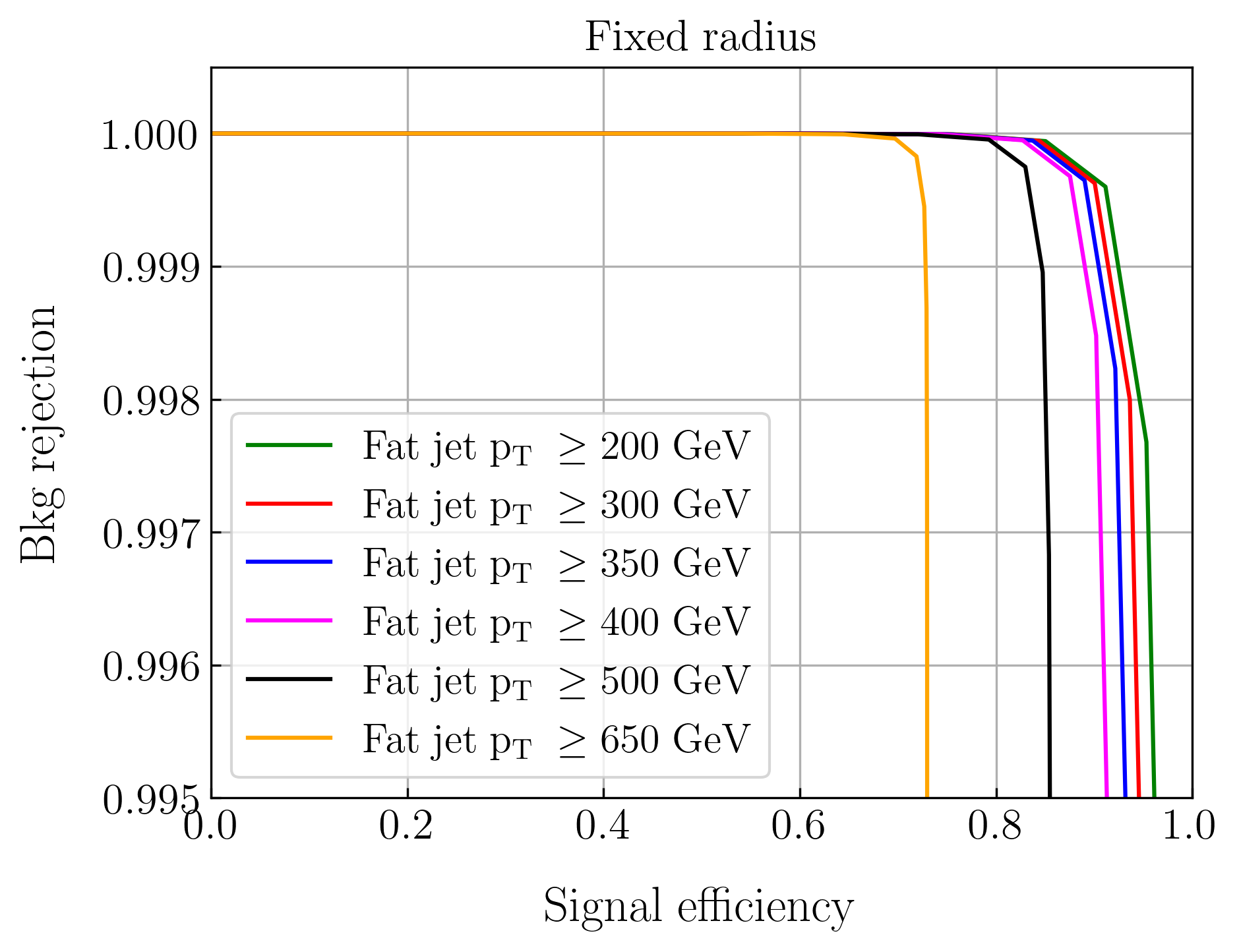}\\
\includegraphics[width=0.47\textwidth]{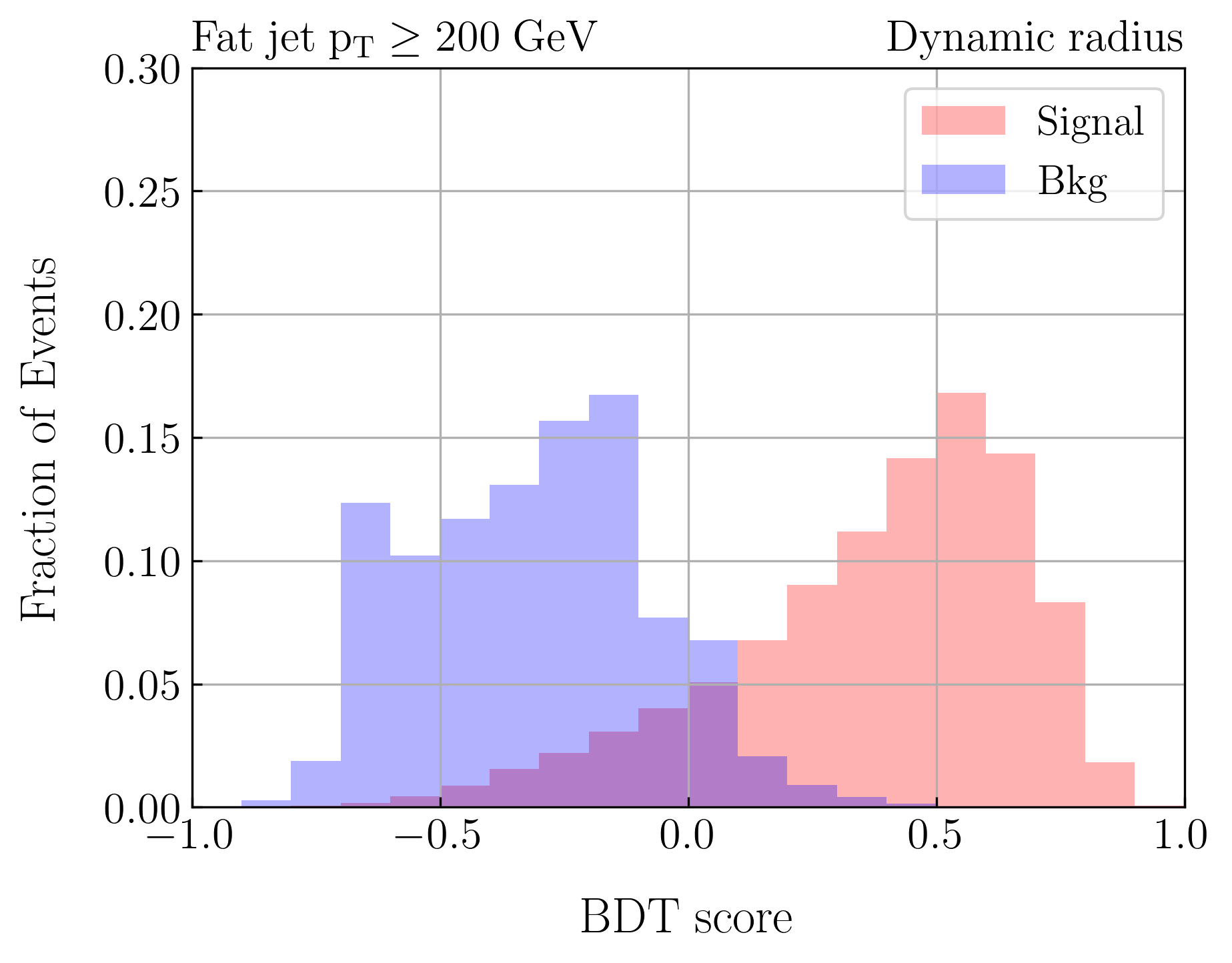}\hfill
\includegraphics[width=0.48\textwidth]{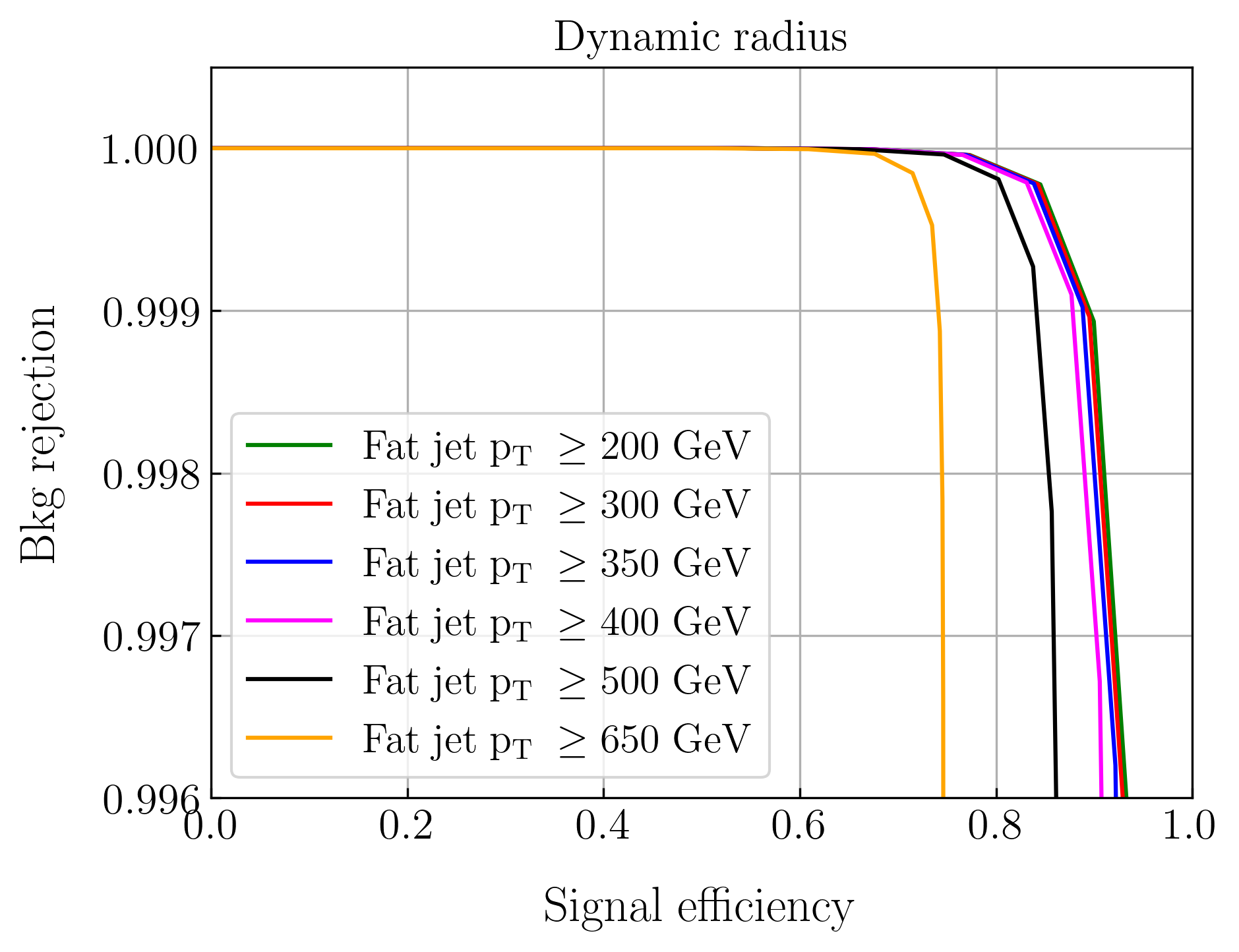}\\
\vspace{-8pt}
\caption{\label{fig:bdt}BDT scores (left) along with the ROC curves (right) for fixed-radius (upper) and dynamic radius (lower) jets. The BDT score distributions are obtained for the nominal fat jet \pt threshold of $200$ GeV (see Section~\ref{sec:evtsel}), which corresponds to the green curves on the right column on each row. The other ROC curves are obtained by applying an additional criterion on the fat jet \pt as indicated in the legend.}
\end{figure}

\begin{description}
\item[Choice of inputs]{
A variety of jet substructure and event observables have been studied for this purpose, some of which are shown in Figures~\ref{fig:substr1}--\ref{fig:1fj_evtshape}. 
However, a subset of these is chosen as inputs to the BDT discriminants since there are large correlations between some of the pairs. 
Observables that were highly correlated with each other were removed from the list of inputs to keep only the less correlated ones. 
For example, none of the ECFs or their ratios have been used as the inputs to the BDT discriminants because they would be highly correlated with $N$-subjettiness and their ratios. 
Different observables have different capabilities of discriminating the signal process from the SM backgrounds, which can be broadly classified into two types, namely reducible and irreducible. 
The backgrounds having top quarks in the final state compose the irreducible type, while the others constitute the reducible type.
Boosted decision trees area fully tree-based, human-readable structure with a finite maximum depth that selects per tree only a few from the input observables list, which provide the best discrimination between signal and backgrounds. 
Thus, using a relatively large number of observables as inputs to the BDTs does not introduce any bias; instead it mostly increases the computational cost due to the residual correlations among the observables, even in the worst case where the performance is not significantly improved by including an observable. 
This strategy has been widely adopted in experimental searches such as Ref.~\cite{CMS:2022fck} as well as in phenomenological studies such as Ref.~\cite{Ghosh:2025gue}.
The input observables are ranked according to their so-called ``importance'' to the BDT discriminants as shown in Appendix~\ref{sec:mvaInputs}. The ``importance'' of an observable is derived by counting how often that observable has been used to split decision tree nodes, and by weighting each split occurrence by the separation gain-squared it has achieved and by the number of events in the node~\cite{Breiman:2017lcz}.
}
\item[Training setup]{
The simulated events are separated into two independent sub-samples, one for training and the other for validation, in $70\%:30\%$ ratio.
An event is picked randomly to belong to any one of these sub-samples.
The adaptive boosting algorithm~\cite{Freund:1997xna} implemented within the TMVA package~\cite{Hocker:2007ht} has been used to train the discriminants, keeping the learning rate, minimum node size, and maximum depth set to $0.5$, $2.5\%$, and $3$, respectively.
Two independent trainings are performed, one for fixed-radius jets and another for DR jets, with exactly the same settings. 
In each training, samples are selected after applying a \pt threshold of $200$~GeV on the leading fat jet as indicated in Section~\ref{sec:evtsel}.
Each process is weighted according to its purity, defined as $\frac{N_i}{\sum_i N_{i}}$ with $i$ running over the signal and background processes, during training.
Checks for overtraining are performed by comparing the respective BDT scores of signal and background between training and validation sub-samples. 
The corresponding Kolmogorov-Smirnov probabilities~\cite{Chakravarti_Laha_Roy_1967} range between $32\%$ -- $97\%$, indicating no overtraining. 
In the left column of Fig.~\ref{fig:bdt}, we show the BDT score distributions of signal and background corresponding to fixed-radius (upper left) and DR (lower left) algorithms for the nominal event selection. 
}
\item[Performance evaluation]{ 
We evaluate the BDT scores also for events passing different thresholds on the leading fat jet \pt, and the performances of the BDT discriminants are then studied via the receiver-operator characteristic (ROC) curves for each of these thresholds separately, along with the nominal case. 
This exercise is primarily aimed at understanding the sensitivity in the high fat jet \pt regime. 
The ROC curves corresponding to different fat jet \pt thresholds are shown in the right column of Fig.~\ref{fig:bdt} for fixed-radius (upper right) and DR (lower right) clustering methods.
The BDT performance falls off less rapidly for tighter thresholds on fat jet \pt in the case of DR jet clustering as indicated  by the ROC curves in Fig.~\ref{fig:bdt}.
}
\end{description}

\begin{figure}[hbpt]\vspace{-8pt}
\centering
\includegraphics[width=0.45\textwidth]{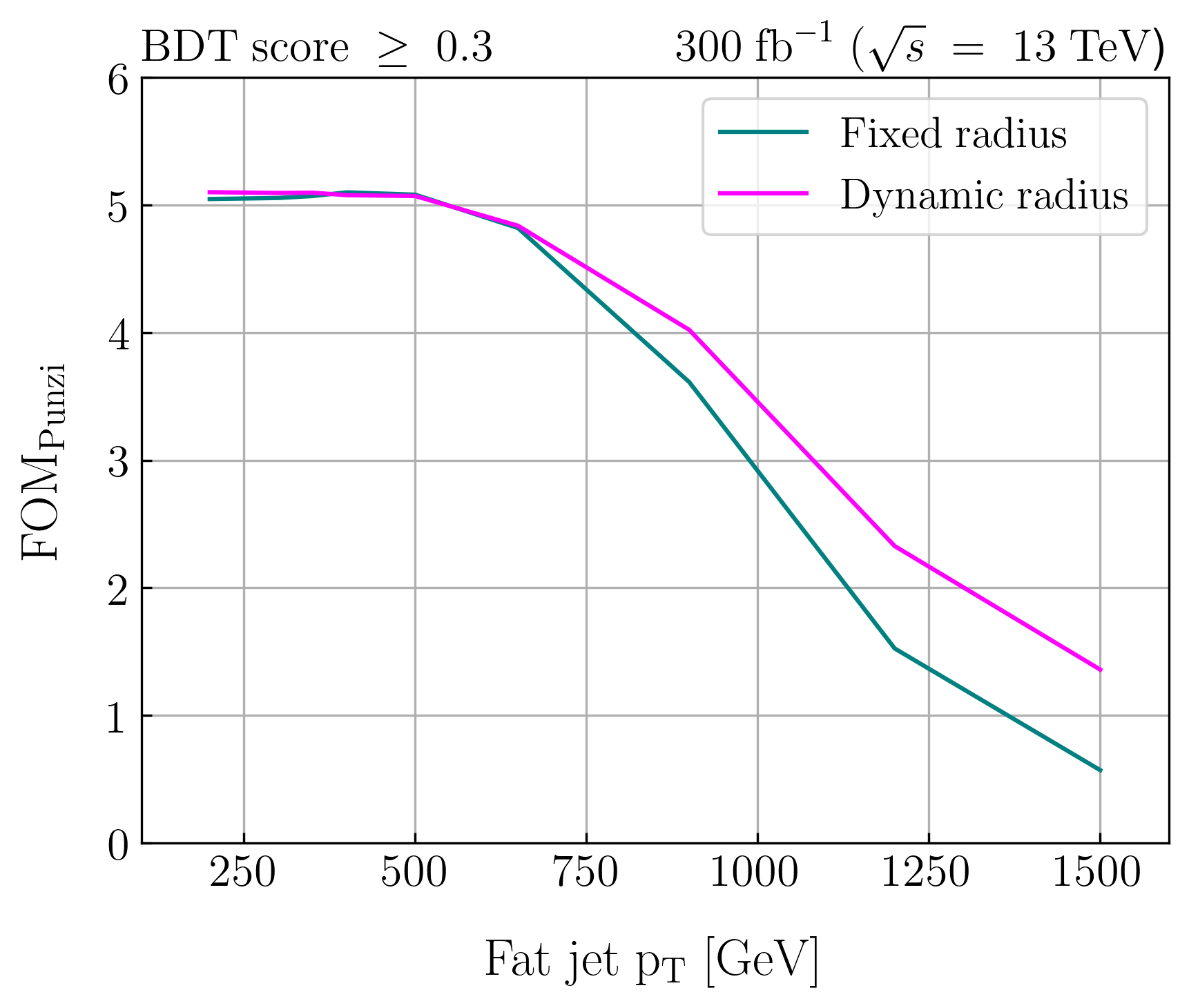}\hfill
\includegraphics[width=0.45\textwidth]{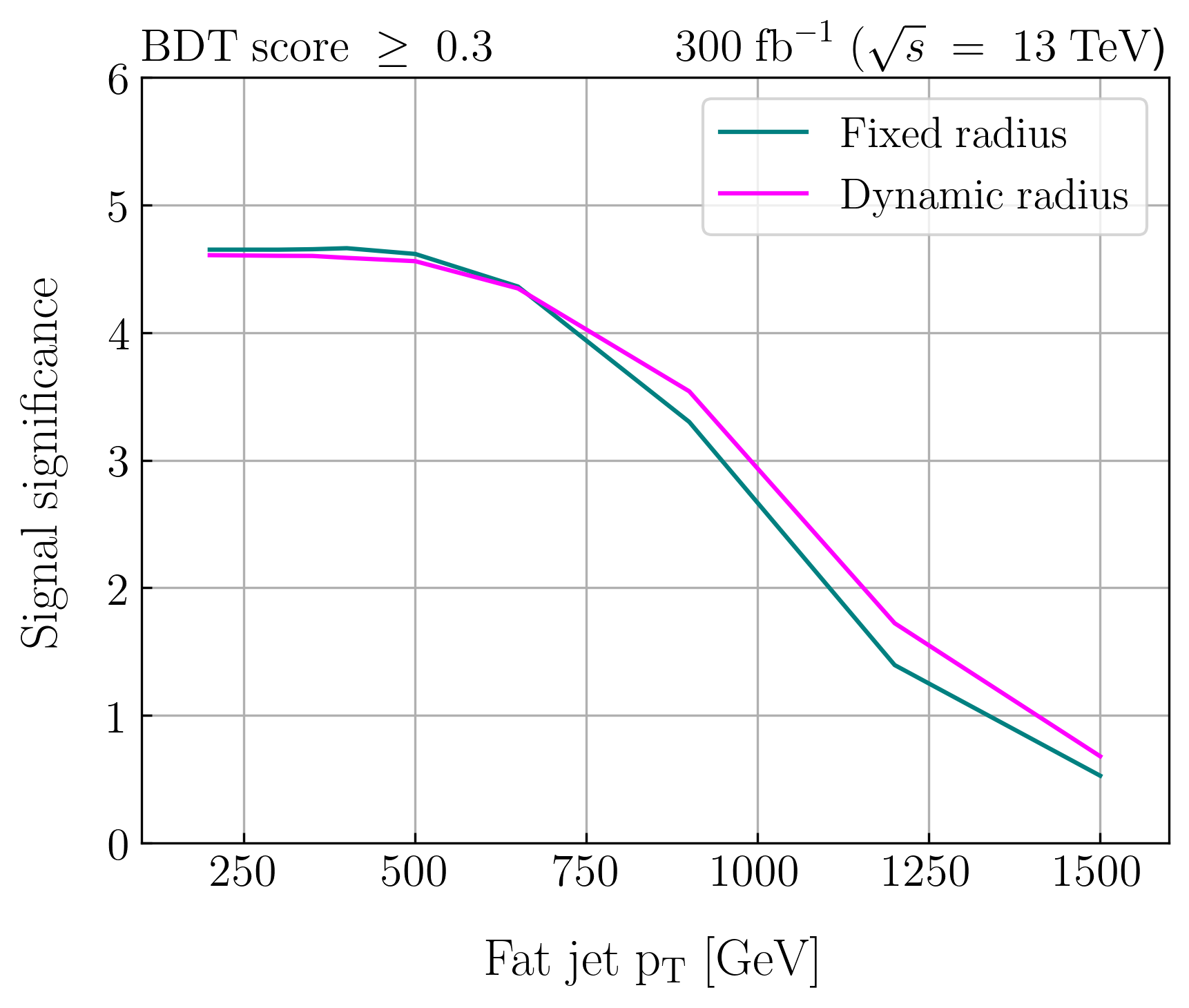}\vspace{-8pt}
\caption{\label{fig:significance}Comparison of $\text{FOM}_{\text{Punzi}}$ (left) and signal significance (right) between fixed-radius and dynamic radius clustering methods as a function of the fat jet \pt threshold for the benchmark scenario defined in Eq.~(\ref{eq:signal_params}).}\vspace{-16pt}
\end{figure}

\subsection{\label{ssec:FOM_limit}Signal sensitivity and interpretation}
\begin{figure}[h]
\vspace{-8pt}
\centering
\includegraphics[width=0.33\textwidth]{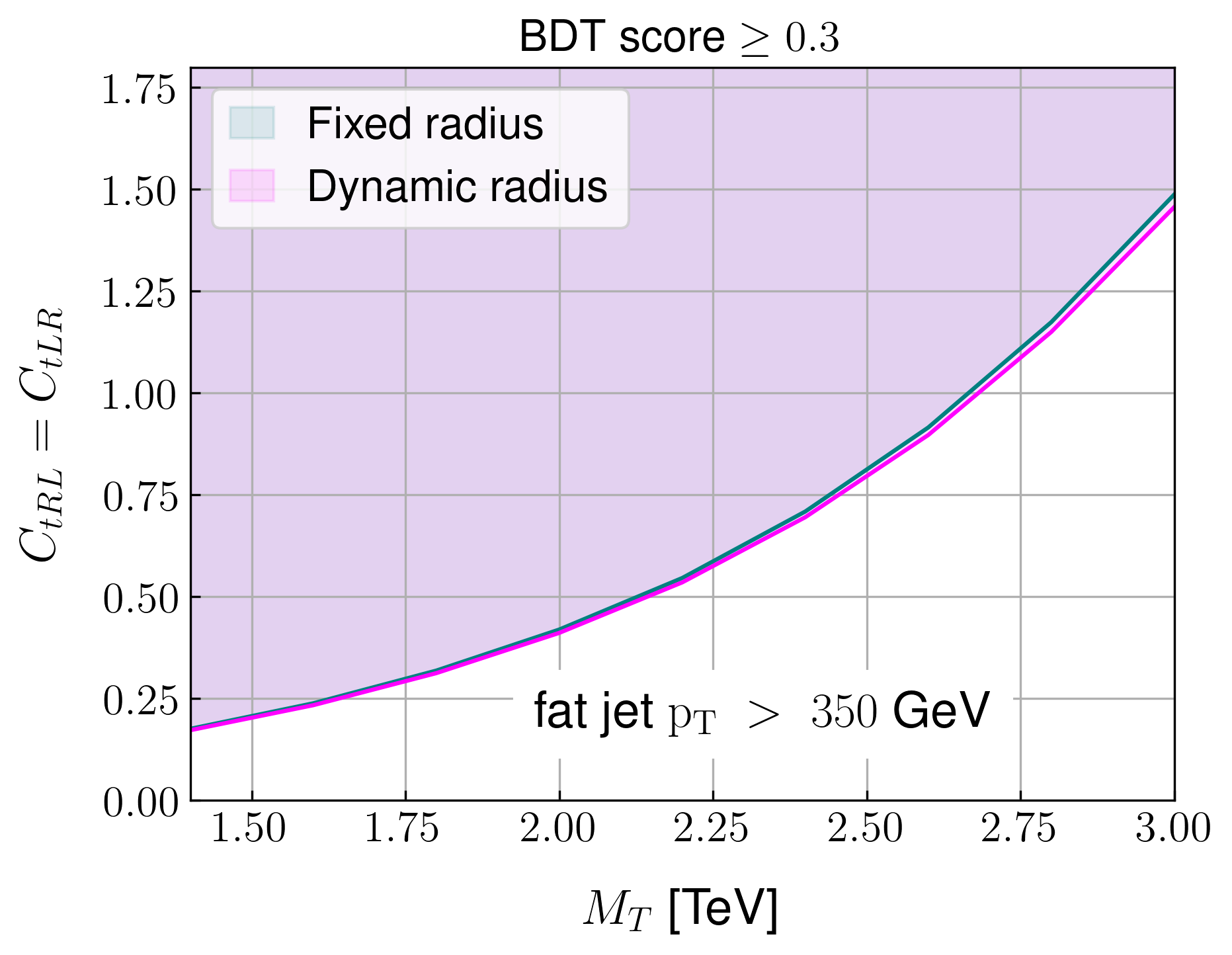}\hfill
\includegraphics[width=0.33\textwidth]{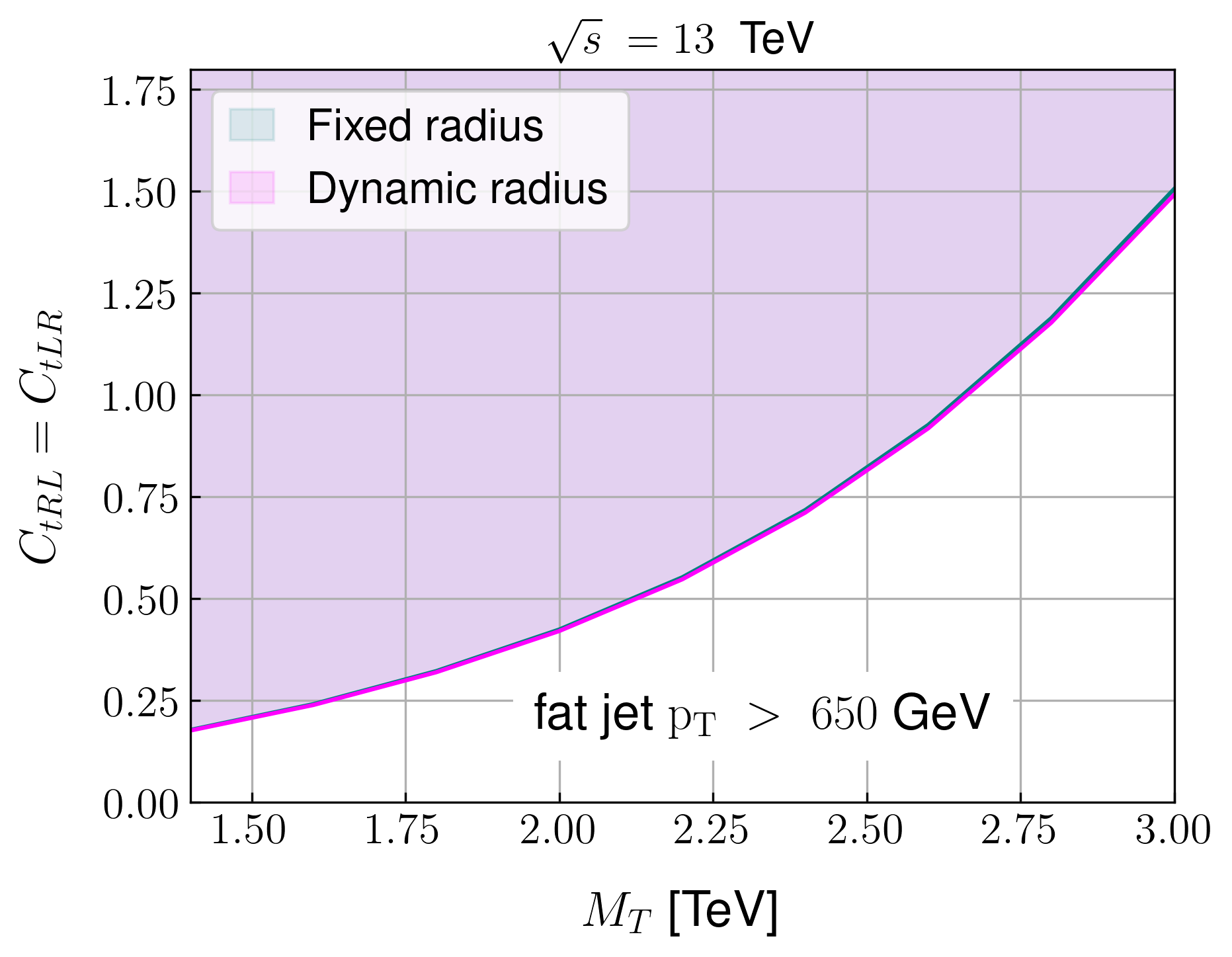}\hfill
\includegraphics[width=0.33\textwidth]{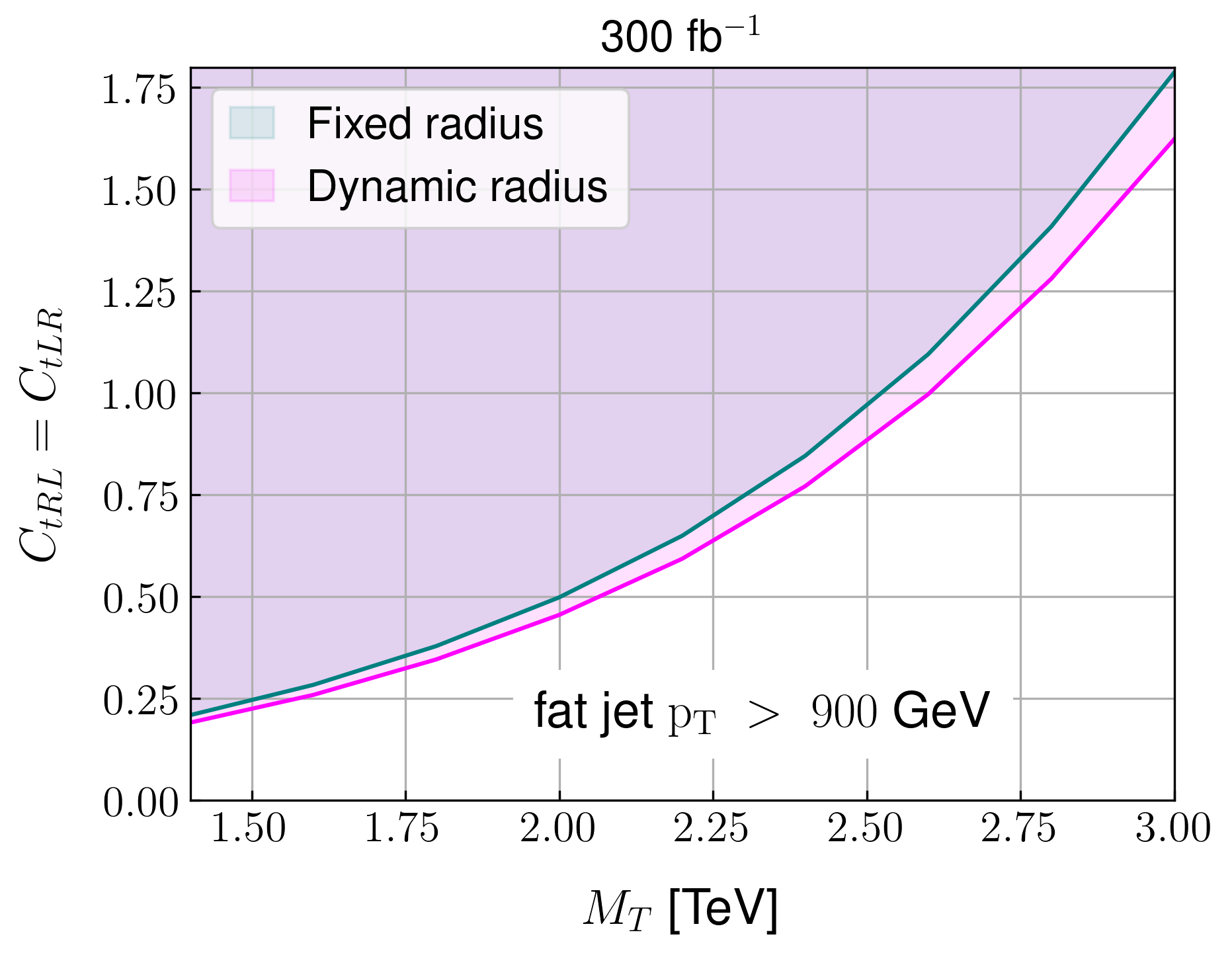}\\\vspace{-8pt}
\caption{\label{fig:upperLimits}Expected $2\sigma$ upper limit on the chromomagnetic coupling $C_{tRL} = C_{tLR}$ as a function of the top partner mass. The three panels correspond to three different fat jet \pt thresholds. For these plots, we use $\Lambda = 8~$TeV, and all other parameters are kept the same as in Eq.~(\ref{eq:signal_params}). Shaded regions indicate the parameter space excluded by our analysis, assuming no signal is observed.}
\end{figure}

\vspace{-2mm}
The BDT discriminants are then utilized for a comparison of signal sensitivities followed by interpretations in terms of expected limits on the plane spanned by $T$ quark mass and its chromomagnetic coupling, corresponding to the two clustering approaches. 
\begin{description}
\item[Comparative study of signal sensitivities]{
Selection thresholds on the BDT scores are optimized based on the so-called ``Punzi figure-of-merit'' ($\text{FOM}_{\text{Punzi}}$) defined in Ref.~\cite{Punzi:2003bu}. 
The signal sensitivity is characterised via two different metrics, namely the signal significance defined as $\frac{S}{\sqrt{S+B}}$ and the $\text{FOM}_{\text{Punzi}}$, to check the consistency and robustness of the conclusion.
The variation of these metrics are studied with the fat jet \pt threshold for the fixed-radius and DR jet clustering methods after applying the optimized selection on the BDT scores $\geqslant 0.3$.  
From Figure~\ref{fig:significance}, it is evident that the DR jet algorithm outperforms the fixed-radius one in the {\bf highly boosted} (fat jet $\pt \geqslant 750~$GeV) region of the phase space. 
The DR algorithm shows higher sensitivity towards large $M_{T}$ values (see Fig.~\ref{fig:top_partner_kin}) compared to the fixed-radius algorithm, allowing for more phase space to hunt for the signal process.
}
\item[Interpretation via expected limits]{ 
The result of the signal sensitivity study is interpreted by calculating the expected $2\sigma$ upper limits on the chromomagnetic couplings, assuming $C_{tRL} = C_{tLR}$, for the top partner masses in the range 1.4 TeV to 3 TeV. 
For each mass point, the expected upper limit has been obtained by varying the coupling strength such that the signal significance reaches $2\sigma$. 
Figure~\ref{fig:upperLimits} shows the resulting upper limits for the three different fat jet \pt thresholds. 
In these plots, the shaded regions correspond to values of the coupling excluded by our analysis at $95\%$ CL under the assumption of no signal observation. 
The upper limits are estimated for both fixed-radius and DR jet clustering methods. 
In all cases, the expected limit for the DR clustering is smaller than that for the fixed radius one, indicating that the analysis with DR jets has higher sensitivity to smaller coupling values. 
The improvement is more prominent for the fat jet \pt threshold of 900 GeV, suggesting that the DR jet clustering is particularly effective for reconstructing signal events in the high fat jet \pt regime, as indicated earlier in Section~\ref{sec:evtsel}.
}
\end{description}

\vspace{-24pt}
\subsection{\label{ssec:syst}Discussions on the dominant systematic effects}\vspace{-2mm}
The dominant systematic effect on the signal sensitivity stems from the inaccuracy of the jet energy estimation.
The jet energy is primarily determined by calorimeter measurements for high \pt jets. 
The jet energy scale (JES) plays a significant role at high \pt due to partial leakage of high-energy hadronic showers beyond the longitudinal depth provided by the calorimeter that leads to an inaccurate estimation of the overall jet energy. 
It is likely to have a similar impact for both jet clustering methods discussed here since it is entirely dependent on the incident energy and the leakage fraction of the hadronic shower produced by the jet constituents, irrespective of the clustering method. 
However, the uncertainty due to jet energy resolution is expected to increase for DR clustering since more particles are likely to be accumulated within the jet, owing to its larger capture radius, thereby leading to higher fluctuations in the measured jet energy.

%% file: summary.tex
In this work, we performed a thorough investigation of collider prospects of a vectorlike top partner produced in association with a standard model top quark via a chromomagnetic coupling. 
The top partner is assumed to decay into the following channels: tZ, th, tg, or bW. 
To effectively suppress the huge QCD background, we focused on semileptonic final states, i.e., events with exactly one isolated lepton arising either from the top quark or from the heavy boson decay. 
Depending on the decay modes of the top partner, the final state could contain one or two fat jets along with a single lepton and b-tagged jet(s). 
We performed the analysis in the event category that contains a single fat jet (1$\ell$\,+\,1 fat jet\,+\,$\geqslant 1$ b-tagged jet) due to its comparatively better signal-to-background ratio.

We carried out a detailed collider analysis at the $\sqrt{s} = 13~$TeV with an integrated luminosity of 300 fb$^{-1}$, incorporating realistic pileup conditions.
For signal events, the {\sc Madgraph5\_aMC@NLO}\,+\,{\sc Pythia8}\,+\,{\sc Delphes} simulation chain has been employed; while for background estimation, appropriate simulated samples from CMS Open Data are utilized.
To enhance signal discrimination, we utilized state-of-the-art jet substructure tools such as soft drop grooming and $N$ subjettiness, along with several event shape observables like Fox-Wolfram moments, Sphericity, and Aplanarity.
Multivariate discriminants based on boosted decision trees are developed for better signal-to-background separation. 
We then optimized a selection thresholds on these discriminants based on the Punzi figure-of-merit to improve signal sensitivity. 
The result is interpreted as the upper limit on the chromomagnetic coupling between the top quark and its heavy vectorlike partner. 
Assuming the cut off scale of $\Lambda = 8$ TeV, the upper limit on the coupling is found to lie within 0.2 to 1.5 for the top partner mass range of 1.4 to 3 TeV.

An important aspect of our work is a comparative performance analysis between conventional fixed-radius vs modern dynamic-radius clustering of anti-\kt jets under realistic pileup conditions.
We observe that the dynamic-radius approach enhances the signal sensitivity, particularly in regions with high jet transverse momentum, allowing for tighter bounds on the chromomagnetic coupling. 
This suggests that the dynamic radius jet clustering is more effective in probing high-energy regimes compared to the conventional fixed-radius method.

The utility of DR clustering is expected to become even more pronounced at the future high-energy colliders such as HL-LHC and FCC-hh. At the HL-LHC, the anticipated $10$ times higher integrated luminosity will significantly enhance the signal event yield in the high-\pt$(T)$ tail. Therefore, the statistical sensitivity in this phase space region is expected to improve at HL-LHC. The FCC-hh collider, which is expected to operate at the proposed centre-of-mass energy of $\sim\! 100~$TeV, will further extend the reach of high-\pt$(T)$ tail along with substantial enhancement in the partonic cross section for pp $\to T$t production.  As a result, the heavy top partner and its decay products are expected to acquire higher transverse momenta, where the DR clustering approach is much more effective compared to the fixed radius approaches. These features make DR jets especially well-suited for future collider environments with jets having extreme boosts. This method, when combined with advanced machine-learning-based tagging algorithms for identifying boosted heavy resonances, becomes increasingly relevant for higher sensitivity.

There remain various scopes for further improvements. Optimization of pile up contamination removal working points, as well as determining jet energy correction and resolution accurately for the dynamic-radius clustering approach, can be extremely useful. 
This might put even tighter bounds on the couplings than the existing ones.
Future directions also include exploring the integration of DR jets with particle-based or graph neural-network-based tagging algorithms for boosted heavy resonance identification, which can reject backgrounds more efficiently and align well with current machine-learning trends in collider physics. Moreover,  adopting a hybrid strategy that combines fixed- and dynamic-radius clustering depending on the fat jet's transverse momentum can potentially enhance the reach and robustness of future studies.

%% file: acknowledgement.tex
The authors sincerely acknowledge the organizers of the WHEPP XVII, 2023, for providing a platform to meet and initiate this work. 
The authors acknowledge the organizers, instructors, and facilitators of the CMS Open Data Workshop at WHEPP XVII, 2023, for the necessary training to access and process CMS Open Data samples. 
The authors are thankful to the computing facility at IISER Kolkata for the computational and data storage resources.

%% file: ref.bbl
\providecommand{\href}[2]{#2}\begingroup\raggedright\endgroup

%% file: app.tex
\section{\label{sec:massPoints}Cross sections and branching ratios for different mass hypotheses}
The signal production rates and decay branching ratios for the chosen mass points are compiled in Table~\ref{tab:vlq_cross_section_br}.
\begin{table}[hbpt]
%\vspace{-20pt}
\caption{\label{tab:vlq_cross_section_br}Cross sections and branching ratios of various decay modes for different top partner masses. All other parameters are kept the same as in the benchmark point defined in Eq.~(\ref{eq:signal_params}).}
%\resizebox{\textwidth}{!}{
\begin{tabular}{c|c|c|c|c|c|c}
\hline\hline
 & & & & & & \\[-8pt]
$M_T$ [TeV] & $\sigma(pp \to T\overline{t} + \overline{T} t)$ [fb]& $\Gamma_T$ [GeV] & $\mathcal{B}(T\to t Z$) & $\mathcal{B}(T\to t h$) & $\mathcal{B}(T\to t g$) & $\mathcal{B}(T\to b W$) \\
 & & & & & & \\[-8pt]
\hline\hline
1.4 & 139.81 &  18.48 & 41.58\% & 9.23\% & 31.68\% & 17.52\% \\
1.6 &  78.24 &  27.67 & 41.95\% & 9.09\% & 31.49\% & 17.47\% \\
1.8 &  45.04 &  39.43 & 42.24\% & 9.00\% & 31.31\% & 17.45\% \\
2.0 &  26.51 &  54.09 & 42.47\% & 8.94\% & 31.13\% & 17.45\% \\
\bf 2.2 & \bf 15.96 & \bf 71.96 & \bf 42.67\% & \bf 8.90\% & \bf 30.97\% & \bf 17.46\% \\
2.4 &   9.72 &  93.36 & 42.84\% & 8.88\% & 30.81\% & 17.47\% \\
2.6 &   6.04 & 118.58 & 42.98\% & 8.86\% & 30.67\% & 17.49\% \\
2.8 &   3.82 & 147.94 & 43.11\% & 8.85\% & 30.53\% & 17.51\% \\
3.0 &   2.45 & 181.76 & 43.22\% & 8.84\% & 30.41\% & 17.53\% \\
\hline\hline
\end{tabular}
%}
\end{table}

%\clearpage
%\newpage
%\vspace{-3mm}
\section{\label{sec:mvaInputs}Input Observables Importance for Multivariate Analysis}
\begin{figure}[hbpt]
\centering
\includegraphics[width=0.5\textwidth]{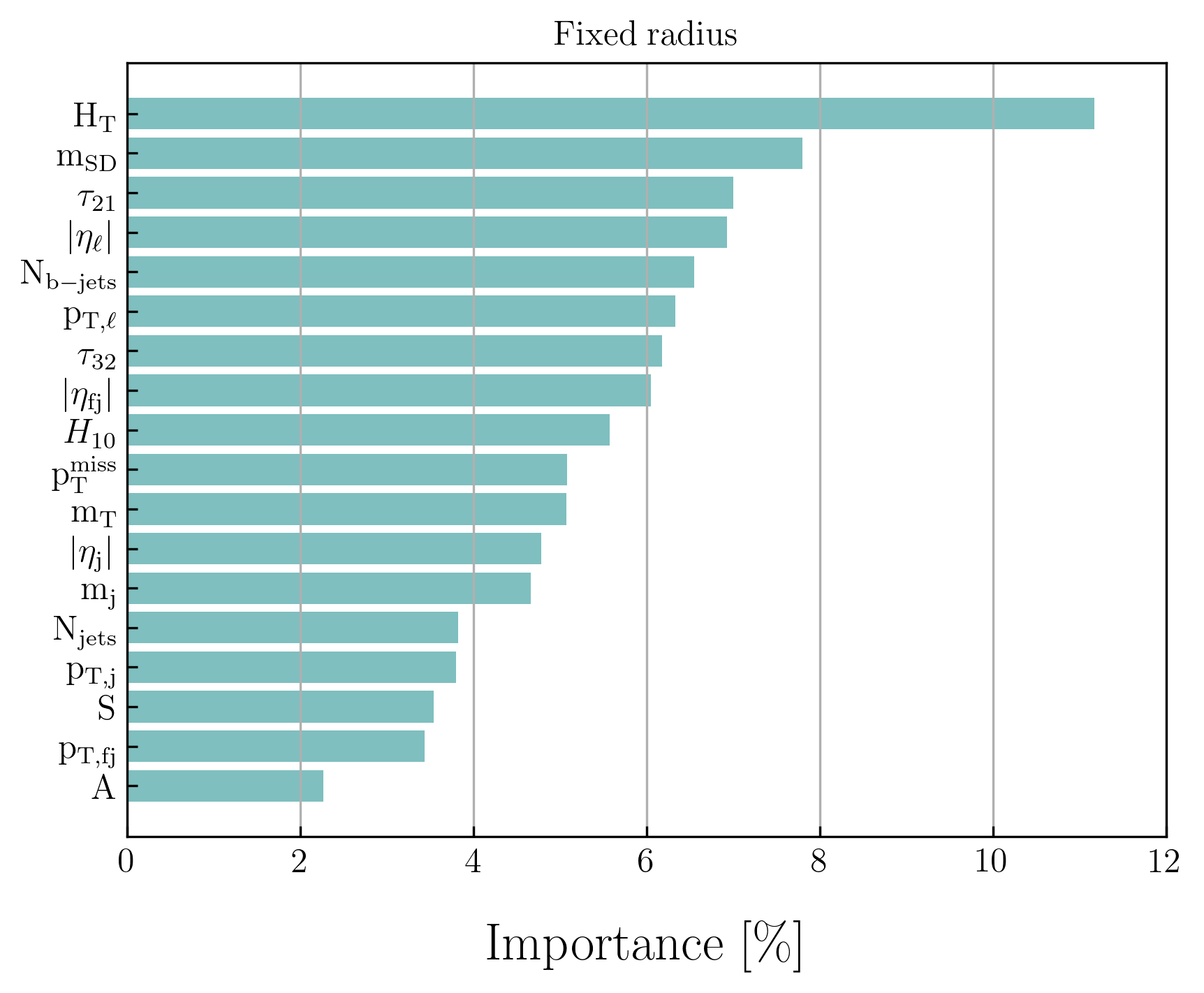}\hfill
\includegraphics[width=0.5\textwidth]{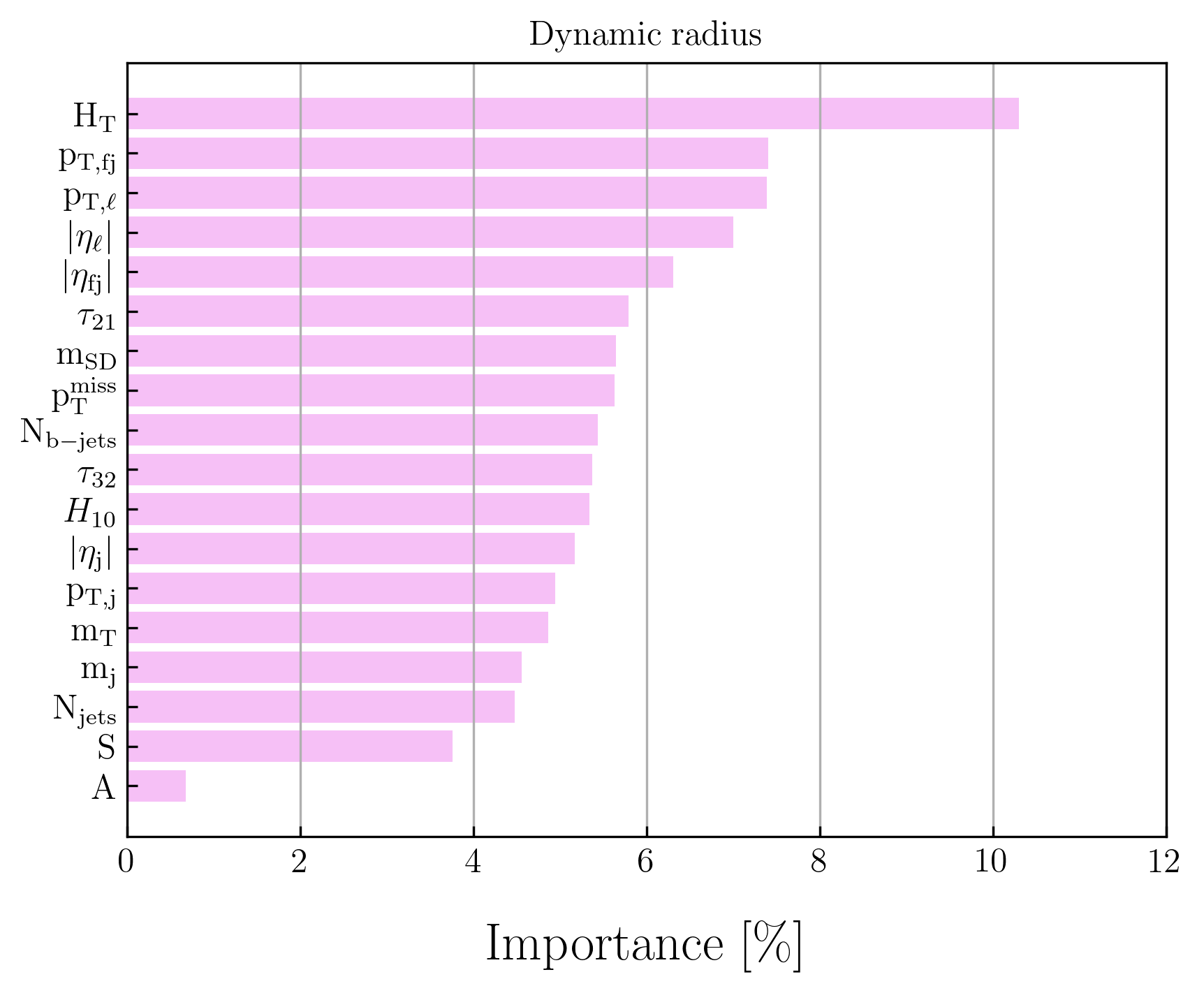}\\
\caption{\label{fig:importance}Importance of input observables to the BDT discriminants corresponding to fixed- (left) and dynamic (right) radius jet clustering.}
\end{figure}

%\vspace{-16pt}
Figure~\ref{fig:importance} illustrates the ranking of the input variables according to their impact on the BDT performance for the fixed and dynamic radius jet clustering approaches.
